\documentclass[a4paper,11pt]{article}
\usepackage{heppub}
\pdfoutput=1
\usepackage{placeins}
\graphicspath{{plots/}}

\newcommand{\abs}[1]{\lvert#1\rvert}
\newcommand{\ord}[1]{\mathcal{O}(#1)}

\newcommand{\ORd}[1]{\mathcal{O}\Bigl(#1\Bigr)}

\newcommand{\Ordsq}[1]{\mathcal{O}\bigl[#1\bigr]}

\newcommand{\df}{\mathrm{d}}
\newcommand{\img}{\mathrm{i}}

\newcommand{\w}{\omega}

\newcommand{\GeV}{\,\mathrm{GeV}}

\newcommand{\nn}{\nonumber}

\newcommand{\bn}{{\bar n}}
\newcommand{\bq}{{\bar q}}

\newcommand{\bt}{{\vec b}_T}
\newcommand{\qt}{{\vec q}_T}

\newcommand{\cC}{\mathcal{C}}
\newcommand{\cI}{\mathcal{I}}
\newcommand{\cJ}{\mathcal{J}}
\newcommand{\cL}{\mathcal{L}}
\newcommand{\cO}{\mathcal{O}}
\newcommand{\cP}{\mathcal{P}}
\newcommand{\cS}{\mathcal{S}}
\newcommand{\cV}{\mathcal{V}}
\newcommand{\Tau}{\mathcal{T}}

\newcommand{\tB}{\tilde{B}}
\newcommand{\tI}{\tilde{I}}
\newcommand{\tS}{\tilde{S}}

\newcommand{\tcI}{\tilde{\cI}}

\newcommand{\as}{\alpha_s}
\newcommand{\Ecm}{E_\mathrm{cm}}
\newcommand{\GammaC}{\Gamma_\mathrm{cusp}}
\newcommand{\lqcd}{\Lambda_\mathrm{QCD}}
\newcommand{\MSbar}{$\overline{\text{MS}}$}

\newcommand{\id}{\mathbf{1}}

\newcommand{\zero}{{(0)}}
\newcommand{\one}{{(1)}}
\newcommand{\two}{{(2)}}
\newcommand{\three}{{(3)}}
\newcommand{\central}{\mathrm{central}}
\newcommand{\cut}{\mathrm{cut}}
\newcommand{\thr}{\mathrm{thr}}
\newcommand{\sing}{\mathrm{sing}}
\newcommand{\sub}{\mathrm{sub}}
\newcommand{\off}{\mathrm{off}}

\newcommand{\scetlib}{{\tt SCETlib}}
\newcommand{\geneva}{{\tt Geneva}}

\newcommand{\WidthTwoSubfigs}{0.48\textwidth}

\allowdisplaybreaks[3]

\title{A Toolbox for $q_{T}$ and 0-Jettiness Subtractions at N$^3$LO\hspace*{-1ex}}

\author[a]{Georgios Billis,}
\emailAdd{georgios.billis@desy.de}

\author[b]{Markus A.~Ebert,}
\emailAdd{ebert@mit.edu}

\author[a]{Johannes K.~L.~Michel,}
\emailAdd{johannes.michel@desy.de}

\author[a]{and Frank J.~Tackmann}
\emailAdd{frank.tackmann@desy.de}

\affiliation[a]{Theory Group, Deutsches Elektronen-Synchrotron (DESY), D-22607 Hamburg, Germany}
\affiliation[b]{Center for Theoretical Physics, Massachusetts Institute of Technology, Cambridge, MA 02139, USA}

\abstract{%
We derive the leading-power singular terms at three loops for both $q_T$ and
0-jettiness, $\Tau_0$, for generic color-singlet processes. Our results provide
the complete set of differential subtraction terms for $q_T$ and $\Tau_0$
subtractions at N$^3$LO, which are an important ingredient for matching N$^3$LO
calculations with parton showers. We obtain the full three-loop structure of the
relevant beam and soft functions, which are necessary ingredients for the
resummation of $q_T$ and $\Tau_0$ at N$^3$LL$'$ and N$^4$LL order, and which
constitute important building blocks in other contexts as well. The
nonlogarithmic boundary coefficients of the beam functions, which contribute to
the integrated subtraction terms, are not yet fully known at three loops. By
exploiting consistency relations between different factorization limits, we
derive results for the $q_T$ and $\Tau_0$ beam function coefficients at N$^3$LO
in the $z\to 1$ threshold limit, and we also estimate the size of the unknown
terms beyond threshold.
}

\date{September 2, 2019}

\preprint{\vbox{%
\hbox{DESY 19-145}
\hbox{MIT--CTP/5140}
}}

\begin{document}

\maketitle

\section{Introduction}
\label{sec:intro}

The ever increasing precision of experimental measurements at the LHC requires
equally precise theoretical predictions in order to be fully exploited.
Color-singlet processes play a central role in the LHC physics program.
The $pp\to Z,W$ Drell-Yan processes are key benchmark processes that have been
measured at the percent level and below~\cite{Aaboud:2017svj, Aaboud:2017ffb, Khachatryan:2016nbe,
CMS-PAS-SMP-17-010}.
Precise measurements of Higgs and diboson processes provide strong
sensitivity to possible contributions beyond the Standard Model~\cite{Sirunyan:2018koj,
ATLAS-CONF-2019-005, ATLAS-CONF-2019-029, Aaboud:2019nkz, Aaboud:2019lgy, Sirunyan:2019gkh}.
They are also important irreducible backgrounds in direct searches for
dark-matter production at the LHC.

The inclusion of higher-order QCD corrections is crucial to obtain reliable predictions.
Depending on the specific process and phase-space region, reducing the current theoretical
uncertainties requires the calculation of the full corrections at the next order in $\alpha_s$
and/or the resummation of the dominant higher-order terms to all orders in $\alpha_s$.
For color-singlet processes, theory predictions are being pushed to the third order
in the fixed-order expansion~\cite{Anastasiou:2014vaa, Li:2014afw, Ahmed:2014cla, Ahmed:2014uya,
Anastasiou:2015ema, Dreyer:2016oyx, Mistlberger:2018etf, Cieri:2018oms, Dulat:2018bfe, Duhr:2019kwi}
as well as in resummed perturbation theory~\cite{Becher:2008cf, Abbate:2010xh,
Hoang:2015hka, Bonvini:2014joa, Schmidt:2015cea, H:2019dcl, Ebert:2017uel,
Chen:2018pzu, Bizon:2018foh}.

A key requirement in both cases is to understand the infrared singular structure of QCD
at N$^3$LO. For fixed-order calculations this is crucial for the cancellation
of infrared divergences between real and virtual emissions, as evidenced by the
variety of methods that have been developed by now at NNLO~\cite{GehrmannDeRidder:2005cm,
Currie:2013vh, Czakon:2010td, Czakon:2014oma, Boughezal:2011jf, Catani:2007vq,
DelDuca:2015zqa, DelDuca:2016csb, Boughezal:2015dva, Gaunt:2015pea, Cacciari:2015jma,
Caola:2017dug, Caola:2019nzf, Herzog:2018ily, Magnea:2018hab, Magnea:2018ebr}.
Resummed predictions are intimately linked to the singular limit, and the
N$^3$LO singular structure is a key ingredient to extend the resummation to
the full three-loop level.

One way to study the infrared singular limit of QCD is to consider a suitable
resolution variable $\tau$, whose differential cross section can be factorized
in the singular limit $\tau\to 0$. In this paper, we consider two such
variables, $0$-jettiness $\Tau_0$ and the total color-singlet transverse
momentum $q_T$, and derive their singular structure to N$^3$LO.
Our results are necessary ingredients for carrying out the resummation for
$q_T$ and $\Tau_0$ at N$^3$LL$'$ and N$^4$LL order.
For the associated $q_T$ and $\Tau_0$ subtraction methods~\cite{Catani:2007vq, Boughezal:2015dva, Gaunt:2015pea}, we provide the complete differential
N$^3$LO subtraction terms in analytic form.
The structure and required ingredients for the $\Tau_0$ subtractions
at N$^3$LO were already discussed in \refcite{Gaunt:2015pea}.
The $q_T$ slicing method at N$^3$LO was also considered in \refcite{Cieri:2018oms}.
Differential $\Tau_0$ subtractions
are the basis of the NNLO$+$PS (parton shower) matching in \geneva~\cite{Alioli:2012fc,
Alioli:2015toa}, and our results are an important ingredient for its extension
to N$^3$LO$+$PS.

To continue our discussion we need to set up some basic notation.
We consider the production of a generic color-singlet final state $L$ in hadronic collisions.
In the singular limit, the only hard interaction process that
contributes is the Born process, which we denote as
\begin{equation} \label{eq:Born_process}
\kappa_a(q_a)\, \kappa_b(q_b)\to L(q)
\quad\text{with}\quad
q_a^\mu + q_b^\mu = q^\mu
\,.\end{equation}
We always use the indices $a$ and $b$ to label the initial states, and
$\kappa_i \in \{g, u, \bar u, d, \bar d, s, \ldots\}$ denotes the parton type and
flavor. When there is no ambiguity we simply identify $\kappa_i \equiv i$,
e.g., we just write $ab\to L$.
The $q_{a,b}^\mu$ are lightlike Born reference (label) momenta given by
\begin{equation} \label{eq:Born}
q_{a,b}^\mu = \w_{a,b} \frac{n_{a,b}^\mu}{2}
\,,\qquad
n_a^\mu \equiv n^\mu = (1, \hat z)
\,, \quad
n_b^\mu \equiv \bn^\mu = (1, -\hat z)
\,,\quad
x_{a,b} = \frac{\w_{a,b}}{\Ecm} = \frac{Q}{\Ecm} e^{\pm Y}
\,,\end{equation}
where $\hat z$ is the beam direction and $E_{\rm cm}$ is the hadronic
center-of-mass energy. The precise definition of the Born momentum fractions
$x_{a,b}$ and the associated $\w_{a,b} = x_{a,b}\Ecm$ depends on how we choose
to parametrize the Born phase space in terms of physical observables. For
definiteness, in \eq{Born} we have chosen the total invariant mass $Q =
\sqrt{q^2}$ and rapidity $Y$ of the color singlet. Other
choices are possible as well, e.g., $\w_{a,b} = q^\mp \equiv \sqrt{Q^2 + q_T^2}
e^{\pm Y}$. In the singular limit, all possible choices are equivalent and yield
the same factorized cross section. The specific choice affects the nonsingular
power-suppressed corrections.

The cross section for color-singlet production for a (suitably factorizable) resolution
variable $\tau$ factorizes for $\tau\to 0$ as~\cite{Collins:1984kg, Stewart:2009yx}
\begin{equation} \label{eq:fact_generic}
\frac{\df\sigma}{\df Q^2 \df Y \df \tau}
= \sum_{a,b} H_{ab}(\w_a \w_b)\, \bigl[B_a(x_a)\otimes B_b(x_b)\otimes S_c \bigr](\tau)
\times \bigl[1 + \ord{\tau}\bigr]
\,.\end{equation}
The convolution structure denoted by $\otimes$ depends on the precise definition of $\tau$.
The key properties of \eq{fact_generic} are that it captures all QCD singularities
in $\tau$ and that it factorizes the dependence on
the underlying process from the dependence on $\tau$.

The process dependence is carried by the hard function
$H_{ab}$, which describes the Born process $ab \to L$, with the sum running over all
relevant parton channels. At lowest order, $H_{ab}^{(0)}$ is equivalent to the
partonic Born cross section for $ab\to L$. At higher orders, it encodes the
finite virtual corrections to the Born process
and thus can be obtained from the corresponding quark or gluon form factors.
Results at three loops are known for $gg \to H$ in the $m_t\to\infty$ limit,
$b\bar{b} \to H$, and Drell-Yan production~\cite{Kramer:1986sg, Matsuura:1987wt, Matsuura:1988sm, Harlander:2000mg, Harlander:2003ai, Gehrmann:2005pd,
Moch:2005tm, Moch:2005id, Ravindran:2006cg, Baikov:2009bg, Lee:2010cga, Gehrmann:2010ue, Anastasiou:2011qx, Gehrmann:2014vha}.
Explicit expressions for the hard functions in our notation can be found in \refcite{Ebert:2017uel}.
The hard function also encodes any additional cuts or measurements on the constituents of $L$, which we keep implicit.

The entire $\tau$ dependence in \eq{fact_generic} is encoded by the beam and soft functions.
The beam functions $B_{a,b}$ describe collinear emissions from the incoming
partons $a$ and $b$, while the soft function $S_c$ encodes soft radiation
between them. They are universal objects that do not depend on the details of the hard process.
Namely, $B_i$ only depends on the type of its incoming parton $i \equiv
\kappa_i$, while $S_c$ only depends on the color channel of the Born
process. In our case, the only possible color channels are $c = \{q\bar q, gg\}$,
which are equivalent to the color representation of the incoming partons,
so we simply label it by $c \equiv i = \{q, g\}$.

The beam and soft functions do depend on the definition of $\tau$, which also
determines their convolution structure in \eq{fact_generic}.
They can be formally defined as renormalized operator matrix elements in
soft-collinear effective theory (SCET)~\cite{Bauer:2000ew, Bauer:2000yr, Bauer:2001ct,
Bauer:2001yt, Bauer:2002nz}.
The beam and soft functions relevant for $\Tau_0$ and $q_T$ are the most basic
of their type, measuring the small light-cone momentum or the total transverse momentum of the
inclusive sum of all collinear and soft emissions, respectively.
For this reason, they are important objects in
their own right, encoding fundamental properties of the singular structure of
QCD, and also appear in a variety of other contexts. In particular, they often
serve as building blocks for constructing the beam and soft functions necessary
for more complicated scenarios or observables, see
e.g.~\refscite{Jouttenus:2011wh, Kasemets:2015uus, Bertolini:2017efs, Bell:2018oqa,
Jain:2011iu, Procura:2014cba, Lustermans:2019plv, Gaunt:2014xxa,
Becher:2013xia, Stewart:2013faa, Gangal:2016kuo, Michel:2018hui, Pietrulewicz:2017gxc}.

In this paper, we derive the analytic structure of the $\Tau_0$ and $q_T$ beam
and soft functions at three loops from their known renormalization group
equations (RGEs). The nonlogarithmic boundary coefficients are not predicted
by the RGE and require an explicit three-loop calculation. While they are not
required for the differential subtraction terms, they are the essential
ingredient required for the integrated subtraction terms. So far, they are known
at three loops for the $q_T$ soft function~\cite{Li:2016ctv}.

The most complicated are the beam function boundary coefficients, because they
are nontrivial functions of a partonic momentum fraction $z$. However, they
drastically simplify in the limit $z \to 1$. In this limit, the energy of
collinear emissions is constrained to be small which means their interactions
with the primary collinear parton can be described in the eikonal approximation
where they only resolve its color charge and direction. This was already pointed
out and exploited at NNLO in \refscite{Gaunt:2014xga, Gaunt:2014cfa}. Here we
exploit this to obtain for the first time the three-loop beam function
coefficients in the $z\to 1$
limit for both $\Tau_0$ and $q_T$ by relating them via appropriate consistency
relations to known soft matrix elements.
The required consistency relations were only partially known so far.
We give their detailed derivation and show explicitly that they hold to all
orders, allowing one to obtain the $z\to 1$ limits of the beam functions also to
higher orders once the relevant soft matrix elements are known.
In case of $q_T$, we provide their general structure for illustration up to six loops.
We find that a previous conjecture for this limit~\cite{Echevarria:2016scs}
only holds up to N$^3$LO but fails starting at N$^4$LO.
Since our results capture the complete
singular structure for $z\to 1$, they can also simplify the full calculation
because it can be carried out strictly for $z < 1$ which reduces the degree of
divergences.
We also employ the obtained eikonal terms of the beam function coefficients
to construct an ansatz for the missing next-to-eikonal coefficients
to estimate their numerical size.

When this paper first appeared, the $\Tau_0$ beam function was only known at NNLO~\cite{Stewart:2010qs,Berger:2010xi,Gaunt:2014xga,Gaunt:2014cfa},
while only partial results were known at N$^3$LO~\cite{Melnikov:2018jxb, Melnikov:2019pdm}.
By now, the complete results for the three-loop beam functions have become available
\cite{Behring:2019quf, Ebert:2020unb}, for which our results provided
important cross checks. In particular, the predicted $z\to1$ limit
was the only available check of the genuine three-loop contribution.
Similarly, the complete results for the three-loop beam functions for $q_T$
have become available~\cite{Luo:2019szz, Ebert:2020yqt}, for which our
results in the $z\to 1$ limit again provided important checks.

For $q_T$, a similar study of the logarithmic structure at N$^3$LO was performed
in \refcite{Cieri:2018oms} to construct an approximate $q_T$ subtraction at this order.
Here, we present a more detailed derivation of its fixed-order structure and
the ensuing $q_T$ subtraction, which differs from the method employed in
\refcite{Cieri:2018oms}.
While the RGEs necessary to derive the three-loop differential subtractions
are in principle known in the literature, we provide here for the first time a
comprehensive account of the complete structure for both $q_T$ and $\Tau_0$.
All required perturbative ingredients are collected in
the Appendix, while the results for the three-loop beam functions can be directly
used together with our results. This provides the complete results for $q_T$
and $\Tau_0$ subtractions for $q\bar q$ and $gg$ processes at N$^3$LO.

In the remainder of this section, we summarize important conventions used throughout this paper.
The three-loop structure of the beam and soft functions and the eikonal limit of
the beam functions are derived for $\Tau_0$ in \sec{tau0} and for $q_T$ in \sec{qT}.
The application to $\Tau_0$ and $q_T$ subtractions at N$^3$LO is discussed in \sec{subtractions}.
Readers primarily interested in this application may directly skip ahead to \sec{subtractions}.
We conclude in \sec{conclusions}.
In \app{plusDist} we collect the needed definitions and relations for plus distributions.
In \apps{threshold_soft}{csoft_function} we discuss in more detail the soft matrix
elements that are involved in extracting the eikonal limits of the beam functions.
Explicit expressions for required perturbative ingredients
are collected in \app{pertubativeingredients}.

\subsection{Notation and conventions}
\label{sec:notation}

Throughout the paper, we denote the perturbative expansion of any function $F(\mu)$
as
\begin{equation} \label{eq:F_expansion}
F(\mu) = \sum_{n = 0}^\infty F^{(n)}(\mu)\, \Bigl[\frac{\alpha_s(\mu)}{4\pi}\Bigr]^n
\,.\end{equation}
All anomalous dimensions $\gamma^i_{x}(\as)$ and the QCD splitting functions
are expanded as
\begin{equation} \label{eq:gamma_expansion}
\gamma^i_x(\as) = \sum_{n = 0}^\infty \gamma_{x\, n}^i\, \Bigl(\frac{\as}{4\pi}\Bigr)^{n+1}
\,, \qquad
P_{ij}(z,\mu) = \sum_{n=0}^\infty P_{ij}^{(n)}(z) \biggl[ \frac{\as(\mu)}{4\pi} \biggr]^{n+1}
\,.\end{equation}
We use the following notation to abbreviate Mellin convolutions and flavor sums
\begin{align} \label{eq:Mellin_conv}
(I^{(m)} P^{(n)})_{ij}(z)
&\equiv \sum_k \int\!\frac{\df z'}{z'}\, I^{(m)}_{ik}\Bigl(\frac{z}{z'}\Bigr) P^{(n)}_{kj}(z')
\,, \nn \\
[\cI(t, \mu) P^{(n)}]_{ij}(z)
&\equiv \sum_k \int\!\frac{\df z'}{z'}\, \cI_{ik}\Bigl(t, \frac{z}{z'}, \mu\Bigr) P^{(n)}_{kj}(z')
\,,\end{align}
where $i,j,k \in \{g, u, \bar u, d, \bar d, s, \ldots\}$ label parton type and flavor.
We also define a corresponding identity operator as
\begin{equation}
\id_{ij}(z) \equiv \delta_{ij}\, \delta(1-z)
\quad\text{with}\quad
(\id P)_{ij}(z) = P_{ij}(z)
\,.\end{equation}

For Fourier-type convolutions, we use the notation
\begin{align} \label{eq:Fourier_conv}
(FG)(k, \mu) &\equiv \int\!\df k'\, F(k - k', \mu)\,G(k', \mu)
\,,\nn \\
\bigl[F\,\cI_{ij}(z)\bigr](t, \mu^2)
&\equiv \int\!\df t'\, F(t - t', \mu^2)\, \cI_{ij}(t', z, \mu^2)
\,.\end{align}
Here, the corresponding identity elements are simply $\delta(k)$ or $\delta(t)$.

We denote logarithmic plus distributions as
\begin{equation}
\cL_{n}(x)
= \biggl[ \frac{\theta(x) \ln^n x}{x}\biggr]_+
\qquad\text{with}\qquad
\int_0^1 \! \df x \, \cL_n(x) = 0
\,.\end{equation}
For dimensionful arguments, we define
\begin{equation}
\cL_{n}(k,\mu) = \frac{1}{\mu} \cL_n\Bigl(\frac{k}{\mu}\Bigr)
\,, \quad
\cL_{n}(t, \mu^2) = \frac{1}{\mu^2} \cL_n\Bigl(\frac{t}{\mu^2}\Bigr)
\,, \quad
\cL_n(\qt, \mu) = \frac{1}{\pi \mu^2} \cL_n\biggl(\frac{q_T^2}{\mu^2}\biggr)
\,.\end{equation}
More details are given in \app{plusDist}.

\section{\texorpdfstring{$\Tau_0$}{Tau0} factorization to three loops}
\label{sec:tau0}

\subsection{Factorization}
\label{sec:tau0_factorization}

The factorization for $N$-jettiness, $\Tau_N$, has been derived using SCET
in \refscite{Stewart:2009yx, Stewart:2010tn, Jouttenus:2011wh}.
Here we focus on $0$-jettiness, $\Tau_0$, which is relevant for color-singlet
production and coincides with beam thrust~\cite{Stewart:2009yx, Berger:2010xi}.
It can be defined in terms of generic measures as~\cite{Stewart:2010tn, Jouttenus:2011wh}
\begin{align} \label{eq:Tau0}
\Tau_0 = \sum_i {\rm min}\Bigl\{ \frac{2 q_a \cdot k_i}{Q_a},\, \frac{2 q_b \cdot k_i}{Q_b}\Bigr\}
\,,\end{align}
where the sum runs over the momenta $k_i$ of all final-state particles excluding
$L$ and any of its constituents.
The measures $Q_{a,b}$ determine different definitions of 0-jettiness.
Two possible choices, corresponding to the original definitions in
\refscite{Stewart:2009yx, Berger:2010xi}, are
\begin{alignat}{4} \label{eq:Tau0_2}
 &\text{leptonic:}\quad & Q_a &= Q_b = \sqrt{\w_a \w_b} = Q \,,\qquad
  & \Tau_0^{\rm lep} &= \sum_i \min \Bigl\{ e^Y n_a \cdot k_i \,,\, e^{-Y} n_b \cdot k_i \Bigr\}
\nn\\
 &\text{hadronic:}\qquad & Q_{a,b} &= \w_{a,b} = Q \, e^{\pm Y} \,,
 & \Tau_0^{\rm cm} &= \sum_i \min \Bigl\{ n_a \cdot k_i \,,\, n_b \cdot k_i \Bigr\}
\,.\end{alignat}
For our present purposes, the precise choice of the $Q_i$ is not important,
so we will simply use the symbol $\Tau_0$.

The factorization for $\Tau_0$ is given by~\cite{Stewart:2009yx}
\begin{align} \label{eq:Tau0_fact}
\frac{\df\sigma}{\df Q^2 \df Y \df \Tau_0}
&= \sum_{a,b} H_{ab}(Q^2, \mu) \int \! \df t_a \, \df t_b \,
   B_a(t_a, x_a, \mu) \, B_b(t_b, x_b, \mu) \,
   S_i\Bigl(\Tau_0 - \frac{t_a}{Q_a} - \frac{t_b}{Q_b}, \mu\Bigr)
\nn \\ &\quad
\times \Bigl[1 + \cO\Bigl(\frac{\Tau_0}{Q}\Bigr)\Bigr]
\,.\end{align}
Explicit definitions of the beam and soft functions for $\Tau_0$ in terms of operator
matrix elements in SCET can be found in \refscite{Stewart:2009yx, Stewart:2010qs}.

The beam function appearing in \eq{Tau0_fact} is the inclusive virtuality-dependent
(SCET$_\mathrm{I}$) beam function. It appears in the $N$-jettiness
factorization for any $N$~\cite{Stewart:2010tn},
including deep-inelastic scattering~\cite{Kang:2013nha}.
Recently, it was shown that it also arises in generalized threshold factorization theorems
for inclusive color-singlet production in hadronic collisions~\cite{Lustermans:2019cau}.
The virtuality-dependent quark and gluon beam functions are known to
NNLO~\cite{Stewart:2010qs, Berger:2010xi, Gaunt:2014xga, Gaunt:2014cfa},
and they are being calculated at N$^3$LO~\cite{Melnikov:2018jxb, Melnikov:2019pdm}.

The soft function in \eq{Tau0_fact} is the hemisphere soft function for incoming
Wilson lines. It is closely related to the hemisphere soft function for $e^+e^-\to $ jets,
which is known to NNLO~\cite{Schwartz:2007ib, Fleming:2007xt, Kelley:2011ng, Monni:2011gb, Hornig:2011iu}.
They have the same anomalous dimensions to all orders~\cite{Stewart:2009yx, Stewart:2010qs},
and are equal to NNLO~\cite{Stewart:2009yx, Kang:2015moa}. It is an open question
whether they remain equivalent at higher fixed orders.

The factorization in \eq{Tau0_fact} receives power corrections suppressed by
$\Tau_0/Q$, as indicated. In addition, starting at N$^4$LO it also
receives contributions from perturbative Glauber-gluon exchanges,
which are not captured by \eq{Tau0_fact}~\cite{Gaunt:2014ska, Zeng:2015iba}.

\subsection{\texorpdfstring{$\Tau_0$}{Tau0} soft function}
\label{sec:tau0_soft}

The beam thrust soft function satisfies the all-order RGE~\cite{Stewart:2009yx, Stewart:2010qs}
\begin{align} \label{eq:Tau0_soft_RGE}
\mu \frac{\df}{\df\mu} S_i(k,\mu)
&= \int\df k' \gamma_S^i(k-k',\mu)\, S_i(k',\mu)
\equiv (\gamma_S^i\, S_i)(k, \mu)
\,, \nn \\[1ex]
\gamma_S^i(k,\mu)
&= 4 \GammaC^i[\as(\mu)]\, \cL_0(k, \mu) + \gamma_S^i[\as(\mu)] \delta(k)
\,,\end{align}
where $\GammaC^i(\as)$ and $\gamma_S^i(\as)$ are the cusp and soft noncusp
anomalous dimensions.

The RGE in \eq{Tau0_soft_RGE} fully predicts the structure of $S_i(k,\mu)$ in
$k$ and $\mu$ to all orders in perturbation theory. By solving it recursively
order by order in $\alpha_s$, we can derive this structure at any given fixed
order.
Expanding both sides of \eq{Tau0_soft_RGE} to fixed order in $\as(\mu)$ and
accounting for the running of $\as(\mu)$,
we obtain a relation for the $(n+1)$-loop term in terms of the terms up to $n$ loops,
\begin{equation} \label{eq:Tau0_soft_RGE_FO}
\mu \frac{\df}{\df\mu} S_i^{(n+1)}(k,\mu)
= \sum_{m=0}^n \Bigl[4 \Gamma^i_{n-m} \bigl(\cL_0 S_i^{(m)}\bigr)(k, \mu)
+ \bigl(2 m \beta_{n-m} + \gamma_{S\,n-m}^i \bigr) S_i^{(m)}(k, \mu) \Bigr]
,\end{equation}
where we used the short-hand notation in \eq{Fourier_conv} for the convolution in $k$.
This can be integrated to give
\begin{align} \label{eq:Tau0_soft_recursive}
S_i^{(n+1)}(k,\mu)
&= \int_{k|_+}^\mu\! \frac{\df\mu'}{\mu'} \sum_{m=0}^n
\Bigl[4 \Gamma^i_{n-m} \bigl(\cL_0 S_i^{(m)}\bigr)(k, \mu)
 + \bigl(2 m \beta_{n-m} + \gamma_{S\,n-m}^i \bigr)S_i^{(m)}(k, \mu) \Bigr]
\nn \\ & \quad
 + \delta(k)\, s_i^{(n+1)}
\,,\end{align}
where the soft function boundary coefficients are defined by
\begin{equation} \label{eq:boundary_def}
S_i^{(n)}(k,\mu = k|_+) = \delta(k)\, s_i^{(n)}
\qquad\text{with}\qquad
s_i^\zero = 1
\,.\end{equation}
Here, we have used distributional scale setting $\mu_0 = k|_+$~\cite{Ebert:2016gcn}, which is defined
such that it effectively allows us to treat the $\mu$ dependence of the
logarithmic distributions like ordinary logarithms. In particular, it satisfies~\cite{Ebert:2016gcn}
\begin{align}
\cL_n(k, \mu_0 = k|_+) &= 0
\,, \nn \\
\delta(k) \ln^{n+1}\frac{\mu_0}{\mu} \bigg\vert_{\mu_0 = k|_+} &= (n+1)\cL_n(k, \mu)
\,, \nn \\
\int_{\mu_0 = k|_+}^\mu\!\frac{\df\mu'}{\mu'}\,\cL_n(k, \mu')
&= -\frac{1}{n+1} \cL_{n+1}(k, \mu)
\,.\end{align}
The first relation is used in \eq{boundary_def} to define the boundary coefficients as the
coefficients of the $\delta(k)$ by setting all logarithmic distributions
in $S_i^{(n)}(k, \mu)$ to zero. The other two relations allow us to easily perform
the $\mu'$ integral in \eq{Tau0_soft_recursive}, essentially turning a $\delta(k)$
into a $\cL_0(k)$ and a $\cL_n(k)$ into a $\cL_{n+1}(k)$.
In addition, to evaluate the cross terms for $m\geq 1$ in \eq{Tau0_soft_recursive},
we need the convolutions~\cite{Ligeti:2008ac}
\begin{align}
(\cL_0 \cL_0)(k, \mu)
&= 2\cL_1(k, \mu) - \zeta_2\, \delta(k)
\,, \nn \\
(\cL_0 \cL_1)(k, \mu)
&= \frac{3}{2}\cL_2(k, \mu) - \zeta_2\cL_0(k, \mu) + \zeta_3\, \delta(k)
\,, \nn \\
(\cL_0 \cL_2)(k, \mu)
&= \frac{4}{3}\cL_3(k, \mu) - 2\zeta_2\cL_1(k, \mu) + 2\zeta_3\cL_0(k, \mu)
 - 2\zeta_4\, \delta(k)
\,, \nn \\
(\cL_0 \cL_3)(k, \mu)
&= \frac{5}{4}\cL_4(k, \mu) - 3\zeta_2 \cL_2(k, \mu) + 6\zeta_3 \cL_1(k, \mu)
 - 6\zeta_4\cL_0(k, \mu) + 6\zeta_5\, \delta(k)
\,.\end{align}

Starting from the LO result, $s_i^\zero = 1$, \eq{Tau0_soft_recursive} yields
up to two loops
\begin{align}
S_i^\zero(k,\mu) &= \delta(k)
\,, \nn \\
S_i^\one(k,\mu)
&= - \cL_1(k,\mu)\,4 \Gamma_0^i - \cL_0(k,\mu)\, \gamma^i_{S\,0}  + \delta(k)\, s_i^\one
\,,\nn\\
S_i^\two(k,\mu)
&= \cL_3(k, \mu)\, 8 (\Gamma^i_0)^2
 + \cL_2(k, \mu)\,
   2 \Gamma^i_0 (2\beta_0 + 3\gamma_{S\,0}^i)
\nn \\ &\quad
 + \cL_1(k, \mu) \Bigl[
   - 16\zeta_2(\Gamma^i_0)^2 + (2\beta_0 + \gamma_{S\,0}^i)\gamma_{S\,0}^i
   - 4 \Gamma^i_1 - 4 \Gamma^i_0\, s_i^\one
\Bigr]
\nn \\ &\quad
 + \cL_0(k, \mu) \Bigl[
   4\Gamma^i_0(4\zeta_3\Gamma^i_0 - \zeta_2 \gamma_{S\,0}^i)
   - \gamma_{S\,1}^i - (2\beta_0 + \gamma_{S\,0}^i)\, s_i^\one
\Bigr]
 + \delta(k)\, s_i^\two
\,,\end{align}
which agrees with \refcite{Gaunt:2015pea}.
Evaluating \eq{Tau0_soft_recursive} at the next order, we obtain the three-loop result,
\begin{equation}
S^\three(k, \mu) = \delta(k)\, s_i^\three + \sum_{\ell = 0}^5 S_{i,\ell}^\three\, \cL_\ell(k, \mu)
\end{equation}
with the coefficients of the logarithmic distributions given by
\begin{align} \label{eq:Tau0_soft_n3lo}
S_{i,5}^\three
&= -8 (\Gamma^i_0)^3
\nn \\
S_{i,4}^\three
&= -\frac{10}{3}(\Gamma^i_0)^2\, (4\beta_0 + 3\gamma_{S\,0}^i)
\nn \\
S_{i,3}^\three
&= 4\Gamma^i_0\Bigl[
   16\zeta_2 (\Gamma^i_0)^2
   - \frac{4}{3}\beta_0^2 - \Bigl(\frac{10}{3}\beta_0 + \gamma_{S\,0}^i\Bigr)\gamma_{S\,0}^i
   + 4\Gamma_1^i + 2 \Gamma_0^i\, s_i^\one
\Bigr]
\nn \\
S_{i,2}^\three
&= 16(\Gamma^i_0)^2 \bigl[-10\zeta_3\Gamma^i_0 + 3\zeta_2 ( \beta_0 + \gamma^i_{S\,0}) \bigr]
 + (4 \beta_0 + 3\gamma^i_{S\,0})(-\beta_0\gamma^i_{S\,0} + 2\Gamma^i_1)
 - \frac{(\gamma^i_{S\,0})^3}{2}
\nn \\ & \quad
 + 2\Gamma^i_0 \Bigl[2\beta_1 + 3 \gamma^i_{S\, 1} + (8 \beta_0 + 3 \gamma^i_{S\,0}) s_i^\one \Bigr]
\nn \\
S_{i,1}^\three
&= 32(\Gamma^i_0)^2 \bigl[\zeta_4\Gamma^i_0 - \zeta_3(3\beta_0 + 2\gamma^i_{S\,0}) \bigr]
 + 8\zeta_2 \Gamma^i_0 \bigl[
   (3\beta_0 + \gamma^i_{S\,0}) \gamma^i_{S\,0}
 - 4\Gamma^i_1
\bigr]
\nn \\ & \quad
 + 4 \beta_0  \gamma^i_{S\,1}
 + 2 \gamma^i_{S\,0} (\beta_1 + \gamma^i_{S\,1})
 + \bigl[-16\zeta_2(\Gamma^i_0)^2 + 8\beta_0^2 + (6\beta_0 + \gamma^i_{S\,0})\gamma^i_{S\,0} - 4 \Gamma^i_1 \bigr] s_i^\one
\nn \\ & \quad
 - 4 \Gamma^i_2
 - 4 \Gamma^i_0\, s_i^\two
\nn \\
S_{i,0}^\three
&= 16 (\Gamma^i_0)^2 \Bigl[
   4\Gamma^i_0 (2 \zeta_2 \zeta_3 - 3 \zeta_5)
   + \zeta_4 \Bigl(2 \beta_0 + \frac{\gamma^i_{S\,0}}{2} \Bigr)
\Bigr]
 - 4\zeta_3\Gamma^i_0 \bigl[(2 \beta_0 + \gamma^i_{S\,0}) \gamma^i_{S\,0} - 8\Gamma^i_1 \bigr]
\nn \\ &\quad
 - 4\zeta_2\bigl(\gamma^i_{S\,0} \Gamma^i_1 + \Gamma^i_0 \gamma^i_{S\,1} \bigr)
 + \Bigl\{
   4\Gamma^i_0 \bigl[4\zeta_3 \Gamma^i_0 - \zeta_2 (2\beta_0 + \gamma^i_{S\,0}) \bigr]
   - (2 \beta_1 + \gamma^i_{S\,1})
\Bigr\} s_i^\one
\nn \\ &\quad
 - \gamma^i_{S\,2}
 - (4 \beta_0 + \gamma^i_{S\,0}) s_i^\two
\,.\end{align}
This agrees with a corresponding numerical expression in \refcite{Abbate:2010xh}.
The required anomalous dimension coefficients up to three loops and boundary
coefficients up to two loops are given in \app{pertubativeingredients}.

\paragraph{Numerical impact}

The soft function $S_i(k, \mu)$ has an explicit dependence on $\mu$, which
cancels against that of the hard and beam functions in \eq{Tau0_fact}. Therefore,
simply varying the scale $\mu$ is not very meaningful for illustrating the numerical
impact of the $\mu$-dependent three-loop terms.
Instead, we consider the resummed soft function,
\begin{equation}
S_i(k, \mu) = \int \! \df k' \, S_i(k', \mu_S) \, U_S^i(k - k', \mu_S, \mu)
\,,\end{equation}
where the evolution factor $U_S^i(k, \mu_S, \mu)$ encodes the
solution of \eq{Tau0_soft_RGE}, with $U_S^i(k, \mu_S, \mu_S) = \delta(k)$.
It can be found e.g. in \refscite{Stewart:2010qs, Berger:2010xi}.
Formally, the $\mu_S$ dependence on the right-hand side cancels, but when the starting
condition $S_i(k, \mu_S)$ is evaluated at fixed order, it only cancels up to higher-order terms.
For ease of presentation, we consider the cumulant of the soft function
\begin{equation}
(S_i \otimes U_S^i)_\cut (\Tau_\cut, \mu)
= \int^{\Tau_\cut} \! \df k \, \int \! \df k' \, S_i(k', \mu_S) \, U_S^i(k - k', \mu_S, \mu)
\,,\end{equation}
for which the distributions turn into ordinary logarithms of $\Tau_\cut$.

\begin{figure*}
\centering
\includegraphics[width=\WidthTwoSubfigs]{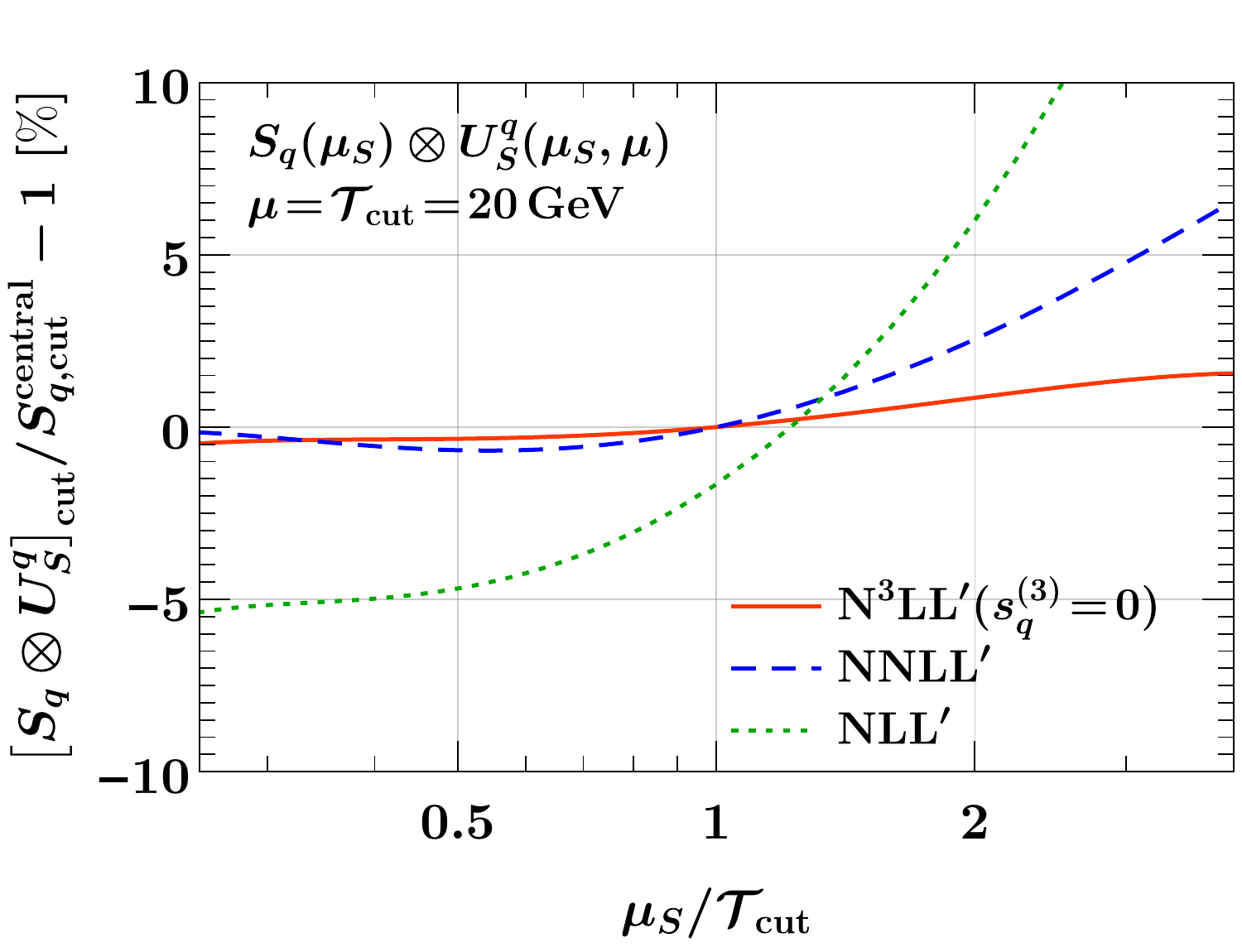}%
\hfill%
\includegraphics[width=\WidthTwoSubfigs]{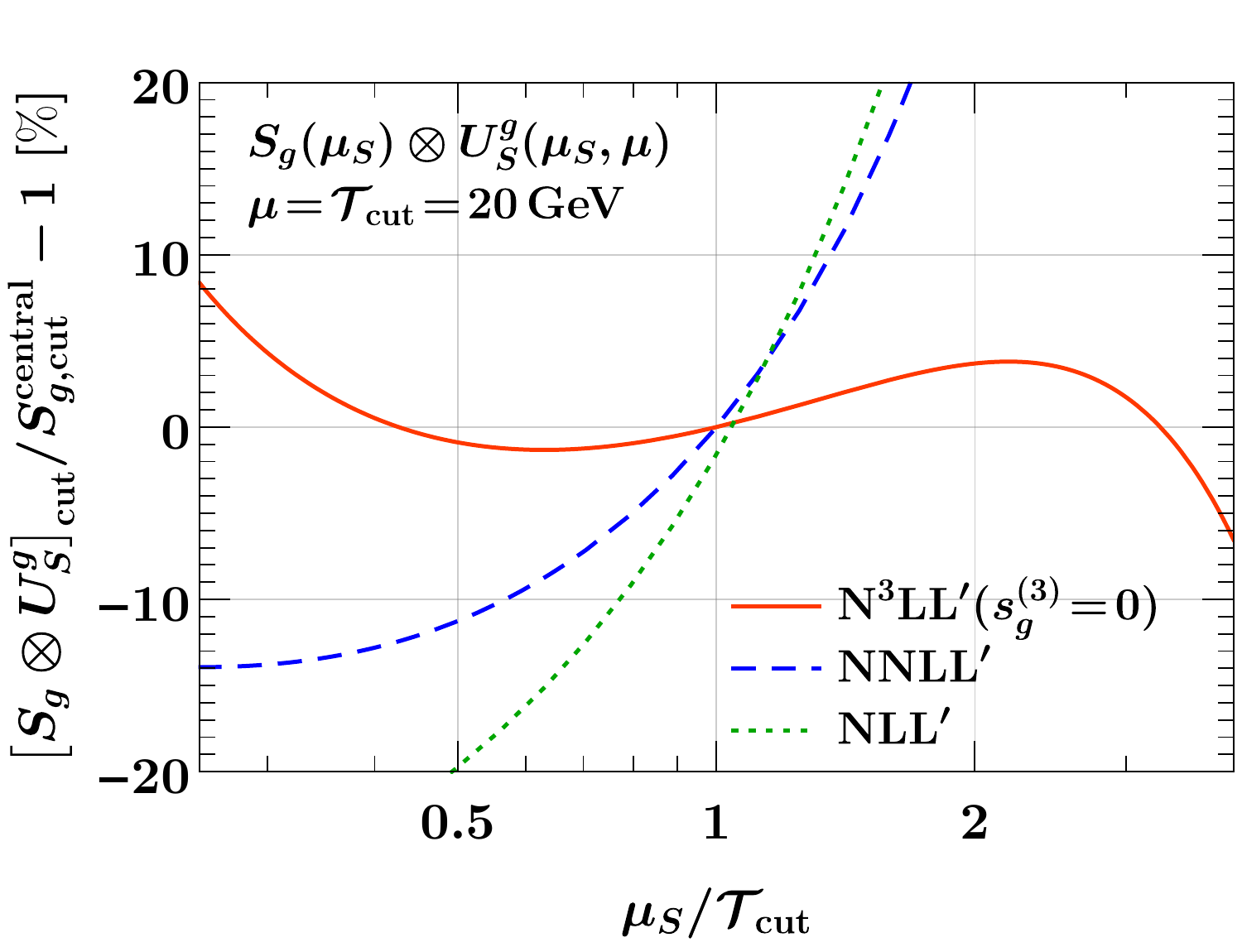}%
\caption{Residual scale dependence of the integrated resummed $\Tau_0$ soft function
for $i = q$ (left) and $i = g$ (right).
Shown are the relative deviations from the NNLL$'$ result $S_{i,\cut}^\central$ at the
central scale $\mu_S = \Tau_\cut$.}
\label{fig:scale_dependence_Tau0_soft}
\end{figure*}

In \fig{scale_dependence_Tau0_soft}, we take as an example $\Tau_\cut = \mu = 20\GeV$ and show the
residual $\mu_S$ dependence of the resummed soft function when varying $\mu_S$
around the canonical central value $\mu_S = \Tau_\cut$
at NLL$'$ (dotted green), NNLL$'$ (dashed blue), and N$^3$LL$'$ (solid orange).
For the latter we set the currently unknown three-loop constant term $s_i^\three = 0$.
In all cases, we show the relative difference to the central value at NNLL$'$.
For simplicity we always use the same four-loop (N$^3$LL) running for $\as$,
which formally amounts to a higher-order effect at (N)NLL$'$.
For the quark soft function (left panel), the $\mu_S$ dependence is more than halved going
from NLL$'$ to NNLL$'$, and roughly halved again at N$^3$LL$'$.
In the gluon case (right panel), the $\mu_S$ dependence is noticeably larger, but also
reduces significantly at each order as it should.
Note that the missing three-loop constant term will add an additional source
of $\mu_S$ dependence due to its $\as^3(\mu_S)$ prefactor, which however should
not change the general picture.

We stress that the residual $\mu_S$ dependence in the resummed soft function by
itself is not necessarily a good indicator of the perturbative uncertainty.
Nevertheless, the reduction in the scale dependence still provides a useful
cross check and an indication of the typical reduction of perturbative uncertainties
one might expect at each order. We also emphasize that the size of the
variations in \fig{scale_dependence_Tau0_soft} does not necessarily reflect the
variations one should expect in the resummed cross section, where the evolution
of the soft function happens in conjunction with the beam and hard functions.

\subsection{\texorpdfstring{$\Tau_0$}{Tau0} beam function}
\label{sec:tau0_beam}

The beam function $B_i(t,x,\mu)$ obeys the all-order RGE~\cite{Stewart:2009yx, Stewart:2010qs}
\begin{align} \label{eq:TauN_beam_RGE}
\mu \frac{\df}{\df \mu} B_i(t, x, \mu)
&= \int\! \df t'\, \gamma_B^i(t-t',\mu)\, B_i(t', x, \mu)
\,, \nn \\[1ex]
\gamma_B^i(t, \mu)
&= -2 \GammaC^i[\alpha_s(\mu)]\,\cL_0(t, \mu^2)
  + \gamma_B^i[\alpha_s(\mu)]\,\delta(t)
\,,\end{align}
where $\GammaC^i(\as)$ and $\gamma^i_B(\as)$ are the cusp and beam noncusp anomalous dimensions.
For $t \gg \lqcd$, the beam function satisfies an OPE in terms
of standard PDFs~\cite{Stewart:2009yx, Stewart:2010qs}
\begin{equation} \label{eq:Tau0_beam_OPE}
B_i(t,x,\mu)
= \sum_j \int\! \frac{\df z}{z} \, \cI_{ij}(t,z,\mu) \, f_j\Bigl(\frac{x}{z},\mu\Bigr) \,
  \biggl[1+\mathcal{O}\biggl(\frac{\lqcd^2}{t}\biggr)\biggr]
\,,\end{equation}
where the $\cI_{ij}(t, z, \mu)$ are perturbatively calculable matching coefficients.
Taking into account the evolution of the PDFs, they obey the RGE~\cite{Stewart:2010qs}
\begin{equation} \label{eq:TauN_beamI_RGE}
\mu\frac{\df}{\df\mu}\cI_{ij}(t,z,\mu)
= \sum_k \int\!\df t'\, \frac{\df z'}{z'}\, \cI_{ik}\Bigl(t - t', \frac{z}{z'},\mu\Bigr)
\Bigl[
   \gamma_B^i(t', \mu)\, \id_{kj}(z')
   -\delta(t')\, 2P_{kj}(z',\mu)
\Bigr]
,\end{equation}
where $\id_{ij}(z) \equiv \delta_{ij}\, \delta(1 - z)$ and
$2P_{ij}(z, \mu)$ are the PDF anomalous dimensions.

By solving the RGE in \eq{TauN_beamI_RGE} recursively order by order, we can derive
the complete structure of $\cI_{ij}(t, z, \mu)$ at any given fixed order,
as was done in \refscite{Berger:2010xi, Gaunt:2014xga} to NNLO.
Following the same procedure as in \sec{tau0_soft}, keeping track of
the flavor indices and Mellin convolutions, the $(n+1)$-loop term is determined
from the up to $n$-loop terms as
\begin{align} \label{eq:Tau0_beam_recursive}
\cI_{ij}^{(n+1)}(t, z, \mu^2)
&= \delta(t)\, I_{ij}^{(n+1)}(z)
+ \int_{t|_+}^{\mu^2}\! \frac{\df\tilde\mu^2}{\tilde\mu^2} \sum_{m=0}^{n}
\biggl\{
   - \Gamma^i_{n-m} \bigl[\cL_0\cI^{(m)}_{ij}(z)\bigr](t, \tilde\mu^2)
\\ \nn & \quad
   + \Bigl( m \beta_{n-m} + \frac{\gamma_{B\,n-m}^i}{2} \Bigr) \cI^{(m)}_{ij}(t, z, \tilde\mu^2)
   - \bigl[\cI^{(m)}(t, \tilde\mu^2) P^{(n-m)} \bigr]_{ij}(z)
\biggr\}
\,,\end{align}
where the $\mu$-independent boundary coefficients are defined as
\begin{equation}
\cI^{(n)}_{ij}(t, z, \mu^2 = t|_+) = \delta(t)\, I^{(n)}_{ij}(z)
\,.\end{equation}

Starting from the LO result, $I^\zero_{ij}(z) = \id_{ij}(z) \equiv \delta_{ij}\, \delta(1-z)$,
we obtain up to two loops
\begin{align} \label{eq:TauN_beamI_NNLO}
\cI_{ij}^\zero(t,z,\mu^2)
&= \delta(t)\, \id_{ij}(z)
\,, \nn \\
\cI_{ij}^\one(t,z,\mu^2)
&= \cL_1(t, \mu^2) \, \Gamma_0^i\, \id_{ij}(z)
 + \cL_0(t, \mu^2)
\Bigl[
   - \frac{\gamma_{B\,0}^i}{2}\, \id_{ij}(z)
   + P^\zero_{ij}(z)
\Bigr]
 + \delta(t)\, I^\one_{ij}(z)
\,, \nn \\
\cI_{ij}^\two(t,z,\mu^2)
&= \cL_3(t, \mu^2)\, \frac{(\Gamma_0^i)^2}{2}\, \id_{ij}(z)
\nn \\ &\quad
 + \cL_2(t, \mu^2)\, \frac{\Gamma_0^i}{2}
\Bigl[
   - \Bigl(\beta_0 + \frac{3}{2} \gamma_{B\,0}^i \Bigr) \id_{ij}(z)
   + 3 P^\zero_{ij}(z)
\Bigr]
\nn \\ &\quad
 + \cL_1(t, \mu^2)
\biggl\{
   \Bigl[
      - \zeta_2 (\Gamma_0^i)^2
      + \Bigl(\beta_0 + \frac{\gamma_{B\,0}^i}{2} \Bigr) \frac{\gamma_{B\,0}^i}{2}
      + \Gamma_1^i
   \Bigr] \id_{ij}(z)
\nn \\ &\qquad\qquad\qquad\quad
   - (\beta_0 + \gamma_{B\,0}^i) P^\zero_{ij}(z)
   + (P^\zero\! P^\zero)_{ij}(z)
   + \Gamma_0^i\, I^\one_{ij}(z)
\biggr\}
\nn \\ &\quad
 + \cL_0(t, \mu^2)
\biggl\{
   \Bigl[
      \Gamma_0^i\Bigl(\zeta_3 \Gamma_0^i + \zeta_2 \frac{\gamma_{B\,0}^i}{2} \Bigr)
      - \frac{\gamma_{B\,1}^i}{2}
   \Bigr] \id_{ij}(z)
   - \zeta_2 \Gamma_0^i\, P^\zero_{ij}(z)
\nn \\ &\qquad\qquad\qquad\quad
   + P^\one_{ij}(z)
   - \Bigl(\beta_0 + \frac{\gamma_{B\,0}^i}{2} \Bigr) I^\one_{ij}(z)
   + (I^\one\! P^\zero)_{ij}(z)
\biggr\}
\nn \\ &\quad
+ \delta(t)\, I^\two_{ij}(z)
\,,\end{align}
which agrees with \refscite{Berger:2010xi, Gaunt:2014xga, Gaunt:2015pea}.
The NLO and NNLO boundary coefficients $I_{ij}^{(1,2)}(z)$ together with the required
Mellin convolutions $(P^\zero P^\zero)_{ij}(z)$ and $(I^\one P^\zero)_{ij}(z)$
can be found in \refscite{Gaunt:2014xga, Gaunt:2014cfa}.%
\footnote{We caution that the functions $P_{ij}(z)$ and $I_{ij}(z)$
in \refscite{Gaunt:2014xga, Gaunt:2014cfa, Gaunt:2015pea} are expanded in powers of $\as/(2\pi)$
while here we expand them in powers of $\as/(4\pi)$.}

Plugging \eq{TauN_beamI_NNLO} back into \eq{Tau0_beam_recursive}, we obtain the N$^3$LO result
\begin{equation}
\cI_{ij}^\three(t, z, \mu^2)
=  \delta(t)\, I_{ij}^\three(z) + \sum_{\ell = 0}^{5} \cI_{ij,\ell}^\three(z)\, \cL_\ell(t, \mu^2)
\end{equation}
with the coefficients
\begin{align} \label{eq:TauN_beamI_N3LO}
\cI_{ij,5}^\three(z)
&= \frac{(\Gamma^i_0)^3}{8} \id_{ij}(z)
\nn \\
\cI_{ij,4}^\three(z)
&= \frac{5}{8}(\Gamma^i_0)^2 \Bigl[
   - \Bigl(\frac{2}{3}\beta_0 + \frac{\gamma_{B\,0}^i}{2} \Bigr) \id_{ij}(z)
   + P_{ij}^\zero
\Bigr]
\nn \\
\cI_{ij,3}^\three(z)
&= \Gamma^i_0 \biggl\{
   \Bigl[
      - \zeta_2 (\Gamma^i_0)^2
      + \frac{\beta_0^2}{3}
      + \Bigl(\frac{5}{3}\beta_0 + \frac{\gamma_{B\,0}^i}{2}\Bigr) \frac{\gamma_{B\,0}^i}{2}
      + \Gamma_1^i
   \Bigr] \id_{ij}(z)
\nn \\ & \quad
   - \Bigl(\frac{5}{3}\beta_0 + \gamma_{B\,0}^i \Bigr) P_{ij}^\zero(z)
   + (P^\zero\! P^\zero)_{ij}(z)
   + \frac{\Gamma_0^i}{2} I_{ij}^\one(z)
\biggr\}
\nn \\
\cI_{ij,2}^\three(z)
&= \biggl\{
   (\Gamma^i_0)^2 \Bigl[
      \frac{5}{2}\zeta_3 \Gamma^i_0
      + \frac{3}{2}\zeta_2 (\beta_0 + \gamma^i_{B\,0})
   \Bigr]
   - \Bigl(\beta_0 + \frac{3}{4}\gamma^i_{B\,0}\Bigr)\Bigl(\beta_0\frac{\gamma^i_{B\,0}}{2} + \Gamma^i_1\Bigr)
   - \frac{(\gamma^i_{B\,0})^3}{16}
\nn \\ & \quad
   - \frac{\Gamma^i_0}{2} \Bigl(\beta_1 + \frac{3}{2}\gamma^i_{B\, 1}\Bigr)
\biggr\}\id_{ij}(z)
 + 3\Bigl[
   - \zeta_2 (\Gamma^i_0)^2
   + \frac{\beta_0^2}{3} + \Bigl(\beta_0 + \frac{\gamma^i_{B\,0}}{4}\Bigr)\frac{\gamma^i_{B\,0}}{2}
   + \frac{\Gamma^i_1}{2}
\Bigr] P_{ij}^\zero(z)
\nn \\ & \quad
 - \frac{3}{2} \Bigl(\beta_0 + \frac{\gamma^i_{B\,0}}{2}\Bigr)(P^\zero\! P^\zero)_{ij}(z)
 + \frac{1}{2} (P^\zero\! P^\zero\! P^\zero)_{ij}(z)
\nn \\ & \quad
 + \frac{3}{2}\Gamma^i_0 \Bigl[
   P_{ij}^\one(z)
   - \Bigl(\frac{4}{3}\beta_0 + \frac{\gamma^i_{B\,0}}{2}\Bigr) I_{ij}^\one(z)
   + (I^\one\! P^\zero)_{ij}(z)
\Bigr]
\nn \\
\cI_{ij,1}^\three(z)
&= \biggl\{
   - (\Gamma^i_0)^2 \Bigl[\frac{\zeta_4}{2}\Gamma^i_0 + \zeta_3(3\beta_0 + 2\gamma^i_{B\,0}) \Bigr]
   - \zeta_2 \Gamma^i_0 \Bigl[
      (3\beta_0 + \gamma^i_{B\,0}) \frac{\gamma^i_{B\,0}}{2}
      + 2\Gamma^i_1
   \Bigr]
\nn \\ & \quad
   + \beta_0 \gamma^i_{B\,1}
   + \frac{\gamma^i_{B\,0}}{2} (\beta_1 + \gamma^i_{B\,1})
   + \Gamma^i_2
\biggr\} \id_{ij}(z)
 + \Bigl\{
   \Gamma^i_0 \bigl[ 4\zeta_3 \Gamma^i_0 + \zeta_2 (3\beta_0 + 2\gamma^i_{B\,0}) \bigr]
\nn \\ & \quad
   - (\beta_1 + \gamma^i_{B\,1})
\Bigr\} P^\zero_{ij}(z)
 - 2\zeta_2 \Gamma^i_0 (P^\zero\! P^\zero)_{ij}(z)
 - (2\beta_0 + \gamma^i_{B\,0}) P^\one_{ij}(z)
\nn \\ & \quad
 + (P^\zero\! P^\one \!+\! P^\one\! P^\zero)_{ij}(z)
 + \Bigl[
   -\zeta_2(\Gamma^i_0)^2
   + 2\beta_0^2
   + \Bigl(3\beta_0 + \frac{\gamma^i_{B\,0}}{2}\Bigr) \frac{\gamma^i_{B\,0}}{2} + \Gamma^i_1
\Bigr] I^\one_{ij}(z)
\nn \\ & \quad
 - (3\beta_0 + \gamma^i_{B\,0}) (I^\one\! P^\zero)_{ij}(z)
 + (I^\one\! P^\zero\! P^\zero)_{ij}(z)
 + \Gamma^i_0\, I^\two_{ij}(z)
\nn \\
\cI_{ij,0}^\three(z)
&= \biggl\{
   (\Gamma^i_0)^2 \Bigl[
      - \Gamma^i_0 (2 \zeta_2 \zeta_3 - 3 \zeta_5)
      + \zeta_4 \Bigl(\beta_0 + \frac{\gamma^i_{B\,0}}{4}\Bigr)
   \Bigr]
 + \zeta_3\Gamma^i_0 \Bigl[
   \Bigl(\beta_0 + \frac{\gamma^i_{B\,0}}{2} \Bigr) \frac{\gamma^i_{B\,0}}{2}
   + 2\Gamma^i_1
\Bigr]
\nn \\ &\quad
 + \frac{\zeta_2}{2}\bigl(\gamma^i_{B\,0} \Gamma^i_1 + \Gamma^i_0 \gamma^i_{B\,1} \bigr)
 - \frac{\gamma^i_{B\,2}}{2}
\biggr\} \id_{ij}(z)
 - \biggl\{
   \Gamma^i_0 \Bigl[\frac{\zeta_4}{2} \Gamma^i_0 + \zeta_3 (\beta_0 + \gamma^i_{B\,0}) \Bigr]
\nn \\ &\quad
   + \zeta_2 \Gamma^i_1
\biggr\} P^\zero_{ij}(z)
 + \Gamma^i_0 \bigl[\zeta_3 (P^\zero\! P^\zero)_{ij}(z) - \zeta_2  P^\one_{ij}(z) \bigr]
 + P^\two_{ij}(z)
\nn \\ &\quad
 + \biggl\{
   \Gamma^i_0 \Bigl[\zeta_3 \Gamma^i_0 + \zeta_2 \Bigl(\beta_0 + \frac{\gamma^i_{B\,0}}{2}\Bigr) \Bigr]
   - \Bigl(\beta_1 + \frac{\gamma^i_{B\,1}}{2}\Bigr)
\biggr\} I^\one_{ij}(z)
 - \Gamma^i_0 \zeta_2 (I^\one\! P^\zero)_{ij}(z)
\nn \\ &\quad
 + (I^\one\! P^\one)_{ij}(z)
 - \Bigl(2\beta_0 + \frac{\gamma^i_{B\,0}}{2}\Bigr) I^\two_{ij}(z)
 + (I^\two\! P^\zero)_{ij}(z)
\,.\end{align}
The required anomalous dimensions and splitting functions up to three loops
are given in \app{pertubativeingredients}.
We have evaluated all Mellin convolutions appearing in \eq{TauN_beamI_N3LO}
using the {\tt MT} package~\cite{Hoeschele:2013gga}.
To calculate $(I^\two P^\zero)_{ij}(z)$ this required employing the identity
\begin{equation} \label{eq:trilogsubst}
\mathrm{Li}_3\Bigl(\frac{1}{1+z}\Bigr)
+ \mathrm{Li}_3(-z) + \mathrm{Li}_3\Bigl(\frac{z}{1+z}\Bigr)
= \zeta_3
 - \zeta_2 \ln(1+z)
 - \frac{1}{2} \ln^2(1+z) \ln z
 + \frac{1}{3} \ln^3(1+z)
\,.\end{equation}

\paragraph{Numerical impact}

As for the soft function above,
to illustrate the numerical impact of the three-loop corrections,
we consider the integrated resummed beam function
\begin{align}
(B_i \otimes U_B^i)_\cut (t_\cut, x, \mu)
&= \int^{t_\cut} \! \df t \, \int \! \df t' \, B_i(t, x, \mu_B) \, U_B^i(t - t', \mu_B, \mu)
\,.\end{align}
The explicit expression for the beam function evolution kernel $U_B^i(t, \mu_B, \mu)$
can be found in \refscite{Stewart:2010qs, Berger:2010xi}.

\begin{figure*}
\centering
\includegraphics[width=\WidthTwoSubfigs]{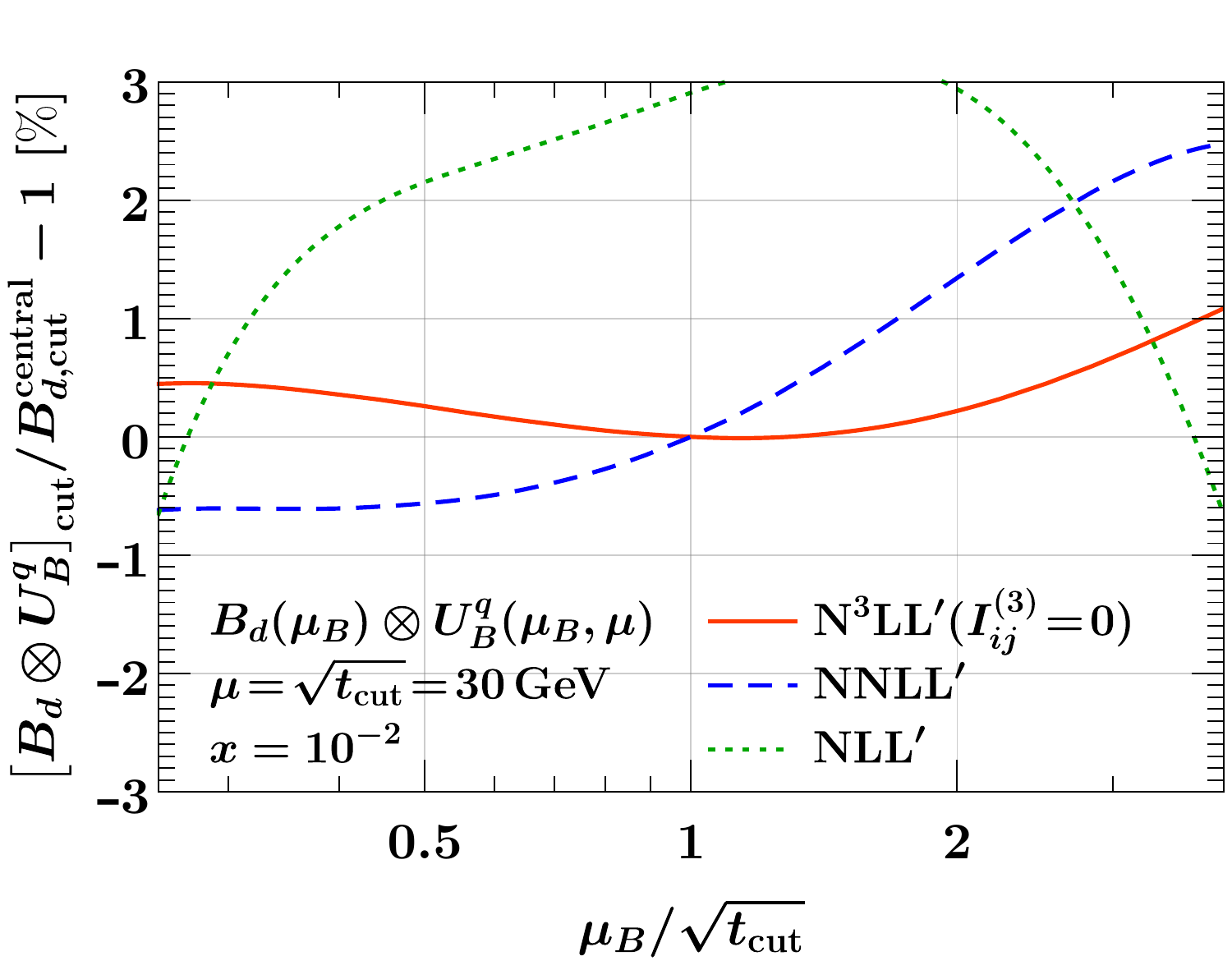}%
\hfill%
\includegraphics[width=\WidthTwoSubfigs]{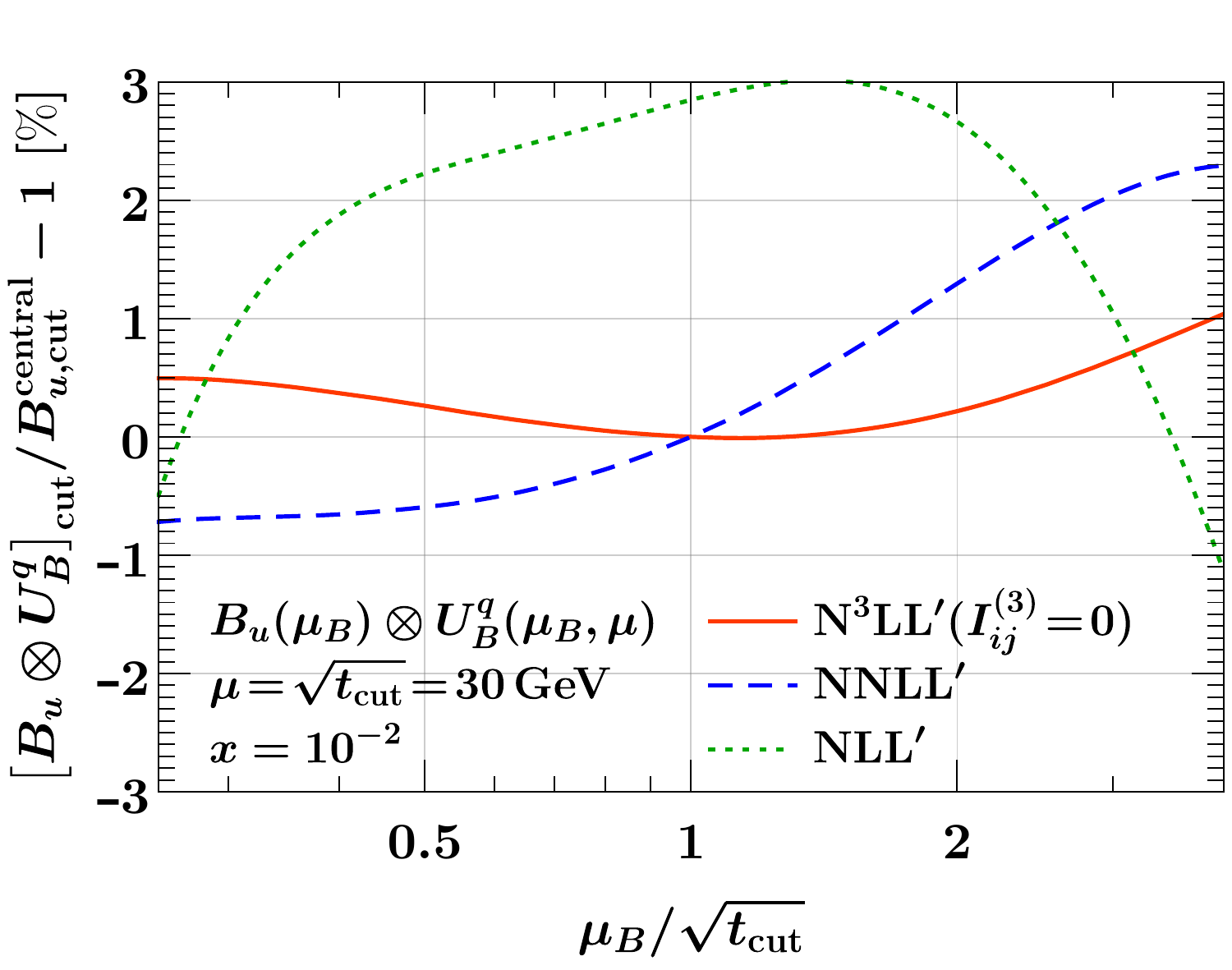}%
\\
\includegraphics[width=\WidthTwoSubfigs]{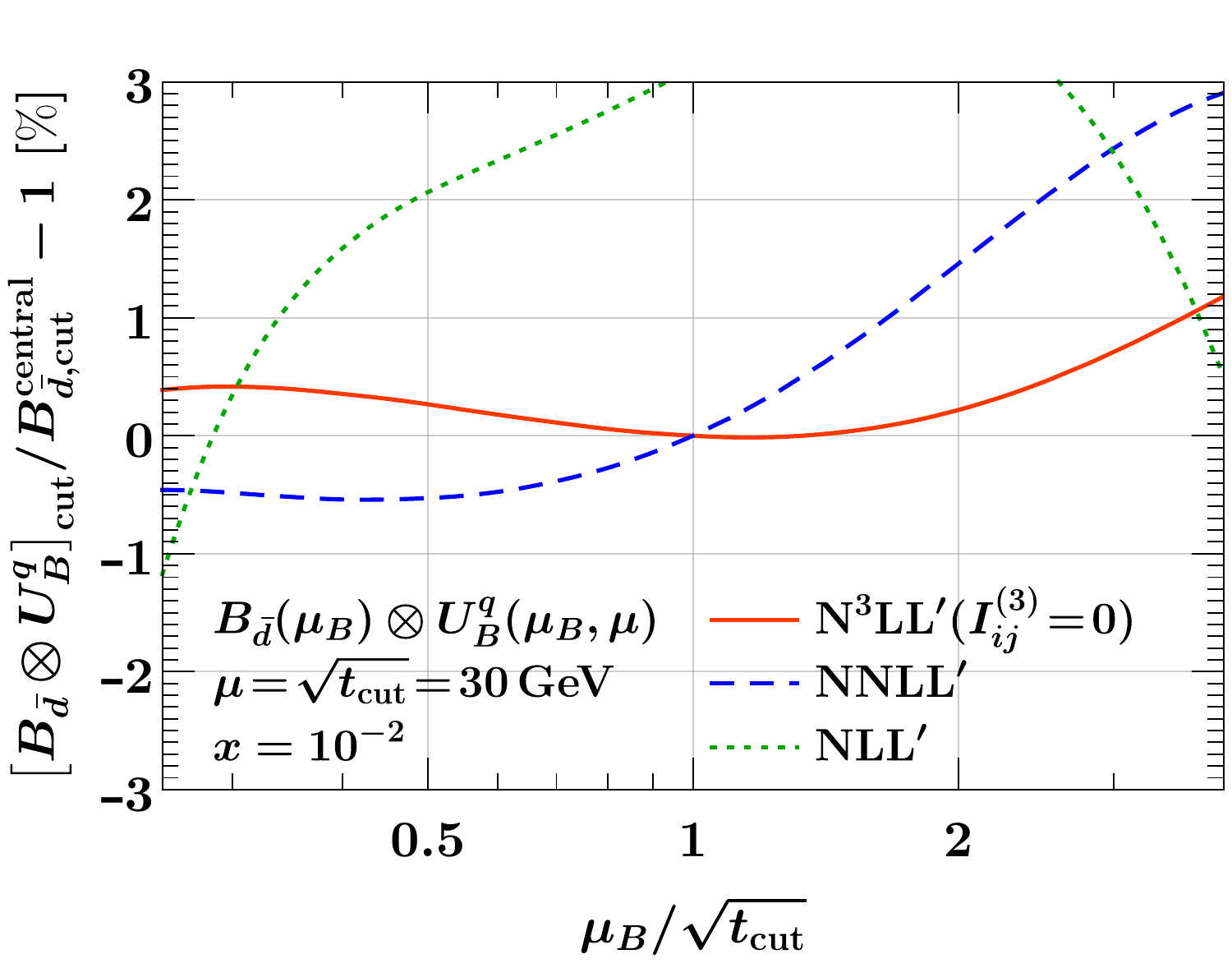}%
\hfill%
\includegraphics[width=\WidthTwoSubfigs]{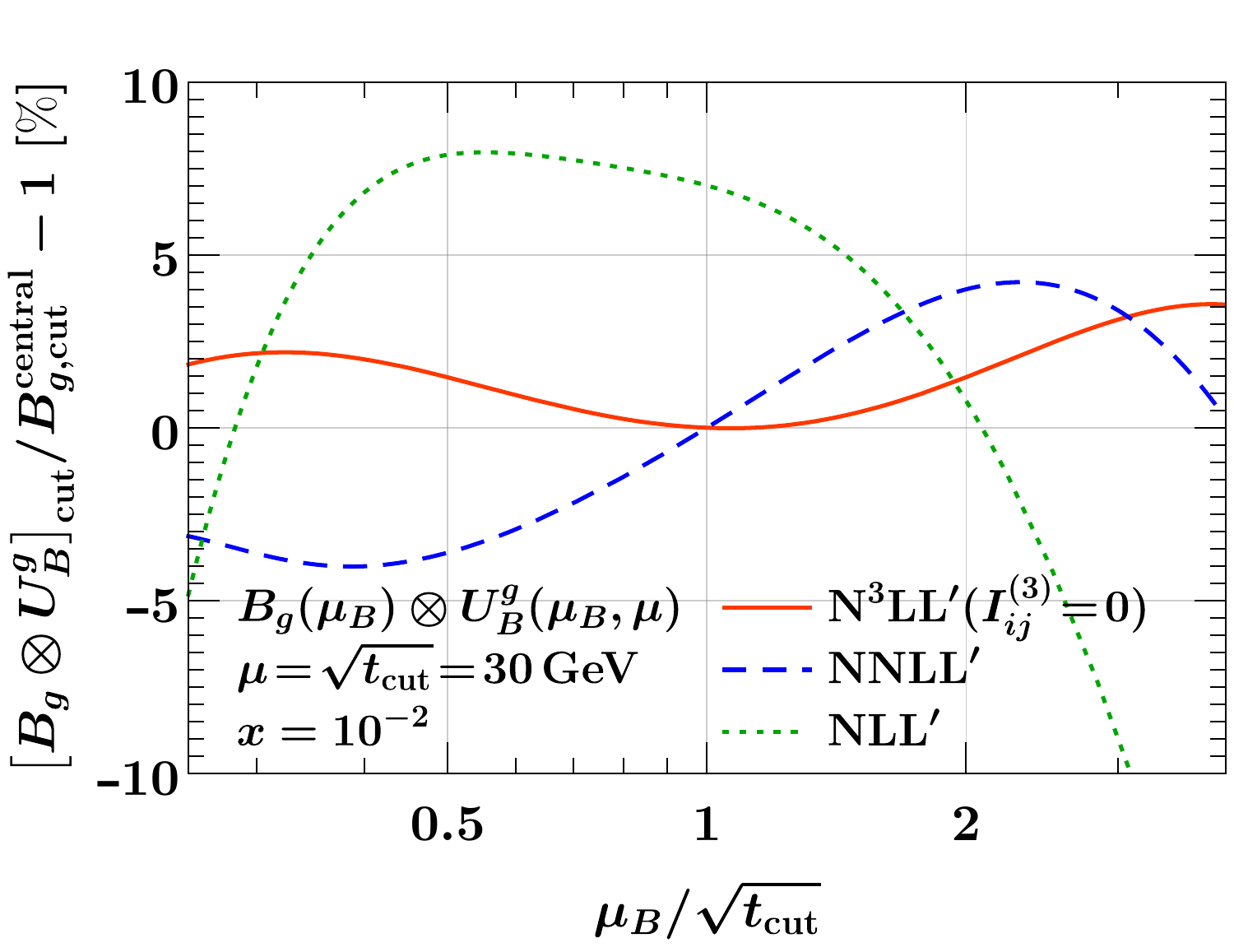}%
\caption{Residual scale dependence of the resummed integrated $\Tau_N$ beam function
for $i = d$ (top left), $u$ (top right), $\bar{d}$ (bottom left), and $g$ (bottom right).
Shown are the relative deviations from the NNLL$'$ result $B_{i,\cut}^\central$ at the
central scale $\mu_B = \sqrt{t_\cut}$.}
\label{fig:scale_dependence_Tau0_beam}
\end{figure*}

In \fig{scale_dependence_Tau0_beam}, we show the residual $\mu_B$ dependence
of the resummed integrated beam function at fixed representative values of $x = 10^{-2}$
and $\sqrt{t_\cut} = \mu = 30 \GeV$.
We again show the relative difference to the NNLL$'$ central result at $\mu_B = \sqrt{t_\cut}$
at NLL$'$ (dotted green), NNLL$'$ (dashed blue), and N$^3$LL$'$ with the unknown three-loop
$I_{ij}^\three(z) = 0$ (solid orange).
We use the \texttt{MMHT2014nnlo68cl}~\cite{Harland-Lang:2014zoa} NNLO PDFs
and four-loop running of $\as$ everywhere.
These evolution orders are sufficient to ensure the formal cancellation of the
$\mu_B$ dependence at N$^3$LL$'$, while at lower orders they amount to a higher-order effect.
Numerical results to N$^3$LL with PDFs and $\as$ evolution at corresponding lower
orders can be found in \refcite{Gaunt:2014cfa}.
The residual dependence on $\mu_B$ is noticeably reduced by about a factor of two
at N$^3$LL$'$ compared to NNLL$'$.
The missing three-loop constant terms will again add an additional source
of $\mu_B$ dependence due to its $\as^3(\mu_B)$ prefactor and also the scale
dependence of the PDFs, which however should not change the qualitative
picture.

\subsection{Beam function coefficients in the eikonal limit}
\label{sec:tau0_beam_eikonal}

We now obtain the beam function coefficients $I_{ij}^{(n)}(z)$
in the $z\to 1$ limit. As was already pointed out and exploited in the NNLO
calculation in \refscite{Gaunt:2014xga, Gaunt:2014cfa}, the beam function in
this limit is effectively determined by a matrix element of eikonal Wilson lines.
Here, we exploit a recently derived consistency relation~\cite{Lustermans:2019cau}
that explicitly relates the $I_{ij}^{(n)}(z\to 1)$ to the threshold soft
function to all orders in $\as$.
Consistency relations of this kind generically arise
from different factorization theorems that apply in different limits of the same
multi-differential cross section.
In particular, a soft or collinear matrix element of several arguments
will refactorize into a product (or convolution) of simpler pieces of fewer
arguments by taking a stronger limit.

We start by defining the color-singlet lightcone momenta $q^\mp$
and corresponding momentum fractions $x_\mp$,
\begin{equation}
q^- \equiv \bn\cdot q = \sqrt{Q^2 + q_T^2}\, e^{+Y}
\,, \qquad
q^+ \equiv n\cdot q = \sqrt{Q^2 + q_T^2}\, e^{-Y}
\,, \qquad
x_{\mp} = \frac{q^{\mp}}{\Ecm}
\,.\end{equation}
As recently shown in \refcite{Lustermans:2019cau}, in the generalized threshold
limit $x_- \to 1$ but generic $x_+$, the inclusive color-singlet cross section differential in $q^\pm$
factorizes as
\begin{equation} \label{eq:factorization_generalized_threshold_qm_qp}
\frac{\df \sigma}{\df q^- \df q^+}
= \sum_{a,b} H_{ab}(q^+ q^-, \mu) \int \! \df t  \, f_a^\thr \Bigl[ x_- \Bigl(1 + \frac{t}{q^+  q^-}\Bigr), \mu\Bigr] B_b(t, x_+, \mu) \,
\bigl[ 1 + \ord{1-x_-} \bigr]
\,.\end{equation}
Here, $H_{ab}$ is the same hard function as in \eq{Tau0_fact}, and $B_b$ is
the same inclusive beam function as in \eq{Tau0_fact}.
The threshold PDF $f^\thr_a(x)$ encodes the extraction of parton $a$ from the proton for $x \to 1$.

On the other hand, in the well-known and stronger soft threshold limit,
where both $x_- \to 1$ and $x_+ \to 1$,
the cross section factorizes as~\cite{Catani:1989ne, Ravindran:2006bu, Westmark:2017uig, Banerjee:2017cfc, Banerjee:2018vvb}
\begin{align} \label{eq:factorization_soft_threshold_qm_qp}
\frac{\df \sigma}{\df q^- \df q^+}
&= \sum_{a,b} H_{ab}(q^+ q^-, \mu) \int \! \df k^- \df k^+ \,
f^\thr_a\Bigl[x_- \Bigl(1 + \frac{k^-}{q^-}\Bigr), \mu\Bigr] \,
f^\thr_b\Bigl[x_+ \Bigl(1 + \frac{k^+}{q^+}\Bigr), \mu\Bigr]
\nn \\ &\quad \times
S^\thr_i(k^-, k^+, \mu) \, \bigl[ 1 + \ord{1-x_-, 1-x_+} \bigr]
\,.\end{align}
The new ingredient is the threshold soft function $S^\thr_i(k^-, k^+, \mu)$.
It describes the process-independent contribution of soft emissions with total
lightcone momenta $k^+ = n\cdot k$ and $k^- = \bn\cdot k$.
It also only depends on the color representation
$c \equiv i = \{q,g\}$ of the incoming partons.

The threshold soft function is defined as a vacuum matrix element of Wilson lines
that are invariant under longitudinal boosts, and therefore satisfies the rescaling property
\begin{align} \label{eq:threshold_soft_rescaling}
S^\thr_i(k^-, k^+, \mu) = S^\thr_i(e^{+y} k^-, e^{-y} k^+, \mu)
\,.\end{align}
More specifically, in the context of SCET, the soft function is invariant under RPI-III transformations~\cite{Manohar:2002fd, Marcantonini:2008qn}.
Exploiting this property, the soft function can be extracted~\cite{Ravindran:2006bu,
Ahmed:2014uya, Li:2016axz, Li:2016ctv, Dulat:2018bfe}
from the soft-virtual limit of the total color-singlet production cross section
$\df \sigma / \df Q^2$, which is known to $\ord{\as^3}$~\cite{Anastasiou:2014vaa, Li:2014afw}.
In \app{threshold_soft}, we review this procedure and give explicit results for
$S^\thr_i(k^-, k^+, \mu)$ to three loops.

The factorization theorems \eqs{factorization_generalized_threshold_qm_qp}{factorization_soft_threshold_qm_qp} describe the same cross section and share a number of common ingredients.
In particular, only the beam function depends on $x_+$ in \eq{factorization_generalized_threshold_qm_qp}.
Further expanding \eq{factorization_generalized_threshold_qm_qp} in the limit $x = x_+ \to 1$,
it must reproduce \eq{factorization_soft_threshold_qm_qp}. As a result,
the eikonal $x\to 1$ limit of the beam function must coincide with the threshold soft function~\cite{Lustermans:2019cau},
\begin{equation} \label{eq:TauN_beam_eikonal_limit_hadronic}
B_i(t, x, \mu)
= \int \! \frac{\df k}{\omega} \, S^\thr_i\Bigl(\frac{t}{\omega}, k, \mu\Bigr)\,
f_i^\thr\Bigl[x \Bigl(1 + \frac{k}{\omega}\Bigr), \mu\Bigr] \,
\bigl[ 1 + \ord{1-x} \bigr]
\,.\end{equation}
Replacing $f^\thr_i[x(1+1-z)]$ by $f_i(x/z)/z$, which is justified at leading power in $1-z$,
yields the corresponding relation for the matching coefficients~\cite{Lustermans:2019cau},
\begin{equation} \label{eq:TauN_beam_eikonal_limit_partonic}
\cI_{ij}(t, z, \mu) = \delta_{ij} \, S_i^\thr\Bigl[\frac{t}{\omega},\omega(1-z), \mu\Bigr] \,
\bigl[ 1 + \ord{1-z} \bigr]
\,.\end{equation}
This relation captures all terms in $\cI_{ij}(t, z, \mu)$ that are singular for $z \to 1$,
while power corrections have at most an integrable singularity for $z \to 1$.
Notably, the beam function becomes flavor diagonal as $z \to 1$,
while offdiagonal channels are $\ord{1-z}$ suppressed.
By \eq{TauN_beam_eikonal_limit_partonic}, the matching coefficients also inherit
the rescaling property in \eq{threshold_soft_rescaling},
i.e., in the limit $z \to 1$, they become invariant under a simultaneous
rescaling $t \mapsto e^{+y} t$ and $1-z \mapsto e^{-y}(1-z)$. In other words,
they are symmetric in $t/\w$ and $\w(1-z)$ such that the dependence on $\w$
cancels on the right-hand side.

In \refcite{Lustermans:2019cau}, \eq{TauN_beam_eikonal_limit_partonic} was explicitly
confirmed at two loops by comparison to \refscite{Gaunt:2014xga,Gaunt:2014cfa}.
We now use it to predict the beam function coefficients in the eikonal limit at three loops.
They are given by the coefficient of $\delta(k^-)$ in the threshold soft function
upon identifying $\delta(k^+) \mapsto \delta(1-z)$ and $\cL_n(k^+, \mu) \mapsto \cL_n(1-z)$.
Including the one-loop and two-loop results for reference, we find
\begin{align} \label{eq:TauN_beam_eikonal_limit_result}
I_{ij}^\one(z)
&= \delta_{ij} \Bigl[
   \cL_1(1-z)\, \Gamma_0^i
   + \delta(1-z)\, s^{\thr\one}_i
\Bigr]
 + \Ordsq{(1-z)^0}
\,, \nn \\
I_{ij}^\two(z)
&= \delta_{ij}
\biggl\{
   \cL_3(1-z) \, \frac{(\Gamma_0^i)^2}{2}
   - \cL_2(1-z) \, \frac{\Gamma_0^i}{2} \beta_0
   + \cL_1(1-z) \Bigl[
      - 2\zeta_2 (\Gamma_0^i)^2
      + \Gamma_1^i
      + \Gamma_0^i \, s^{\thr\one}_i
   \Bigr]
\nn \\ &\quad \qquad
   + \cL_0(1-z) \Bigl[
      2\zeta_3 (\Gamma_0^i)^2
      + \frac{\gamma_{S\,1}^{i}}{2}
      - \beta_0\, s^{\thr\one}_i
   \Bigr]
   + \delta(1-z) \, s^{\thr\two}_i
\biggr\}
 + \Ordsq{(1-z)^0}
\,, \nn \\
I_{ij}^\three(z)
&= \delta_{ij}
\biggl\{
   \cL_5(1-z)\, \frac{(\Gamma^i_0)^3}{8}
   - \cL_4(1-z)\, \frac{5}{12}(\Gamma^i_0)^2 \beta_0
\nn \\ & \quad \qquad
   + \cL_3(1-z)\, \Gamma^i_0 \Bigl[
      - 2\zeta_2 (\Gamma^i_0)^2
      + \frac{\beta_0^2}{3}
      + \Gamma_1^i
      + \frac{\Gamma_0^i}{2} s_i^{\thr\one}
   \Bigr]
\nn \\ & \quad \qquad
   + \cL_2(1-z)\,
   \Bigl[
      (\Gamma^i_0)^2 (5\zeta_3 \Gamma^i_0 + 3\zeta_2 \beta_0)
      - \beta_0 \Gamma^i_1
      - \frac{\Gamma^i_0}{2} \Bigl(
         \beta_1
         - \frac{3}{2}\gamma^i_{S\, 1}
         + 4\beta_0\, s_i^{\thr\one}
      \Bigr)
   \Bigr]
\nn \\ & \quad \qquad
   + \cL_1(1-z)\,
   \Bigl[
      (\Gamma^i_0)^2 (4\zeta_4\Gamma^i_0 - 6\zeta_3\beta_0)
      - 4\zeta_2 \Gamma^i_0 \Gamma^i_1
      - \beta_0 \gamma^i_{S\,1}
      + \Gamma^i_2
\nn \\ & \quad \qquad \hspace{13ex}
      + \bigl(
         -2\zeta_2(\Gamma^i_0)^2
         + 2\beta_0^2
         + \Gamma^i_1
      \bigr) s_i^{\thr\one}
      + \Gamma^i_0\, s_i^{\thr\two}
   \Bigr]
\nn \\ & \quad \qquad
   + \cL_0(1-z)\,
   \Bigl[
      (\Gamma^i_0)^2 \bigl(
         - \Gamma^i_0 (8 \zeta_2 \zeta_3 - 6 \zeta_5)
         + 2\zeta_4 \beta_0
      \bigr)
      + \Gamma^i_0 (4\zeta_3\Gamma^i_1 - \zeta_2 \gamma^i_{S\,1})
      + \frac{\gamma^i_{S\,2}}{2}
\nn \\ & \quad \qquad \hspace{13ex}
      + \Bigl(
         (\Gamma^i_0)^2 2\zeta_3
         + \Gamma^i_0 2\zeta_2 \beta_0
         - \beta_1 + \frac{\gamma^i_{S\,1}}{2}
      \Bigr) s_i^{\thr\one}
      - 2\beta_0\, s_i^{\thr\two}
   \Bigr]
\nn \\ & \quad \qquad
   + \delta(1-z) \, s^{\thr\three}_i
\biggr\}
 + \Ordsq{(1-z)^0}
\,.\end{align}
The boundary coefficients $s^{\thr(n)}_i$ of the threshold soft function are given in \eq{threshold_soft_finite_terms}.
We have exploited that the noncusp anomalous dimension of the threshold soft function
is given by $-\gamma_S^i(\as)$, see \app{threshold_soft_three_loops}.
For brevity, we also used that $\gamma_{S\,0}^{i} = 0$.
The result for generic $\gamma_{S\,0}^{i}$ can be read off from
the full expression for the threshold soft function in \eq{threshold_soft_n3lo}.

The three-loop result in \eq{TauN_beam_eikonal_limit_result} is new and a
genuine prediction of the consistency relation in
\eq{TauN_beam_eikonal_limit_partonic}. We stress that the information provided
by it goes beyond the RGE predicted three-loop structure in
\eq{TauN_beamI_N3LO}. The fact that the leading $z\to 1$ terms must be symmetric
in $t/\w$ and $\w(1-z)$ allows one to directly determine (or check) the
$\delta(t)\cL_n(1-z)$ terms from the RGE-predicted $\cL_n(t)\delta(1-z)$ terms,
which was already noted in \refscite{Procura:2011aq, Gaunt:2014xga}. However,
the $\delta(t)\delta(1-z)$ coefficient cannot be predicted in this way, and
\eq{TauN_beam_eikonal_limit_partonic} explicitly identifies it with the
threshold soft function coefficients $s^{\thr\three}_i$.

As was shown in \refcite{Lustermans:2019cau}, a factorization theorem analogous
to \eq{factorization_generalized_threshold_qm_qp} also holds for the inclusive
cross section differential in $Q$ and $Y$, with $B_i$ replaced by a closely
related, modified beam function $\tilde{B}_i(t, x, \mu)$.%
\footnote{Not to be confused with the $q_T$ beam function $\tilde B_i(x, \bt, \mu, \nu)$
in the following section.}
Note that in contrast to
\eqs{fact_generic}{Tau0_fact}, here the difference between $q^\pm$ and $(Q, Y)$
matters. The RGE for $\tilde{B}_i(t, x, \mu)$ is the same as for $B_i(t, x,
\mu)$ in \eq{TauN_beam_RGE}, and hence \eq{TauN_beamI_N3LO} also holds for
$\tilde{B}_i$ just with different boundary coefficients $\tilde
I_{ij}^{(n)}(z)$. In the limit $z \to 1$, the difference between $B_i$ and
$\tilde{B}_i$ becomes power suppressed in $1-z$. As a result, the $z\to 1$ limit
of the modified $\tilde I_i^{(n)}$ is also given by
\eq{TauN_beam_eikonal_limit_result}.

\subsection{Estimating beam function coefficients beyond the eikonal limit}
\label{sec:tau0_beam_beyond_eikonal}

Having the eikonal limit of the beam function coefficients at hand, we can study
to what extent it can be used to approximate the full result and/or estimate
the uncertainty due to the missing terms beyond the eikonal limit.

\begin{figure*}
\centering
\includegraphics[width=\WidthTwoSubfigs]{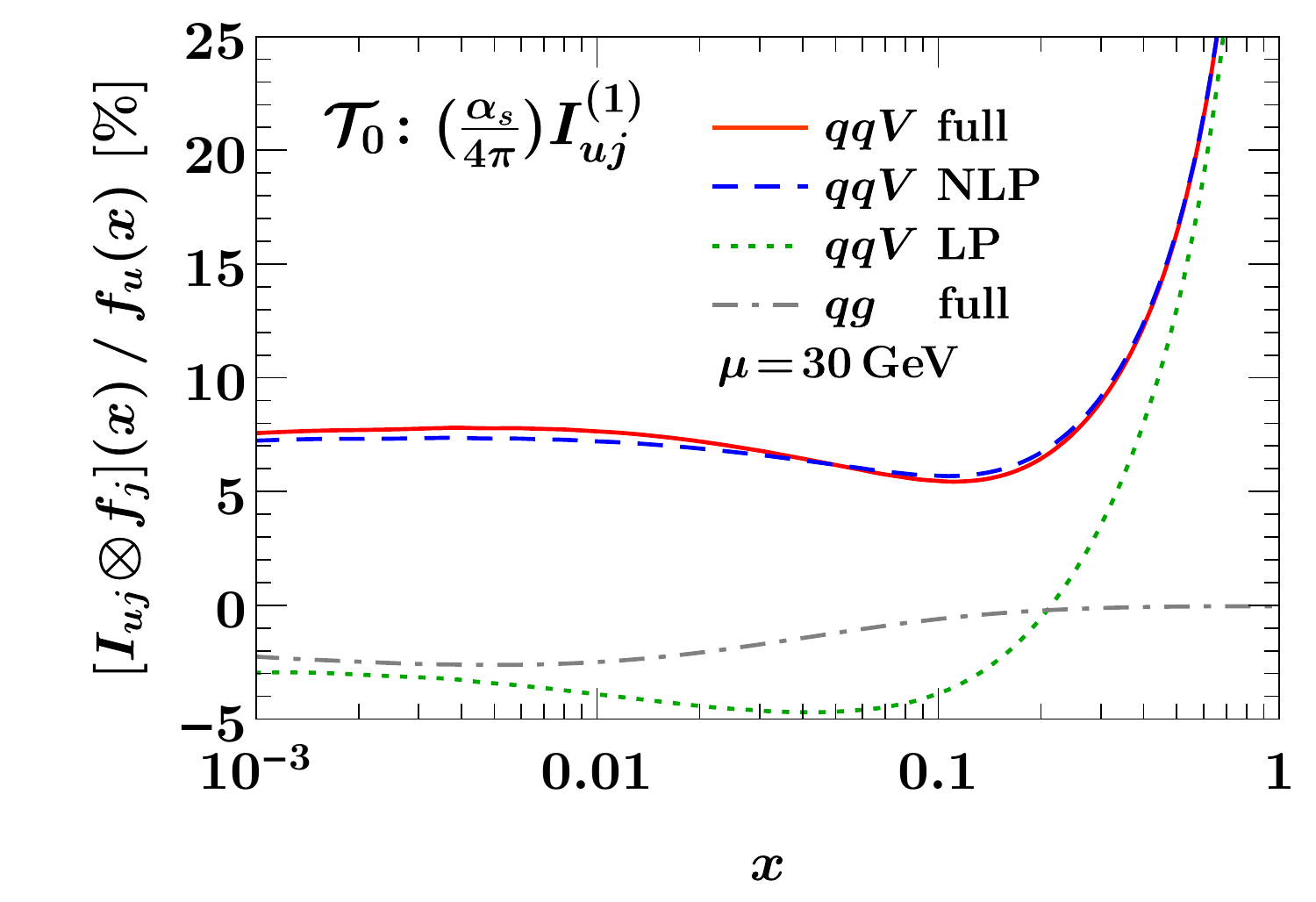}%
\hfill%
\includegraphics[width=\WidthTwoSubfigs]{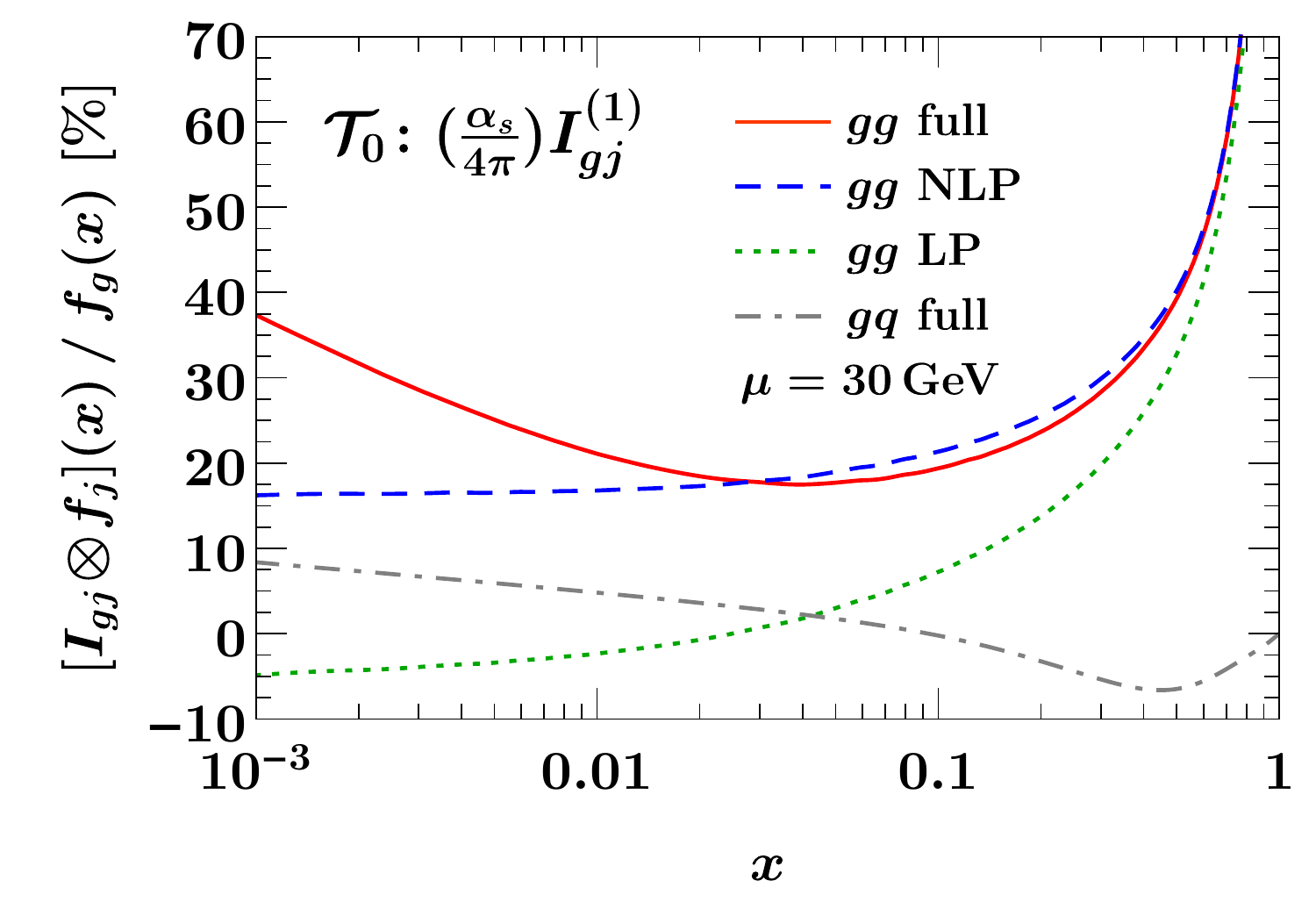}%
\\
\includegraphics[width=\WidthTwoSubfigs]{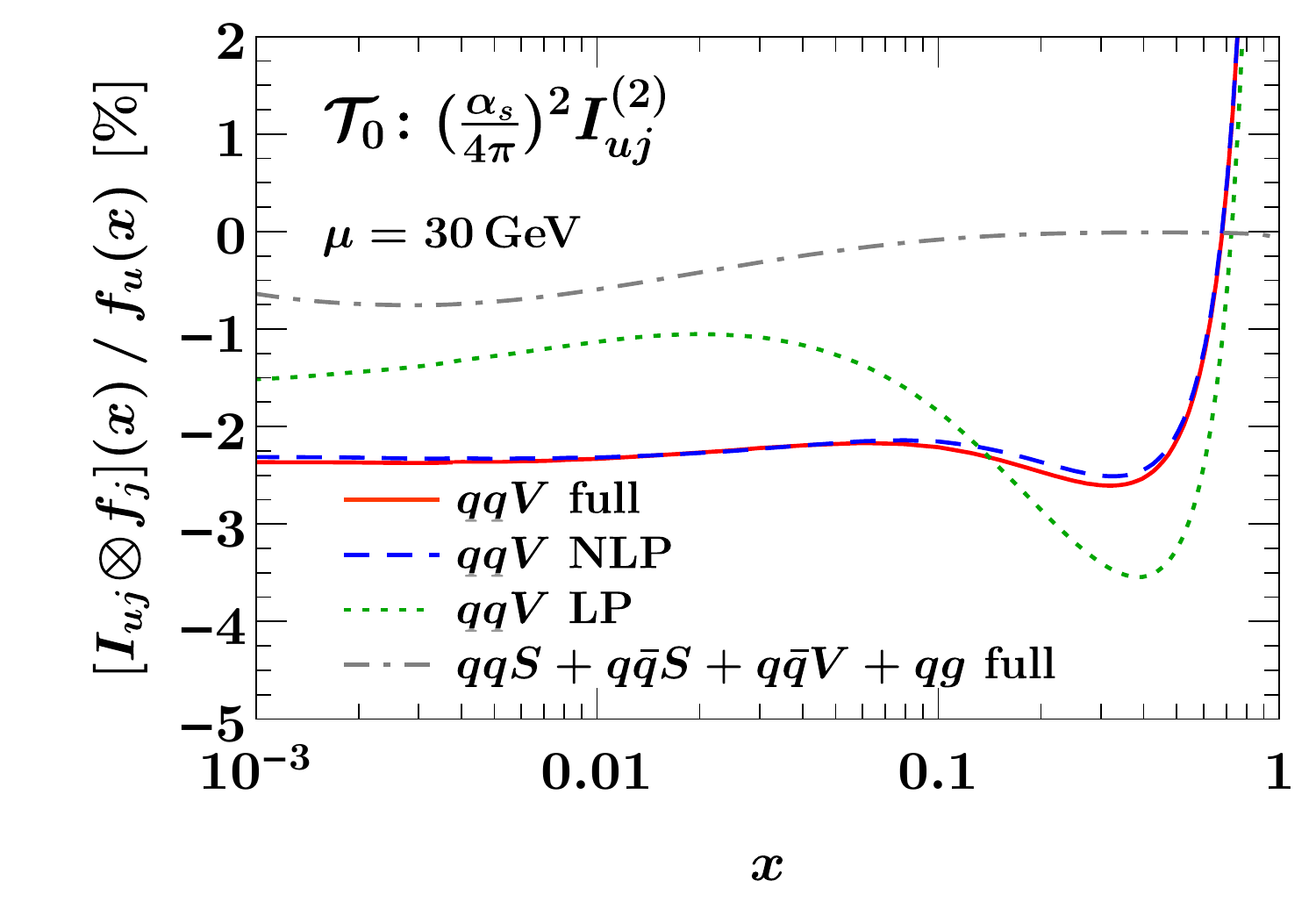}%
\hfill%
\includegraphics[width=\WidthTwoSubfigs]{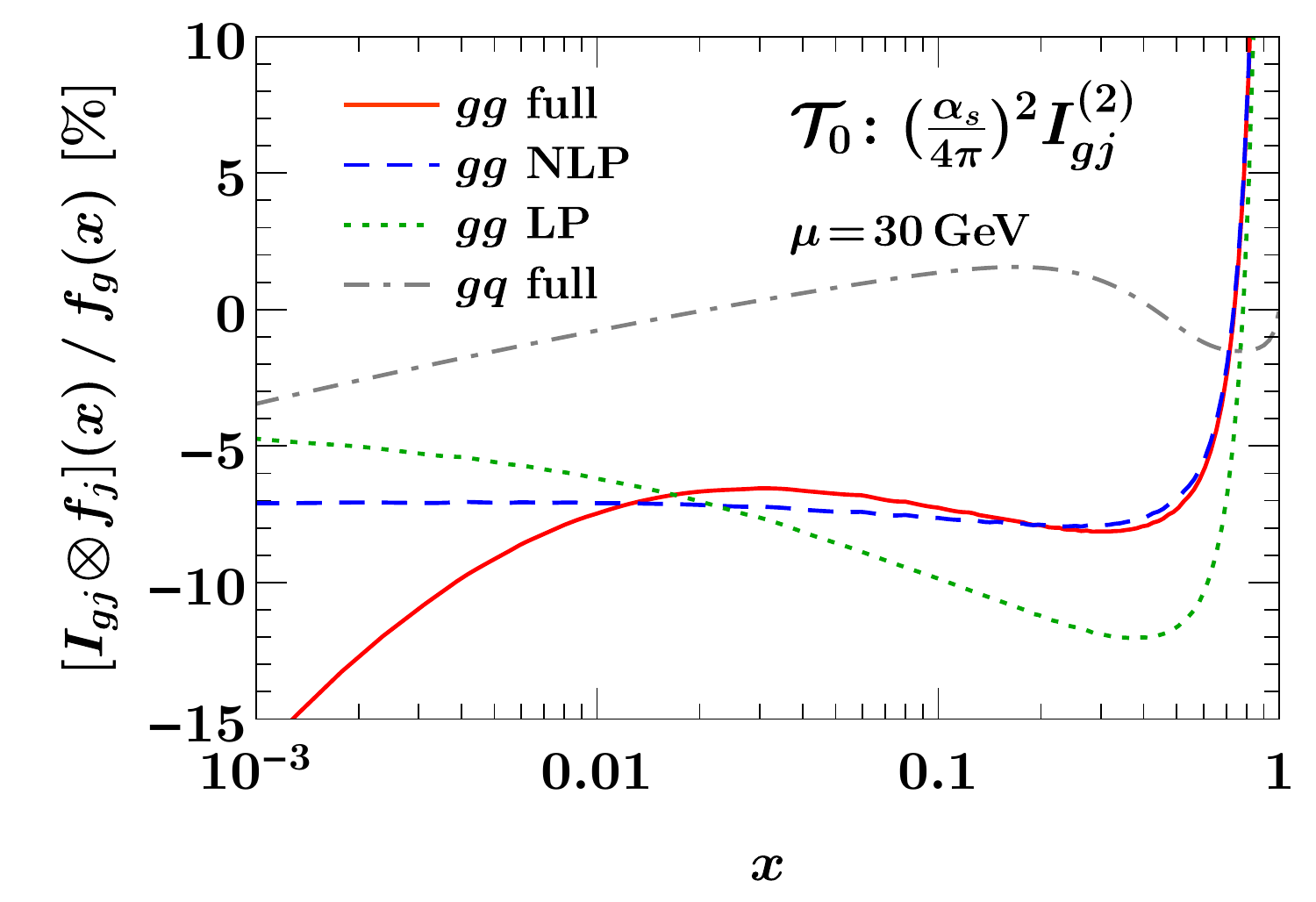}%
\caption{Comparison of the full beam function coefficients to their leading eikonal (LP)
and next-to-eikonal (NLP) expansion at NLO (top) and NNLO (bottom). The
$u$-quark channel is shown on the left and the gluon channel on the right. In
each case we also show the sum of all nondiagonal partonic channels for comparison.
}
\label{fig:TauKernelpowerexp}
\end{figure*}

In \fig{TauKernelpowerexp}, we compare the full $\Tau_0$ beam function coefficient (solid)
to its eikonal (LP dotted green) and next-to-eikonal (NLP dashed blue) expansions
at NLO and NNLO for the $u$ quark and gluon channels. We always show the convolution
$(I_{ij}\otimes f_j)(x) / f_i(x)$ with the appropriate PDF $f_j$ and normalize to the PDF $f_i(x)$,
corresponding to the LO result, where $i=u$ for the $u$-quark case and $i=g$ for the gluon case.
With this normalization,
the shape gives an indication of the rapidity dependence of the beam function coefficient
relative to the LO rapidity dependence induced by the shape of the PDFs.
We also include the appropriate powers of $\alpha_s/(4\pi)$ at each order, so the overall
normalization shows the percent impact relative to the LO result.
For definiteness, we choose $\mu = 30\,\GeV$ for the scale entering the PDFs and $\alpha_s$.

The eikonal
approximation reproduces the correct divergent behavior of the full flavor-diagonal
contributions, denoted as $qqV$ and $gg$, toward large $x$ but is
off away from large $x$. On the other hand, including the next-to-eikonal terms
yields an excellent approximation for all $x$, particularly for the quark beam function.
The rise at very small $x$ for the gluon, which is not reproduced at NLP, is due to
the $z\to 0$ divergent behavior in the gluon coefficient, which is not reproduced by
its $z\to 1$ expansion. If desired, it can be captured by including the leading
$z\to 0$ behavior of the coefficients, which for simplicity we refrain from doing here.
For illustration, we also show the total
contribution from all other corresponding nondiagonal channels (gray dot-dashed).
In each case, they are numerically subdominant to the flavor-diagonal channel
and also much flatter in $x$, since they only start at NLP.

The fact that the NLP result reproduces the full result very well,
motivates us to construct an approximate ansatz for it, which we can then use
at three loops to get a good estimate of the size of the unknown three-loop
beam function coefficient beyond the eikonal limit.

We consider the following ansatz to approximate the coefficient,
\begin{align} \label{eq:tau0_ansatz}
I_{ij,\mathrm{approx}}^{(n)}(z)
&= I_{ij}^{\mathrm{LP}(n)}(z) + I_{ij,\mathrm{approx}}^{\mathrm{NLP}(n)}(z)
   + X_2 \, (1-z) I_{ij,\mathrm{approx}}^{\mathrm{NLP}(n)}(z)
\,,\end{align}
where the NLP coefficient itself is approximated as
\begin{align} \label{eq:tau0_NLP}
I_{ij,\mathrm{approx}}^{\mathrm{NLP}(n)}(z)
&= - (1-z) I_{ij}^{\mathrm{LP}(n)}(z)
   + X_1 \, (1-z) \frac{\df}{\df(1-z)} \Bigl[(1-z) I_{ij}^{\mathrm{LP}(n)}(z)\Bigr]
\,,\end{align}
and $X_1$ and $X_2$ are free parameters that can be varied to estimate residual uncertainties.
This ansatz is motivated by the known general logarithmic structure at NLP
\begin{align}
 I_{ij}^{\mathrm{NLP}(n)}(z) = \sum_{k=0}^{2n-1} c_{n,k}^\mathrm{NLP} \ln^k(1-z)
\,.\end{align}
By multiplying the LP term by $(1-z)$ in \eq{tau0_NLP}, we generate the appropriate logarithmic
structure at NLP. The first term in \eq{tau0_NLP} reproduces the correct NLO and NNLO coefficients
for the leading logarithm at NLP $c_{1,1}^\mathrm{NLP} = - 4 C_i$
and $c_{2,3}^\mathrm{NLP} = -8 C_i^2$ for both quarks and gluons. Here, the additional
double logarithm is determined by the same power of $(\Gamma_0^i)^n$ as at LP, and this pattern
can be expected to hold at higher orders. The second term in \eq{tau0_NLP}
generates a next-to-leading logarithmic NLP series. We fix the central value for $X_1 = 1$
to reproduce the NLL constant term at NLO $c_{1,0} = 4 C_i$. Interestingly, we
find that this choice also reproduces all NNLO coefficients $c_{2,k}$ very well,
typically to within 10\%, for both quarks and gluons and also independently of
the choice of $n_f$. This provides a very nontrivial check and so we expect that
\eq{tau0_NLP} provides a very good model of the true NLP structure also at higher
orders. To estimate the uncertainties, we vary $X_1$ by $\pm 0.5$, which
effectively varies the coefficients of the subleading terms. At NNLO, this variation
covers the exact value for all coefficients.
In addition, the last term in \eq{tau0_ansatz} estimates the possible effect of
terms beyond NLP. Here, we simply take the central value $X_2 = 0$ and vary it by $\pm 1$.

Since $X_i$ probe independent structures, we can consider them as
uncorrelated. Hence, we add the impacts $\Delta_i$ on the final result of their
variation in quadrature
\begin{align}
\Delta = \Delta_{1} \oplus \Delta_{2} = \sqrt{\Delta_{1}^2 + \Delta_{2}^2}
\,.
\end{align}

\begin{figure*}
\centering
\includegraphics[width=\WidthTwoSubfigs]{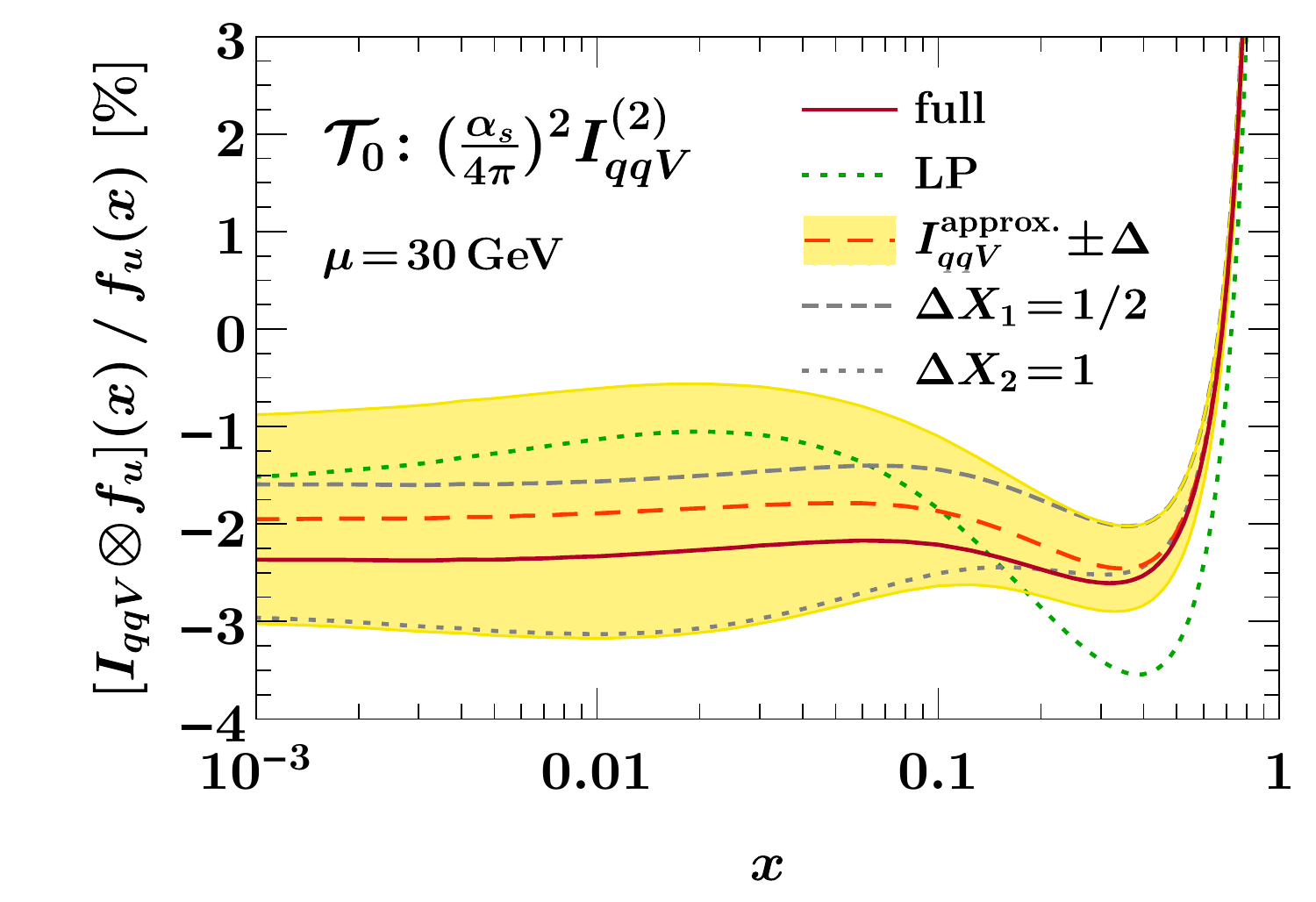}%
\hfill%
\includegraphics[width=\WidthTwoSubfigs]{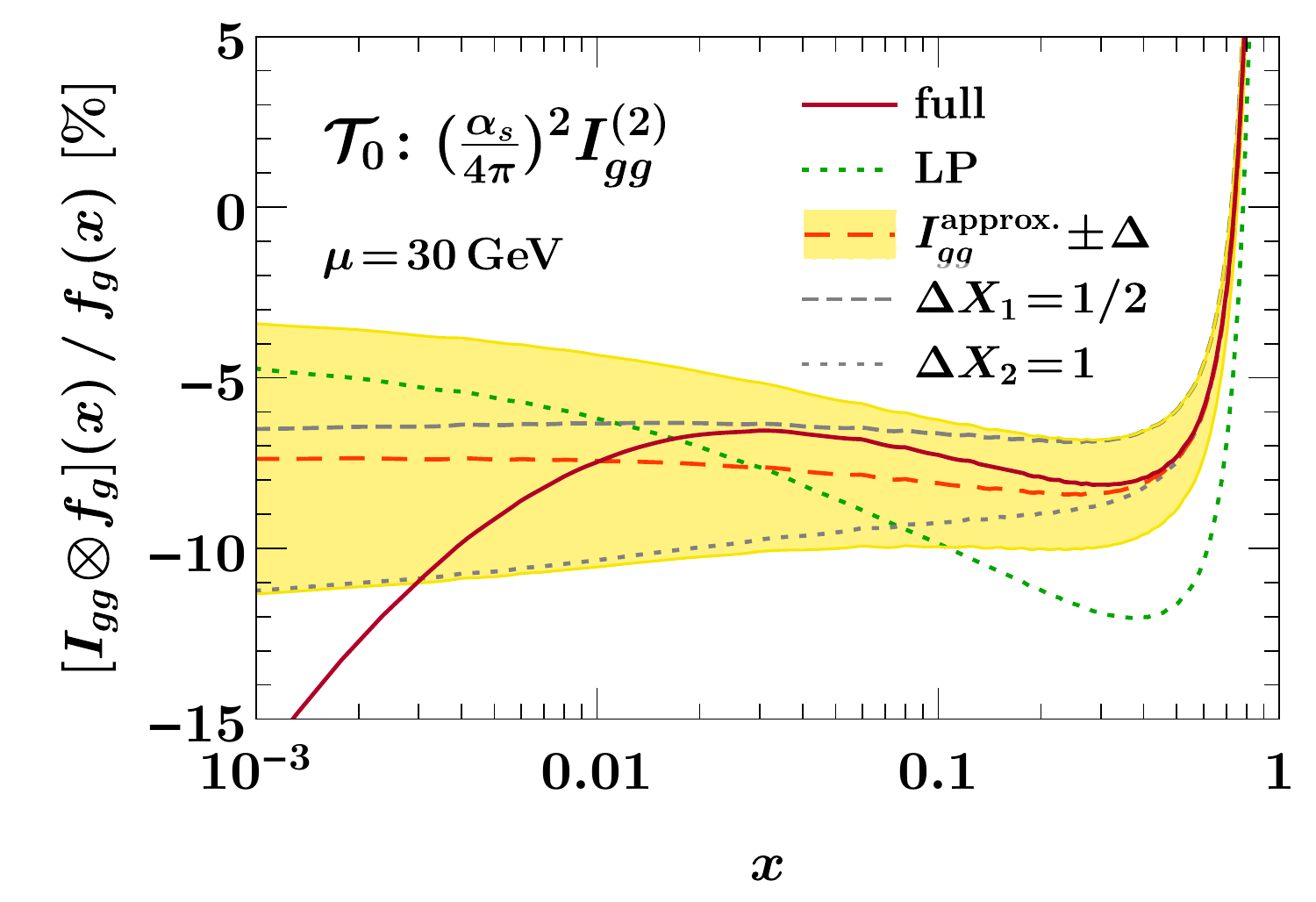}%
\\
\includegraphics[width=\WidthTwoSubfigs]{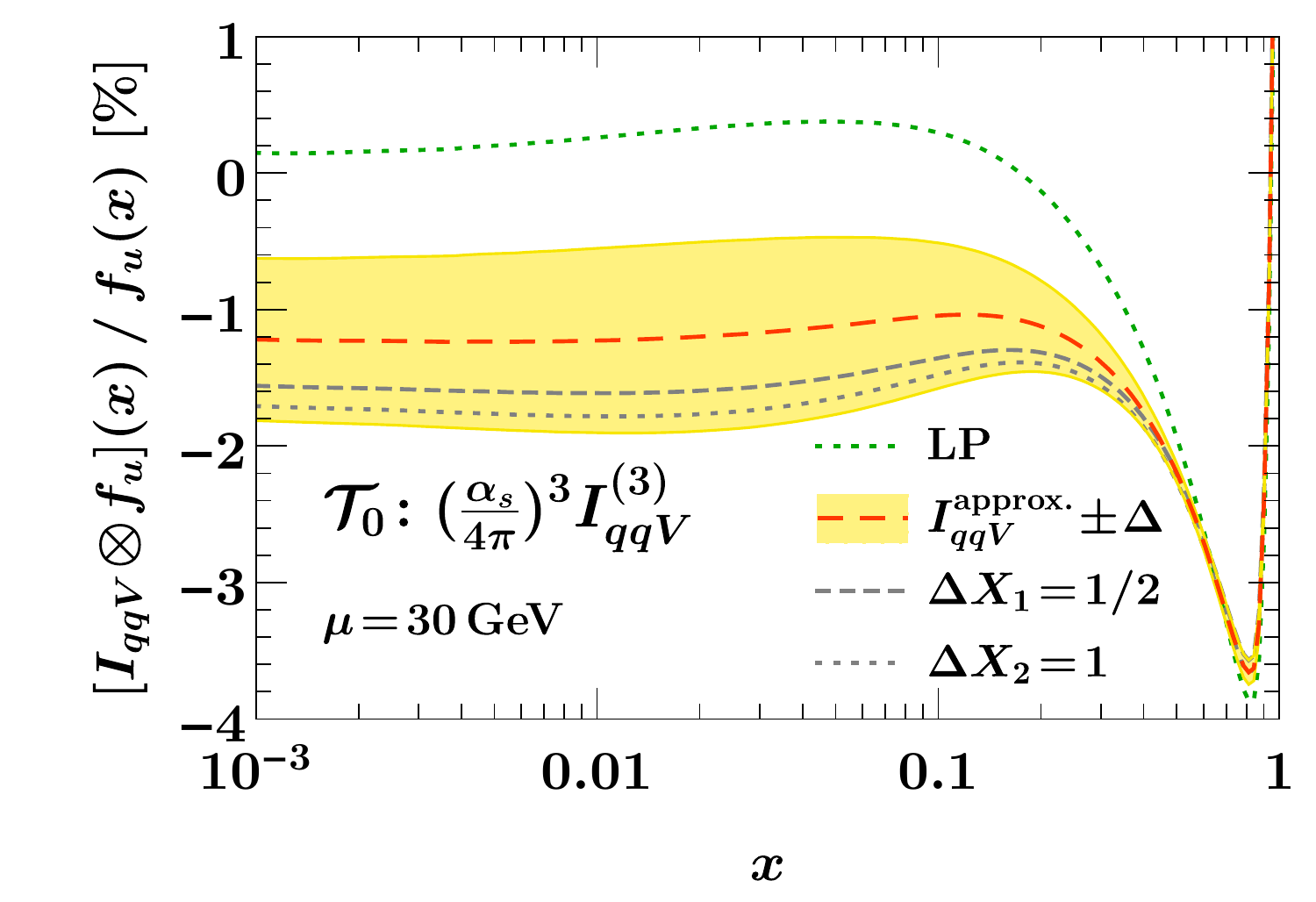}%
\hfill%
\includegraphics[width=\WidthTwoSubfigs]{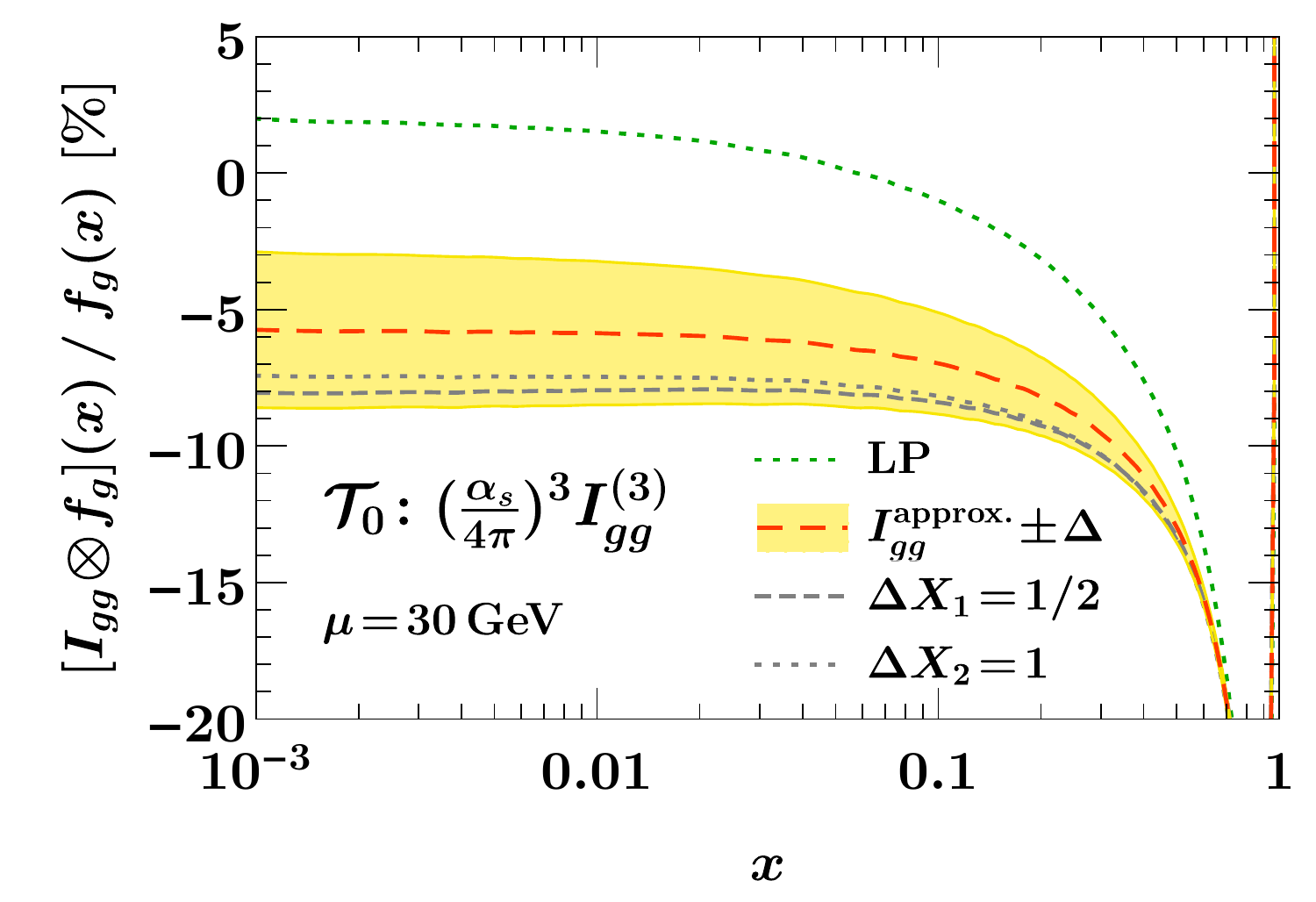}%
\caption{Approximate ansatzes for the NNLO (top) and N$^3$LO (bottom) kernels, in the $u$-quark (left) and gluon (right) channels.}
\label{fig:TauKernelApprox}
\end{figure*}

In \fig{TauKernelApprox}, we show the approximate kernel at NNLO (top)
and N$^3$LO (bottom) for the $u$-quark (left) and gluon (right) channels.
The dashed orange line shows the central result from our ansatz and the yellow band
its estimated uncertainty. The gray lines show the impact of the individual
variations of the $X_i$ as indicated.
In the top panel (NNLO), we also show the known
full two-loop results (red solid). It shows that the ansatz including uncertainties
approximates the true result very well, except for the gluon at very small $x$
where we do not expect it to hold.

At N$^3$LO, we see that the approximate result gives rise to a sizable shift
from the pure eikonal limit, which we believe to be genuine. Hence, we expect the
full three-loop coefficients to have a nontrivial impact in the one to few percent range.
The uncertainties at N$^3$LO are reduced compared to NNLO as expected, but
are still sizable, which adds motivation for the exact calculation of the
full three-loop coefficients.

\FloatBarrier

\section{\texorpdfstring{$q_T$}{qT} factorization to three loops}
\label{sec:qT}

\subsection{Factorization theorem}
\label{sec:qT_factorization}

The factorization of the $\qt$ distribution in the limit $q_T \equiv \abs{\qt} \ll Q$ was first
established by Collins, Soper, and Sterman (CSS)~\cite{Collins:1981uk,
Collins:1981va, Collins:1984kg}, and was later elaborated on in
\refscite{Catani:2000vq, deFlorian:2001zd, Catani:2010pd, Collins:1350496}.
The factorization for $q_T$ was also shown within the framework of SCET in
\refscite{Becher:2010tm, GarciaEchevarria:2011rb, Chiu:2012ir, Li:2016axz}.
Sometimes it is also referred to as transverse-momentum dependent (TMD) factorization.
We write the factorized cross section as
\begin{align} \label{eq:qt_factorization}
\frac{\df \sigma}{\df Q^2 \df Y \df^2 \qt}
&= \int\!\df^2\bt \, e^{\img\,\qt \cdot \bt} \,
\frac{\df\tilde\sigma^\mathrm{sing}(\vec{b}_T)}{\df Q^2 \df Y}
\Bigl[ 1 + \cO\Bigl(\frac{q_T^2}{Q^2}\Bigr)\Bigr]
\,, \nn \\
\frac{\df \tilde\sigma^\mathrm{sing}(\vec{b}_T)}{\df Q^2 \df Y}
&= \sum_{a,b} H_{ab}(Q^2,\mu)
\tilde B_a(x_a, \bt, \mu, \nu)\, \tilde B_b(x_b, \bt, \mu, \nu)\,
\tilde S_i(b_T, \mu, \nu)
\,.\end{align}
It receives power corrections suppressed by $q_T^2/Q^2$ as indicated.
As is common, we consider the factorized singular cross section in Fourier-conjugate $\bt$ space,
where convolutions in $\qt$ space turn into simple products. In particular, Fourier transforming
the $\qt$-dependent plus distributions $\cL_n(\qt, \mu)$ turns them into powers of the canonical
$\bt$-space logarithms, which we denote as
\begin{equation} \label{eq:LogB}
L_b = \ln\frac{b_T^2 \mu^2}{b_0^2} \,,\qquad b_0 = 2 e^{-\gamma_E}
\,.\end{equation}
More details on their Fourier transformation are given in \app{plus_distr_qt}.

The $q_T$ factorization is affected by rapidity divergences that must be regulated
by a dedicated rapidity regulator. This gives rise to an additional rapidity scale,
denoted as $\nu$ in \eq{qt_factorization}.
We use the exponential regulator of \refcite{Li:2016axz}, which up to two loops
gives results equivalent to the $\eta$ regulator of \refscite{Chiu:2011qc, Chiu:2012ir}.

The beam function appearing in \eq{qt_factorization} is the inclusive transverse-momentum dependent
(SCET$_{\rm II}$) beam function, which also appears in the $q_T$ factorization of $Z+j$
and $\gamma+j$~\cite{Buffing:2018ggv, Chien:2019gyf}.
The $q_T$-dependent soft function in \eq{qt_factorization} is the renormalized
vacuum matrix element of two incoming soft Wilson lines.
Note that for simplicity, we generically refer to them as $q_T$ beam and soft functions, even though
we mostly consider their $b_T$-dependent Fourier conjugates.
The $q_T$ beam and soft functions are known at two loops for several regulators
\cite{Catani:2011kr, Catani:2012qa, Gehrmann:2012ze, Gehrmann:2014yya, Echevarria:2015byo, Luebbert:2016itl, Echevarria:2016scs, Luo:2019hmp}.
The soft function is known at three loops using the exponential regulator~\cite{Li:2016ctv}.

We also note that one can equivalently define $\nu$-independent TMDPDFs as
\begin{equation}
 \tilde f_i(x,\bt,\mu,\zeta) = \tilde B_i(x, \bt, \mu, \nu) \sqrt{\tilde S_i(b_T,\mu,\nu)}
\,,\qquad
 \zeta = \omega^2 = (x \Ecm)^2
\,,\end{equation}
as is done e.g.\ in \refscite{Collins:1984kg, Catani:2000vq, deFlorian:2001zd, Catani:2010pd, Collins:1350496, Becher:2010tm, GarciaEchevarria:2011rb}.
Here, the Collins-Soper scale~\cite{Collins:1981uk,Collins:1981va} $\zeta = \omega^2$ is given
in terms of the lightcone momentum $\omega = x P^-$ carried by the struck parton.%

\subsection{Rapidity anomalous dimension}
\label{sec:gamma_nu}

The $\nu$ dependence of the beam and soft functions is encoded in their rapidity RGEs~\cite{Chiu:2012ir},
\begin{align}
\nu \frac{\df}{\df\nu} \tilde B_i(x, \bt, \mu, \nu)
&= \tilde\gamma_{\nu,B}^i(b_T,\mu)\, \tilde B_i(x, \bt, \mu, \nu)
\,,\nn\\
\nu \frac{\df}{\df\nu} \tilde S_i(b_T, \mu, \nu)
&= \tilde\gamma_{\nu,S}^i(b_T,\mu)\, \tilde S_i(b_T, \mu, \nu)
\,,\end{align}
where $\tilde\gamma_{\nu,B}^i$ and $\tilde\gamma_{\nu, S}^i$ are the beam and soft rapidity anomalous dimensions, which are closely related to the Collins-Soper kernel \cite{Collins:1981uk,Collins:1981va}.
Because the cross section in \eq{qt_factorization} is independent of $\nu$, they are related by
\begin{align}
\tilde\gamma_\nu^{i}(b_T,\mu) \equiv \tilde\gamma_{\nu,S}^i(b_T,\mu) = -2 \tilde\gamma_{\nu,B}^i(b_T,\mu)
\,,\end{align}
and we will simply refer to $\tilde\gamma_\nu^{i}(b_T,\mu)$ as the rapidity anomalous dimension.

An important property of $\tilde\gamma_\nu^{i}(b_T,\mu)$ is that like the soft function
it only depends on the color representation $i=\{q,g\}$ but not on the specific massless
quark flavor. While we only need its fixed-order expansion,
we note that it becomes genuinely nonperturbative for $b_T^{-1} \lesssim \Lambda_{\rm QCD}$,
and recently a proposal was made to calculate it nonperturbatively using lattice QCD \cite{Ebert:2018gzl,Ebert:2019okf}.

The rapidity anomalous dimension itself satisfies an RGE in $\mu$,
\begin{align} \label{eq:gNu_RGE}
\mu \frac{\df}{\df\mu} \tilde\gamma^i_\nu(b_T,\mu) = - 4 \GammaC^i[\as(\mu)]
\,,\end{align}
which predicts its all-order structure in $b_T$ and $\mu$.
Similar to the $\Tau_0$ soft function in \sec{tau0_soft}, it can be solved recursively
order by order in $\as$. Expanding both sides of \eq{gNu_RGE} to fixed order in $\as(\mu)$
and accounting for the running of $\as(\mu)$, the $(n+1)$-loop term is related to the lower-order terms by
\begin{equation}
\tilde\gamma_\nu^{i\,(n+1)}(b_T, \mu)
= -2 \Gamma^i_{n+1} L_b + \sum_{m=0}^{n} 2 (m+1) \,
\beta_{n-m} \int_{b_0/b_T}^\mu\! \frac{\df\mu'}{\mu'}\, \tilde\gamma_\nu^{i\,(m)}(b_T,\mu')
+ \tilde\gamma_{\nu\,n+1}^i
\,,\end{equation}
where the nonlogarithmic boundary coefficients are defined as
\begin{equation}
\tilde\gamma_{\nu\,n}^i = \tilde\gamma_\nu^{i\,(n)}(b_T, \mu=b_0/b_T)
\,.\end{equation}
The result up to three loops is
\begin{align} \label{eq:gammaNu_FO}
\tilde\gamma_\nu^{i\,(0)}(b_T,\mu)
&= -L_b\, 2\Gamma_0^i + \tilde\gamma_{\nu\,0}^i
\,,\nn\\
\tilde\gamma_\nu^{i\,(1)}(b_T,\mu)
&= - L_b^2\, \Gamma_0^i \beta_0
 + L_b \bigl(\beta_0 \tilde\gamma_{\nu\,0}^i -2\Gamma_1^i \bigr) + \tilde\gamma_{\nu\,1}^i
\,,\nn\\
\tilde\gamma_\nu^{i\,(2)}(b_T,\mu)
&= -L_b^{3}\, \frac{2}{3}\Gamma_0^i \beta_0^2
 + L_b^2 \bigl(\beta_0^2 \tilde\gamma_{\nu\,0}^i -2 \Gamma_1^i \beta_0 - \Gamma_0^i \beta_1 \bigr)
+ L_b \bigl(2 \beta_0 \tilde\gamma_{\nu\,1}^i + \beta_1 \tilde\gamma_{\nu\,0}^i - 2\Gamma_2^i \bigr)
\nn \\ & \quad
+ \tilde\gamma_{\nu\,2}^i
\,.\end{align}
The boundary coefficients $\tilde\gamma_{\nu\,n}^i$ are known up to three loops~\cite{Luebbert:2016itl, Li:2016ctv, Vladimirov:2016dll} and are summarized in \eq{gammaNuConstants}.

\subsection{Soft function}
\label{sec:qTsoft}

The soft function is explicitly known to three loops~\cite{Li:2016ctv}.
For completeness, we explicitly derive its fixed-order structure
to illustrate the joint solution of its $\mu$ and $\nu$ RGEs,
\begin{align} \label{eq:qt_soft_RGEs}
\mu\frac{\df}{\df\mu} \tilde S_i(b_T,\mu,\nu)
&= \tilde\gamma_S^i(\mu,\nu)\, \tilde S_i(b_T,\mu,\nu)
\,,\nn\\
\nu\frac{\df}{\df\nu} \tilde S_i(b_T,\mu,\nu)
&= \tilde\gamma_\nu^i(b_T, \mu)\, \tilde S_i(b_T,\mu,\nu)
\,.\end{align}
The perturbative structure of $\gamma_\nu^i$ is discussed in \sec{gamma_nu} above.
The $\mu$ anomalous dimension has the all-order structure
\begin{equation}
 \tilde\gamma_S^i(\mu,\nu) = 4 \GammaC^i[\as(\mu)] \ln\frac{\mu}{\nu} + \tilde\gamma_S^i[\as(\mu)]
\,,\end{equation}
where $\GammaC^i(\as)$ and $\tilde\gamma_S^i(\as)$ are the cusp and the soft
noncusp anomalous dimensions.
Expanding both sides of \eq{qt_soft_RGEs} order by order in $\as$, we obtain the
coupled RGEs
\begin{align} \label{eq:qt_soft_RGE_FO}
\mu\frac{\df}{\df\mu}{\tilde S_i^{(n+1)}(b_T,\mu,\nu)}
&= \sum_{m=0}^{n} \Bigl( 4 \Gamma^i_{n-m} \ln\frac{\mu}{\nu} + 2 m \beta_{n-m} + \tilde\gamma_{S\,n-m}^i \Bigr)
\tilde S_i^{(m)}(b_T,\mu,\nu)
\,, \nn \\
\nu\frac{\df}{\df\nu}{\tilde S_i^{(n+1)}(b_T,\mu,\nu)}
&= \sum_{m=0}^{n} \tilde\gamma_\nu^{i\,(n-m)}(b_T,\mu)\, \tilde S_i^{(m)}(b_T,\mu,\nu)
\,.\end{align}
These are easily integrated to give
\begin{align} \label{eq:qt_soft_rec}
\tilde S_i^{(n+1)}(b_T,\mu,\nu)
&= \sum_{m=0}^{n} \biggl[
   \int_{b_0/b_T}^\mu\! \frac{\df\mu'}{\mu'}\,
   \Bigl( 4 \Gamma^i_{n-m} \ln\frac{\mu'}{\nu} + 2 m \beta_{n-m} + \tilde\gamma_{S\,n-m}^i\Bigr)
   \tilde S_i^{(m)}(b_T,\mu',\nu)
\nn \\ & \quad
   + \int_{b_0/b_T}^\nu\! \frac{\df\nu'}{\nu'}\, \tilde\gamma_{\nu\,n-m}^i\, \tilde S_i^{(m)}(b_T,b_0/b_T,\nu')
\biggr]
 + \tilde s_{i}^{(n+1)}
\,,\end{align}
where we first integrated the $\nu$ RGE at fixed $\mu = b_0/b_T$ and then the $\mu$ RGE at
arbitrary $\nu$. In this way, the rapidity anomalous dimension reduces
to its boundary coefficients $\gamma_{\nu\,n}^i$.
The soft boundary coefficients in \eq{qt_soft_rec} are defined as
\begin{equation}
\tilde s_{i}^{(n)} = \tilde S_i^{(n)}(b_T, \mu = b_0/b_T, \nu = b_0/b_T)
\,.\end{equation}

Starting from the LO result, $\tilde s_i^\zero = 1$, and expressing the results in terms of
\begin{equation}
 L_b = \ln\frac{b_T^2 \mu^2}{b_0^2} \,,\quad b_0 = 2 e^{-\gamma_E}
\,,\qquad
 L_\nu = \ln\frac{\mu}{\nu}
\,,\end{equation}
\eq{qt_soft_rec} yields up to two loops
\begin{align} \label{eq:qt_soft_NNLO}
\tilde S_i^\zero(b_T,\mu,\nu) &= 1
\,, \nn \\
\tilde S_i^\one(b_T,\mu,\nu)
&= - L_b^2\, \frac{\Gamma_0^i}{2}
 + L_b \Bigl( L_\nu\, 2\Gamma_0^i + \frac{\tilde\gamma_{S\,0}^i}{2} + \frac{\tilde\gamma_{\nu\,0}^i}{2} \Bigr)
 - L_\nu \tilde\gamma_{\nu\,0}^i  + \tilde s_i^\one
\,, \nn \\
\tilde S_i^\two(b_T,\mu,\nu)
&= L_b^4\, \frac{(\Gamma_0^i)^2}{8}
 - L_b^3\, \Gamma_0^i \Bigl(
   L_\nu \Gamma_0^i
   + \frac{\beta_0}{3} + \frac{\tilde\gamma_{S\,0}^i}{4} + \frac{\tilde\gamma_{\nu\,0}^i}{4}
   \Bigr)
\nn \\ & \quad
+ L_b^2 \biggl[
   L_\nu^2\, 2(\Gamma_0^i)^2
   + L_\nu \Gamma_0^i \Bigl( \beta_0 + \tilde\gamma_{S\,0}^i + \frac32 \tilde\gamma_{\nu\,0}^i \Bigr)
\nn \\ & \quad \hspace{6ex}
   + \beta_0\Bigl(\frac{\tilde\gamma_{S\,0}^i}{4} + \frac{\tilde\gamma_{\nu\,0}^i}{2}\Bigr)
   + \frac{1}{8}(\tilde\gamma_{S\,0}^i + \tilde\gamma_{\nu\,0}^i)^2
   - \frac{\Gamma_1^i}{2}
   - \frac{\Gamma_0^i}{2}\, \tilde s_i^\one
\biggr]
\nn\\&\quad
+ L_b \biggl\{
   - L_\nu^2\, 2\Gamma_0^i \tilde\gamma_{\nu\,0}^i
   + L_\nu \Bigl[
      - \Bigl(\beta_0 + \frac{\tilde\gamma_{S\,0}^i}{2} + \frac{\tilde\gamma_{\nu\,0}^i}{2} \Bigr) \tilde\gamma_{\nu\,0}^i
      + 2 \Gamma_1^i
      + 2 \Gamma_0^i\, \tilde s_i^\one
   \Bigr]
\nn \\ & \quad \hspace{6ex}
   + \frac{\tilde\gamma_{S\,1}^i}{2} + \frac{\tilde\gamma_{\nu\,1}^i}{2}
   + \Bigl(\beta_0 + \frac{\tilde\gamma_{S\,0}^i}{2} + \frac{\tilde\gamma_{\nu\,0}^i}{2}\Bigr)
      \tilde s_i^\one
\biggr\}
\nn\\&\quad
 + L_\nu^2\, \frac{(\tilde\gamma_{\nu\,0}^i)^2}{2}
 - L_\nu \bigl(\tilde\gamma_{\nu\,1}^i + \tilde\gamma_{\nu\,0}^i\, \tilde s_i^\one \bigr)
 + \tilde s_i^\two
\,.\end{align}
At three loops, we write the result as
\begin{equation}
 \tilde S_i^\three(b_T,\mu,\nu) = \sum_{\ell=0}^6 \tilde S_{i,\ell}^\three(L_\nu) \, L_b^\ell
\,,\end{equation}
where the $\tilde S_{i,k}^\three$ coefficients themselves are polynomials in $L_\nu$.
Inserting $\tilde\gamma_{S\,0}^i = \tilde\gamma_{\nu\,0}^i = 0$ for brevity, they
are given by
\begin{align} \label{eq:qt_soft_N3LO}
\tilde S_{i,6}^\three(L_\nu)
&= -\frac{(\Gamma^i_0)^3}{48}
\,,\nn\\
\tilde S_{i,5}^\three(L_\nu)
&= L_\nu\, \frac{(\Gamma^i_0)^3}{4} + \frac{(\Gamma^i_0)^2}{6} \beta_0
\,,\nn\\
\tilde S_{i,4}^\three(L_\nu)
&= -L_\nu^2\, (\Gamma^i_0)^3
 - L_\nu\, \frac{7}{6} (\Gamma^i_0)^2 \beta_0
 + \frac{\Gamma_0^i}{4} \Bigl( -\beta_0^2 + \Gamma^i_1 + \frac{\Gamma^i_0}{2}\, \tilde s_i^\one \Bigr)
\,,\nn\\
\tilde S_{i,3}^\three(L_\nu)
&= L_\nu^3\, \frac{4}{3} (\Gamma^i_0)^3
 + L_\nu^2\, 2(\Gamma^i_0)^2 \beta_0
 + L_\nu \Gamma_0^i \Bigl(\frac{2}{3} \beta_0^2 - 2\Gamma^i_1 - \Gamma^i_0\, \tilde s_i^\one\Bigr)
\nn \\ & \quad
 - \frac{2}{3} \Gamma^i_1 \beta_0
 - \Gamma_0^i \Bigl(
   \frac{\beta_1}{3}
   + \frac{\tilde\gamma_{S\,1}^i}{4}
   + \frac{\tilde\gamma_{\nu\,1}^i}{4}
   + \frac{5}{6} \beta_0\, \tilde s_i^\one
\Bigr)
\,,\nn\\
\tilde S_{i,2}^\three(L_\nu)
&= L_\nu^2\, \Gamma^i_0 \bigl(4 \Gamma^i_1 +2 \Gamma^i_0\, \tilde s_i^\one \bigr)
 + L_\nu \Bigl[
   2 \Gamma^i_1 \beta_0
   + \Gamma^i_0 \Bigl(
      \beta_1
      + \tilde\gamma_{S\,1}^i
      + \frac{3}{2} \tilde\gamma_{\nu\,1}^i
      + 3 \beta_0\, \tilde s_i^\one
   \Bigr)
\Bigr]
\nn \\ & \quad
 + \beta_0 \Bigl(\frac{\tilde\gamma_{S\,1}^i}{2} + \tilde\gamma_{\nu\,1}^i \Bigr)
 - \frac{\Gamma^i_2}{2}
 + \Bigl(\beta_0^2 - \frac{\Gamma^i_1}{2} \Bigr) \tilde s_i^\one
 - \frac{\Gamma^i_0}{2}\, \tilde s_i^\two
\,,\nn\\
\tilde S_{i,1}^\three(L_\nu)
&= - L_\nu^2\, 2\Gamma^i_0 \tilde\gamma_{\nu\,1}^i
 + L_\nu\, 2\bigl(
   - \beta_0 \tilde\gamma_{\nu\,1}^i
   + \Gamma^i_2
   + \Gamma^i_1\, \tilde s_i^\one
   + \Gamma^i_0\, \tilde s_i^\two
\bigr)
 \nn\\&\quad
 + \frac{\tilde\gamma_{S\,2}^i}{2} + \frac{\tilde\gamma_{\nu\,2}^i}{2}
 + \Bigl(\beta_1 + \frac{\tilde\gamma_{S\,1}^i}{2} + \frac{\tilde\gamma_{\nu\,1}^i}{2} \Bigr) \tilde s_i^\one
 + 2 \beta_0\, \tilde s_i^\two
\,,\nn\\
\tilde S_{i,0}^\three(L_\nu)
&= -L_\nu \bigl(\tilde\gamma_{\nu\,2}^i + \tilde\gamma_{\nu\,1}^i\,\tilde s_i^\one \bigr)
 + \tilde s_i^\three
\,.\end{align}
\Eqs{qt_soft_NNLO}{qt_soft_N3LO} agree with \refscite{Luebbert:2016itl, Li:2016ctv}.
The required anomalous dimension and boundary coefficients
up to three loops are given in \app{ingredients_qt}.

\paragraph{Numerical impact}
The soft function $\tilde S_i(b_T,\mu,\nu)$ has an explicit dependence on the scales $\mu$ and $\nu$
that cancels against that of the hard and beam functions in \eq{qt_factorization}.
Therefore, varying $\mu$ and $\nu$ is not very meaningful for illustrating the numerical impact
of the scale-dependent three-loop terms. Instead, we consider the resummed soft function,
\begin{align} \label{eq:S_qt_resum}
\tilde{S}_i(b_T, \mu, \nu)
&= \tilde{S}_i(b_T, \mu_S, \nu_S) \, \tilde U_S^i(b_T, \mu_S, \mu, \nu_S, \nu)
\,,\nn\\
\tilde U_S^i(b_T, \mu_S, \mu, \nu_S, \nu)
&= \exp\biggl[ \ln\frac{\nu}{\nu_S}\, \tilde\gamma_\nu^i(b_T,\mu_S) \biggr]
\exp\biggl[ \int_{\mu_S}^\mu\! \frac{\df \mu'}{\mu'}\, \tilde \gamma_S^i(\mu',\nu)\biggr]
\,,\end{align}
where we have chosen to first evolve in $\nu$ and then in $\mu$.

\begin{figure*}
\centering
\includegraphics[width=\WidthTwoSubfigs]{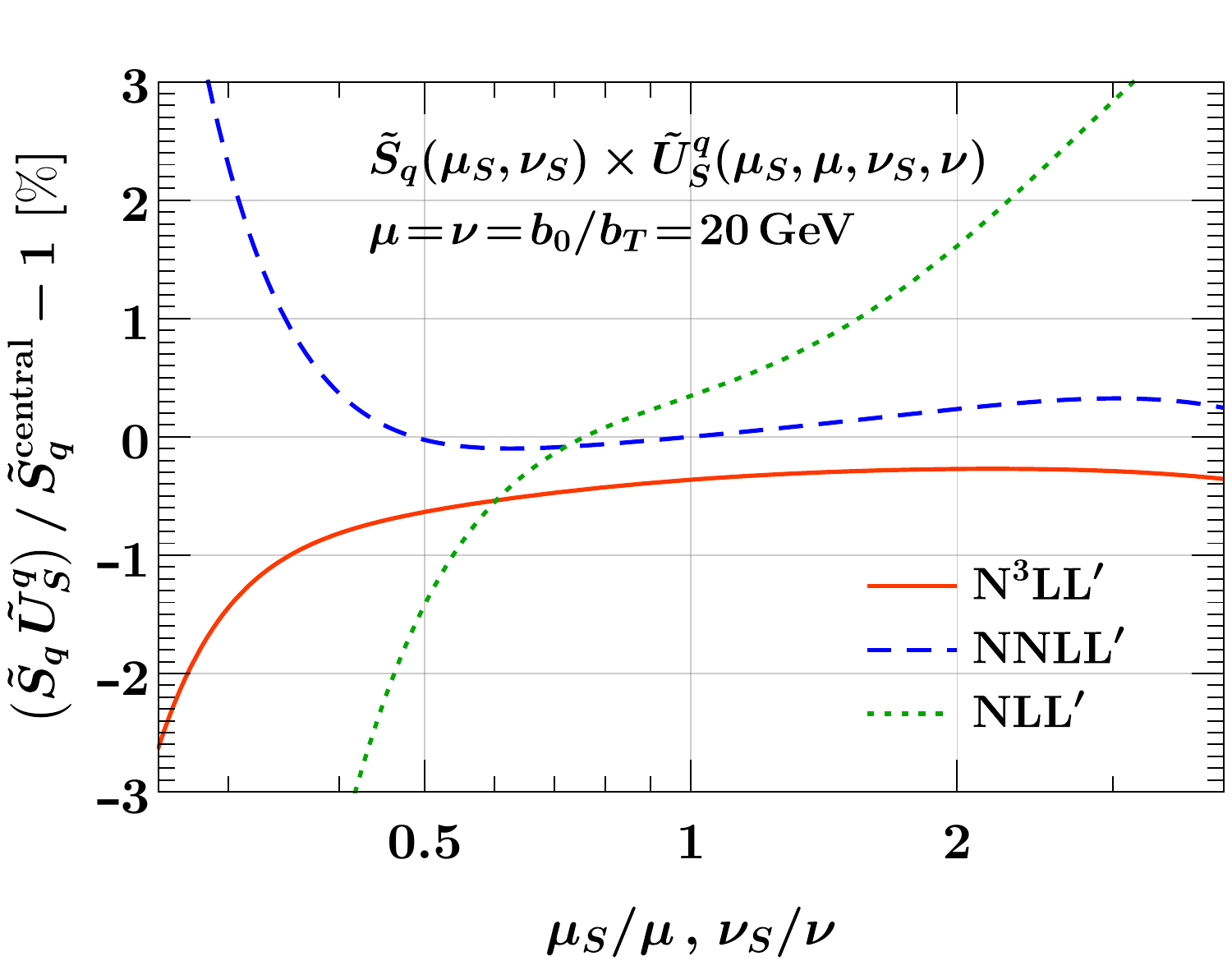}%
\hfill%
\includegraphics[width=\WidthTwoSubfigs]{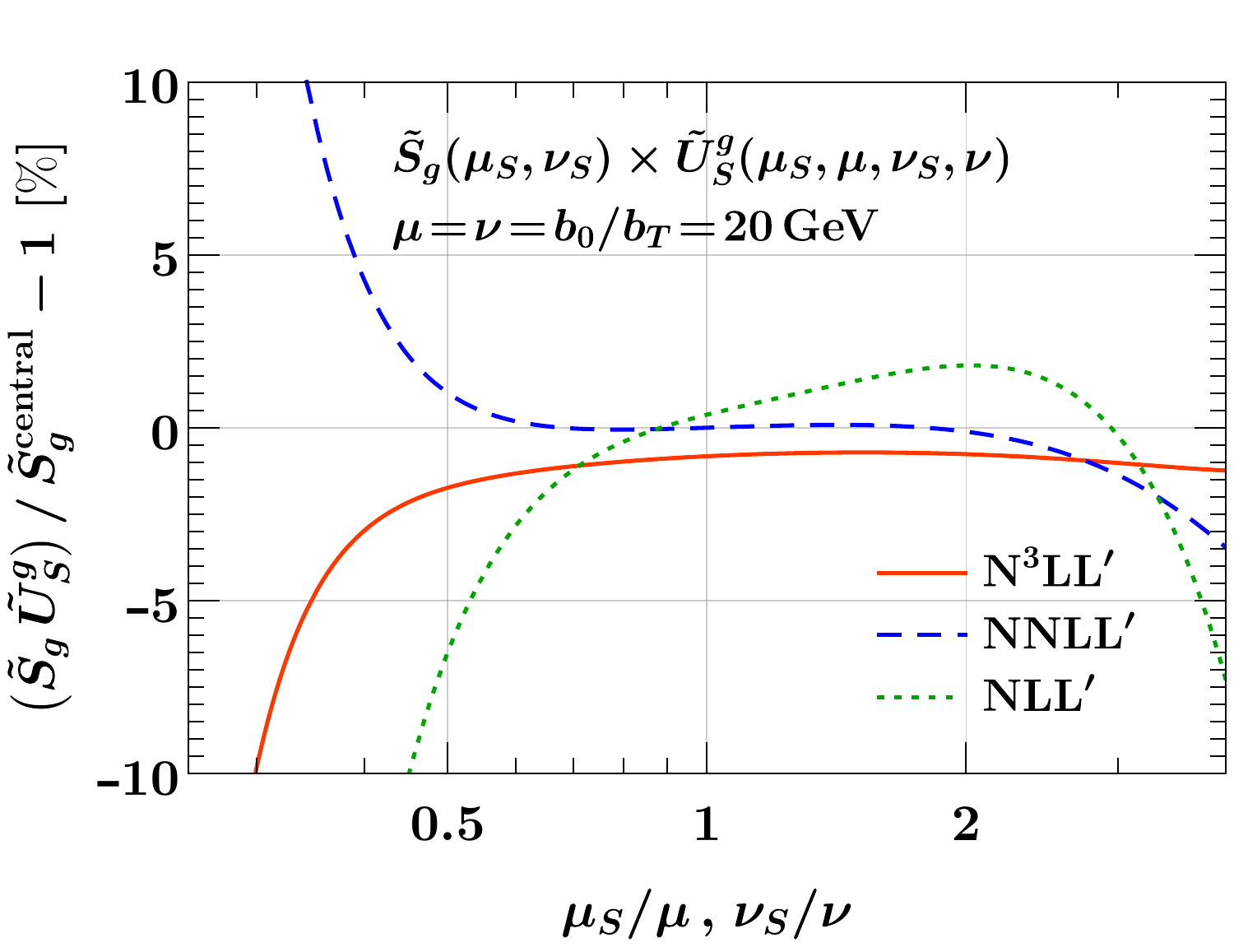}%
\caption{Residual scale dependence of the resummed $q_T$ soft function in
Fourier space for $i = q$ (left) and $i = g$ (right).
Shown are the relative deviations from the NNLL$'$ result $\tS_i^\central$ at the central scales $(\mu_S, \nu_S) = (\mu, \nu) = (b_0/b_T, b_0/b_T)$.}
\label{fig:scale_dependence_qT_soft}
\end{figure*}

To probe the full set of terms in the fixed-order expansion of $\tilde{S}_i(b_T, \mu_S, \nu_S)$,
we consider simultaneous variations of $(\mu_S,\nu_S)$ around the canonical central scales
$\mu_S = \nu_S = \mu = \nu = b_0 / b_T$.
In \fig{scale_dependence_qT_soft} we show the residual scale dependence of the resummed soft function
at the representative value $b_0/b_T = 20 \GeV$ at NLL$'$ (dotted green), NNLL$'$ (dashed blue),
and N$^3$LL$'$ (solid orange), normalized to the NNLL$'$ result at the central scale.
The three-loop finite term is included in
\fig{scale_dependence_qT_soft}, so the NNLL$'$ and N$^3$LL$'$ results do not
coincide at the central scales. We use four-loop running of $\as$ throughout,
which formally amounts to a higher-order effect at (N)NLL$'$.
As expected, the scale dependence reduces from NLL$'$ to NNLL$'$, where it is already
quite small. At N$^3$LL$'$, it further stabilizes over a wider range of scales.
As in \sec{tau0_soft}, we stress that the residual scale dependence in the resummed soft function by
itself is not necessarily a good indicator of the perturbative uncertainty, but gives an indication of
the typical reduction of perturbative uncertainties one might expect at each order.

\subsection{Beam function}
\label{sec:qTbeam}

The beam function obeys the coupled RGEs
\begin{align} \label{eq:qt_beam_RGE}
\mu \frac{\df}{\df\mu}{\tilde B_i(x,\bt,\mu,\nu)}
&= \tilde\gamma_B^i(\mu,\nu/\omega)\, \tilde B_i(x,\bt,\mu,\nu)
\,, \nn \\
\nu \frac{\df}{\df\nu}{\tilde B_i(x,\bt,\mu,\nu)}
&= -\frac{1}{2}\tilde\gamma_\nu^i(b_T,\mu)\, \tilde B_i(x,\bt,\mu,\nu)
\,,\end{align}
where the $\nu$ anomalous dimension was discussed in \sec{gamma_nu}, and
the $\mu$ anomalous dimension has the all-order form
\begin{equation}
\tilde\gamma_B^i(\mu,\nu/\omega)
= 2 \GammaC^i[\as(\mu)] \ln\frac{\nu}{\omega} + \tilde\gamma_B^i[\as(\mu)]
\,,\end{equation}
where $\GammaC^i(\as)$ and $\tilde\gamma_B^i(\as)$ are the cusp and the beam
noncusp anomalous dimensions.

For perturbative $b_0/b_T \gg \lqcd$, the TMD beam function satisfies an OPE in terms
of standard PDFs~\cite{Collins:1984kg},
\begin{align} \label{eq:qt_beam_matching}
\tilde B_q(x,b_T,\mu,\nu)
&= \sum_j \int\! \frac{\df z}{z}\, \tilde\cI_{qj}(z,b_T,\mu,\nu/\w) f_j\Bigl(\frac{x}{z},\mu\Bigr)
   \bigl[1 + \cO(b_T \lqcd)\bigr]
\,,\nn\\
\tilde B^{\rho\lambda}_g(x,\bt,\mu,\nu)
&= \sum_j \int\! \frac{\df z}{z} \biggl[
   \frac{g_\perp^{\rho\lambda}}{2}\, \tilde\cI_{gj}(z,b_T,\mu,\nu/\w)
   + \biggl(\frac{g_\perp^{\rho\lambda}}{2} - \frac{b_\perp^\rho b_\perp^\lambda}{b_\perp^2} \biggr) \tilde\cJ_{gj}(z,b_T,\mu,\nu/\w)
\biggr]
\nn \\ & \quad \hspace{10ex} \times
f_j\Bigl(\frac{x}{z},\mu\Bigr)
\bigl[1 + \cO(b_T \lqcd)\bigr]
\,.\end{align}
For the gluon beam function, we have made its dependence on the gluon helicity
explicit, and decomposed it into two orthogonal structures, namely the
polarization-independent piece $\tilde\cI_{gj}$ and the polarization-dependent
piece $\tilde\cJ_{gj}$, where $g_\perp^{\rho\lambda} = g^{\rho\lambda} -
(n^\rho \bn^\lambda + \bn^\rho n^\lambda)/2$ is the transverse metric and
$b_\perp^\rho$ is the transverse four vector with $b_\perp^2 = -\bt^2$. Due to
the multiplicative structure of \eq{qt_beam_RGE}, both $\tilde\cI_{gj}$ and
$\tilde\cJ_{gj}$ obey the same RGE, and in the following we will only consider
the RGEs for $\tilde\cI_{ij}$.

The $\tilde\cI_{ij}$ are perturbatively calculable matching coefficients,
whose RGEs follow from \eq{qt_beam_RGE} by taking the evolution of the PDFs into account,
\begin{align}
\mu \frac{\df}{\df\mu}{\tilde\cI_{ij}(z,b_T,\mu,\nu/\w)}
&= \sum_k \int\! \frac{\df z'}{z'}\, \tilde\cI_{ik}\Bigl(\frac{z}{z'},b_T,\mu,\nu/\w\Bigr)
   \bigl[ \tilde\gamma_B^i(\mu,\nu/\w) \id_{kj}(z') - 2  P_{kj}(z',\mu) \bigr]
\,,\nn\\
\nu \frac{\df}{\df\nu}{\tilde\cI_{ij}(z,b_T,\mu,\nu/\w)}
&= -\frac{1}{2}\tilde\gamma_\nu^i(b_T,\mu)\, \tilde\cI_{ij}(z,b_T,\mu,\nu/\w)
\,.\end{align}
Similar to the soft function, these coupled RGEs can be solved recursively as
\begin{align}
\tilde\cI_{ij}^{(n+1)}(z,b_T,\mu,\nu/\w)
&= \sum_{m=0}^n \biggl[
   \int_{b_0/b_T}^\mu \! \frac{\df\mu'}{\mu'}
   \Bigl( 2 \Gamma^i_{n-m} \ln\frac{\nu}{\omega} + \tilde\gamma_{B\,n-m}^i + 2 m \beta_{n-m} \Bigr)
   \tilde\cI_{ij}^{(m)}(z,b_T,\mu',\nu/\w)
\nn \\ & \quad
 - 2 \int_{b_0/b_T}^\mu \! \frac{\df\mu'}{\mu'}\, \bigl[\tilde\cI^{(m)}(b_T,\mu',\nu/\w) P^{(n-m)}\bigr]_{ij}(z)
\nn \\ & \quad
 - \int_{\omega}^\nu\! \frac{\df\nu'}{\nu'}\, \frac{\tilde\gamma_{\nu\,n-m}^i}{2}\, \tilde\cI_{ij}^{(m)}(z,b_T,b_0/b_T,\nu'/\w)
 \biggr] + \tilde I_{ij}^{(n+1)}(z)
\,,\!\end{align}
where the nonlogarithmic boundary coefficients are defined as
\begin{align}
\tilde I_{ij}^{(n+1)}(z) = \tilde\cI_{ij}^{(n+1)}(z,b_T,\mu = b_0/b_T,\nu/\w = 1)
\,.\end{align}
Starting from the LO result, $\tilde I_{ij}^\zero(z) = \id_{ij}(z) \equiv \delta_{ij}\,\delta(1-z)$,
we obtain up to two loops
\begin{align} \label{eq:qT_I}
\tilde\cI_{ij}^\zero(z,b_T,\mu,\nu/\w)
&= \id_{ij}(z)
\,,\nn\\
\tilde\cI_{ij}^\one(z,b_T,\mu,\nu/\w)
&= L_b \Bigl[
   \Bigl(L_\w \Gamma^i_0 + \frac{\tilde\gamma_{B\,0}^i}{2} \Bigr) \id_{ij}(z)
   - P^\zero_{ij}(z)
\Bigr]
 - L_\w \frac{\tilde\gamma_{\nu\,0}^i}{2}\, \id_{ij}(z)
 + \tilde I_{ij}^\one(z)
\,,\nn\\
\tilde\cI_{ij}^\two(z,b_T,\mu,\nu/\w)
&= L_b^2 \biggl\{
   \Bigl[
      L_\w^2\, \frac{(\Gamma_0^i)^2}{2}
      + L_\w\, \frac{\Gamma_0^i}{2} (\beta_0+\tilde\gamma_{B\,0}^i)
      + \Bigl(\beta_0 + \frac{\tilde\gamma_{B\,0}^i}{2}\Bigr) \frac{\tilde\gamma_{B\,0}^i}{4}
   \Bigr] \id_{ij}(z)
\nn \\ & \qquad\quad
   - \Bigl(L_\w \Gamma_0^i + \frac{\beta_0}{2} + \frac{\tilde\gamma_{B\,0}^i}{2} \Bigr) P^\zero_{ij}(z)
   + \frac12 (P^\zero\! P^\zero)_{ij}(z)
\biggr\}
\nn \\ & \quad
 + L_b \biggl\{
   \Bigl[
      - L_\w^2\, \Gamma_0^i \frac{\tilde\gamma_{\nu\,0}^i}{2}
      + L_\w \Bigl[-\Bigl(\beta_0 + \frac{\tilde\gamma_{B\,0}^i}{2}\Bigr)\frac{\tilde\gamma_{\nu\,0}^i}{2} + \Gamma_1^i \Bigr]
      + \frac{\tilde\gamma_{B\,1}^i}{2} \Bigr] \id_{ij}(z)
\nn \\ & \qquad\qquad
   + L_\w\, \frac{\tilde\gamma_{\nu\,0}^i}{2}\, P^\zero_{ij}(z)
   - P^\one_{ij}(z)
\nn \\ & \qquad\qquad
   + \Bigl( L_\w \Gamma_0^i + \beta_0 + \frac{\tilde\gamma_{B\,0}^i}{2} \Bigr) \tilde I^\one_{ij}(z)
   - (\tilde I^\one\! P^\zero)_{ij}(z)
\biggr\}
\nn \\ & \quad
  + \Bigl[
     L_\w^2\, \frac{(\tilde\gamma_{\nu\,0}^i)^2}{8}
     - L_\w\, \frac{\tilde\gamma_{\nu\,1}^i}{2}
  \Bigr] \id_{ij}(z)
  - L_\w\, \frac{\tilde\gamma_{\nu\,0}^i}{2}\, \tilde I_{ij}^\one(z)
  + \tilde I_{ij}^\two(z)
\,,\end{align}
where we abbreviated
\begin{equation}
L_b = \ln\frac{b_T^2\mu^2}{b_0^2} \,, \quad b_0 = 2 e^{-\gamma_E}
\,,\qquad
L_\w = \ln\frac\nu\omega
\,.\end{equation}
Note that $L_\w$ differs from the characteristic logarithm
of the soft function in the previous section.
The $\tilde{I}_{ij}^{(n)}(z)$ are given in \refcite{Luebbert:2016itl} for quark and gluon beam functions
in terms of the results of \refcite{Gehrmann:2014yya},
and were directly calculated at NNLO using the exponential regulator for the quark case in \refcite{Luo:2019hmp}.

At three loops we write
\begin{equation}
\tilde\cI_{ij}^\three(z,b_T,\mu,\nu/\w)
= \sum_{\ell=0}^3 \tilde\cI_{ij,\ell}^\three(z, L_\w) \, L_b^\ell
\,,\end{equation}
and using $\tilde\gamma_{\nu\,0}^i = 0$ for brevity, the coefficients are
\begin{align} \label{eq:qT_beamI_N3LO}
\tilde\cI_{ij,3}^\three(z,L_\w)
&= \biggl\{
   L_\w^3\, \frac{(\Gamma^i_0)^3}{6}
   + L_\w^2\, \frac{(\Gamma^i_0)^2}{2} \Bigl(\beta_0 + \frac{\tilde\gamma_{B\,0}^i}{2} \Bigr)
   + L_\w \Gamma^i_0 \Bigl[
      \frac{\beta_0^2}{3}
      + \Bigl(\beta_0 + \frac{\tilde\gamma_{B\,0}^i}{4}\Bigr) \frac{\tilde\gamma_{B\,0}^i}{2}
   \Bigr]
\nn \\ & \qquad
   + \Bigl(\beta_0 + \frac{\tilde\gamma_{B\,0}^i}{2}\Bigr)\Bigl(\beta_0 + \frac{\tilde\gamma_{B\,0}^i}{4}\Bigr) \frac{\tilde\gamma_{B\,0}^i}{6}
\biggr\} \id_{ij}(z)
\nn \\ & \quad
 - \biggl[
   L_\w^2\, \frac{(\Gamma^i_0)^2}{2}
   + L_\w \Gamma^i_0 \Bigl(\beta_0 + \frac{\tilde\gamma_{B\,0}^i}{2}\Bigr)
   + \frac{\beta_0^2}{3}
   + \Bigl(\beta_0 + \frac{\tilde\gamma_{B\,0}^i}{4}\Bigr)\frac{\tilde\gamma_{B\,0}^i }{2}
\biggr] P^\zero_{ij}(z)
\nn \\ & \quad
  + \Bigl(
     L_\w\, \frac{\Gamma^i_0}{2}
     + \frac{\beta_0}{2} + \frac{\tilde\gamma_{B\,0}^i}{4}
\Bigr) (P^\zero\! P^\zero)_{ij}(z)
  - \frac{1}{6} (P^\zero\! P^\zero\! P^\zero)_{ij}(z)
\,,\nn\\
\tilde\cI_{ij,2}^\three(z,L_\w)
&= \biggl\{
   L_\w^2\, \Gamma^i_0 \Gamma^i_1
   + L_\w \Bigl[
      \Gamma_1^i\Bigl(\beta_0 + \frac{\tilde\gamma_{B\,0}^i}{2}\Bigr)
      + \frac{\Gamma_0^i}{2} (\beta_1 + \tilde\gamma_{B\,1}^i)
   \Bigr]
\nn \\ & \qquad
   + \beta_0 \frac{\tilde\gamma_{B\,1}^i}{2}
   + \frac{\tilde\gamma_{B\,0}^i}{4}(\beta_1 + \tilde\gamma_{B\,1}^i)
\biggr\} \id_{ij}(z)
 - \Bigl( L_\w \Gamma^i_1 + \frac{\beta_1}{2} + \frac{\tilde\gamma_{B\,1}^i}{2} \Bigr) P_{ij}^\zero(z)
\nn \\ & \quad
 - \Bigl( L_\w \Gamma^i_0 +\beta_0+\frac{\tilde\gamma_{B\,0}^i}{2} \Bigr) P_{ij}^\one(z)
 + \frac12 (P^\zero\! P^\one + P^\one\! P^\zero)_{ij}(z)
\nn \\ & \quad
 + \Bigl[
   L_\w^2\, \frac{(\Gamma^i_0)^2}{2}
   + L_\w\, \frac{\Gamma^i_0}{2} (3 \beta_0 + \tilde\gamma_{B\,0}^i)
   + \Bigl(\beta_0 + \frac{\tilde\gamma_{B\,0}^i}{2}\Bigr)\Bigl(\beta_0 + \frac{\tilde\gamma_{B\,0}^i}{4}\Bigr)
\Bigr] \tilde I_{ij}^\one(z)
\nn \\ & \quad
 - \Bigl(L_\w \Gamma^i_0  + \frac{3}{2} \beta_0 + \frac{\tilde\gamma_{B\,0}^i}{2}\Bigr) (\tilde I^\one\! P^\zero)_{ij}(z)
 + \frac12 (\tilde I^\one\! P^\zero\! P^\zero)_{ij}(z)
\,,\nn\\
\tilde\cI_{ij,1}^\three(z,L_\w)
&= \biggl\{
   - L_\w^2\, \Gamma^i_0 \frac{\tilde\gamma_{\nu\,1}^i}{2}
   + L_\w \Bigl[
      - \Bigl(\beta_0  + \frac{\tilde\gamma_{B\,0}^i}{4} \Bigr) \tilde\gamma_{\nu\,1}^i
      + \Gamma^i_2
   \Bigr]
   + \frac{\tilde\gamma_{B\,2}^i}{2}
\biggr\} \id_{ij}(z)
\nn \\ & \quad
  + L_\w \frac{\tilde\gamma_{\nu\,1}^i}{2} P_{ij}^\zero(z)
  - P_{ij}^\two(z)
  + \Bigl( L_\w \Gamma^i_1 + \beta_1 + \frac{\tilde\gamma_{B\,1}^i}{2} \Bigr) \tilde I_{ij}^\one(z)
  - (\tilde I^\one\! P^\one)_{ij}(z)
\nn \\ & \quad
  + \Bigl( L_\w \Gamma^i_0 + 2 \beta_0 + \frac{\tilde\gamma_{B\,0}^i}{2} \Bigr) \tilde I_{ij}^\two(z)
  - (\tilde I^\two\! P^\zero)_{ij}(z)
\,,\nn\\
\tilde\cI_{ij,0}^\three(z,L_\w)
&= - L_\w\, \frac{\tilde\gamma_{\nu\,2}^i}{2}\, \id_{ij}(z)
 - L_\w\, \frac{\tilde\gamma_{\nu\,1}^i}{2}\, \tilde I_{ij}^\one(z)
 + \tilde I_{ij}^\three(z)
\,,\end{align}
where the three-loop boundary coefficients $\tilde I_{ij}^\three(z)$ are currently unknown.
We have evaluated all Mellin convolutions appearing in \eqs{qT_I}{qT_beamI_N3LO}
with the help of the {\tt MT} package~\cite{Hoeschele:2013gga}.
In contrast to \refcite{Cieri:2018oms}, we were able to perform all required
convolutions in terms of standard harmonic polylogarithms without encountering multiple polylogarithms,
after using the identity in \eq{trilogsubst} to simplify some of the inputs.

The polarization-dependent kernels $\tilde\cJ_{gj}$ have a simpler structure than the $\tilde\cI_{ij}$
because their LO contribution vanishes.
For unpolarized gluon-fusion processes,
the accompanying tensor structures are only contracted with each other,
and hence we only require their NNLO expressions for the N$^3$LO cross section.
They are given by
\begin{align}
\tilde\cJ_{gj}^\zero(z,b_T,\mu,\nu/\w) &= 0
\,,\nn\\
\tilde\cJ_{gj}^\one(z,b_T,\mu,\nu/\w)
&= \tilde J_{gj}^\one(z) = 4 C_j\, \frac{1-z}{z}
\,,\nn\\
\tilde\cJ_{gj}^\two(z,b_T,\mu,\nu/\w)
&= L_b \Bigl[
   \Bigl(L_\w \Gamma^g_0 + \beta_0 + \frac{\tilde\gamma_{B\,0}^g}{2} \Bigr) \tilde J_{gj}^\one(z)
   - (\tilde J^\one\! P^\zero)_{gj}(z)
\Bigr]
\nn \\ & \quad
  - L_\w\, \frac{\tilde\gamma_{\nu\,0}^g}{2}\, \tilde J_{gj}^\one(z) + \tilde J_{gj}^\two(z)
\,.\end{align}
The two-loop terms $\tilde J_{gj}^\two$ have recently been calculated
in \refcite{Luo:2019bmw} using the exponential regulator
and in \refcite{Gutierrez-Reyes:2019rug} using the $\delta$ regulator.
They can be converted to our convention via the relation
\begin{align} \label{eq:J_nnlo}
 \tilde J_{gj}^\two(z) &
 = I^{\prime (2)}_{gi}(z)
 \nn\\&
 = -\delta^L C_{g \leftarrow j}^{(2;0,0)}(z) - \frac12 \tilde s_g^{\one} \tilde J_{gj}^\one(z)
\,.\end{align}
In the first line of \eq{J_nnlo}, $I^{\prime (2)}_{gi}(z)$ is the two-loop boundary term
as given in \refcite{Luo:2019bmw}. In the second line of \eq{J_nnlo},
$\tilde s_g^{\one} = - 2 C_A \zeta_2$ is the soft function constant at one loop
and $\delta^L C_{g \leftarrow j}^{(2;,0,0)}(z)$ is the two-loop finite piece
of the TMDPDF given in \refcite{Gutierrez-Reyes:2019rug}.

\paragraph{Numerical impact}
As for the soft function above, we illustrate the numerical impact of the three-loop corrections
for the resummed beam function
\begin{align} \label{eq:B_qt_resum}
 \tilde{B}_i(x, \vec{b}_T, \mu, \nu) &=
 \tilde{B}_i(x, \vec{b}_T, \mu_B, \nu_B) \, \tilde U_B^i(\omega, b_T, \mu_B, \mu, \nu_B, \nu)
\,,\nn\\
 \tilde U_B^i(\omega, b_T, \mu_B, \mu, \nu_B, \nu) &=
 \exp\biggl[ -\frac12 \ln\frac{\nu}{\nu_B} \tilde\gamma_\nu^i(b_T,\mu_B) \biggr]
 \exp\biggl[ \int_{\mu_B}^\mu \frac{\df \mu'}{\mu'} \tilde \gamma_B^i(\mu',\nu/\w)\biggr]
\,.\end{align}
For $i = g$, we restrict to the polarization-independent piece $\tcI_{gj}$
and write $\tB_g \equiv -g_{\perp,\rho \lambda} \tB_g^{\rho\lambda}$ for short.
As for the soft function, we restrict to simultaneous variations of $\mu_B$ and $\nu_B$.

\begin{figure*}
\centering
\includegraphics[width=\WidthTwoSubfigs]{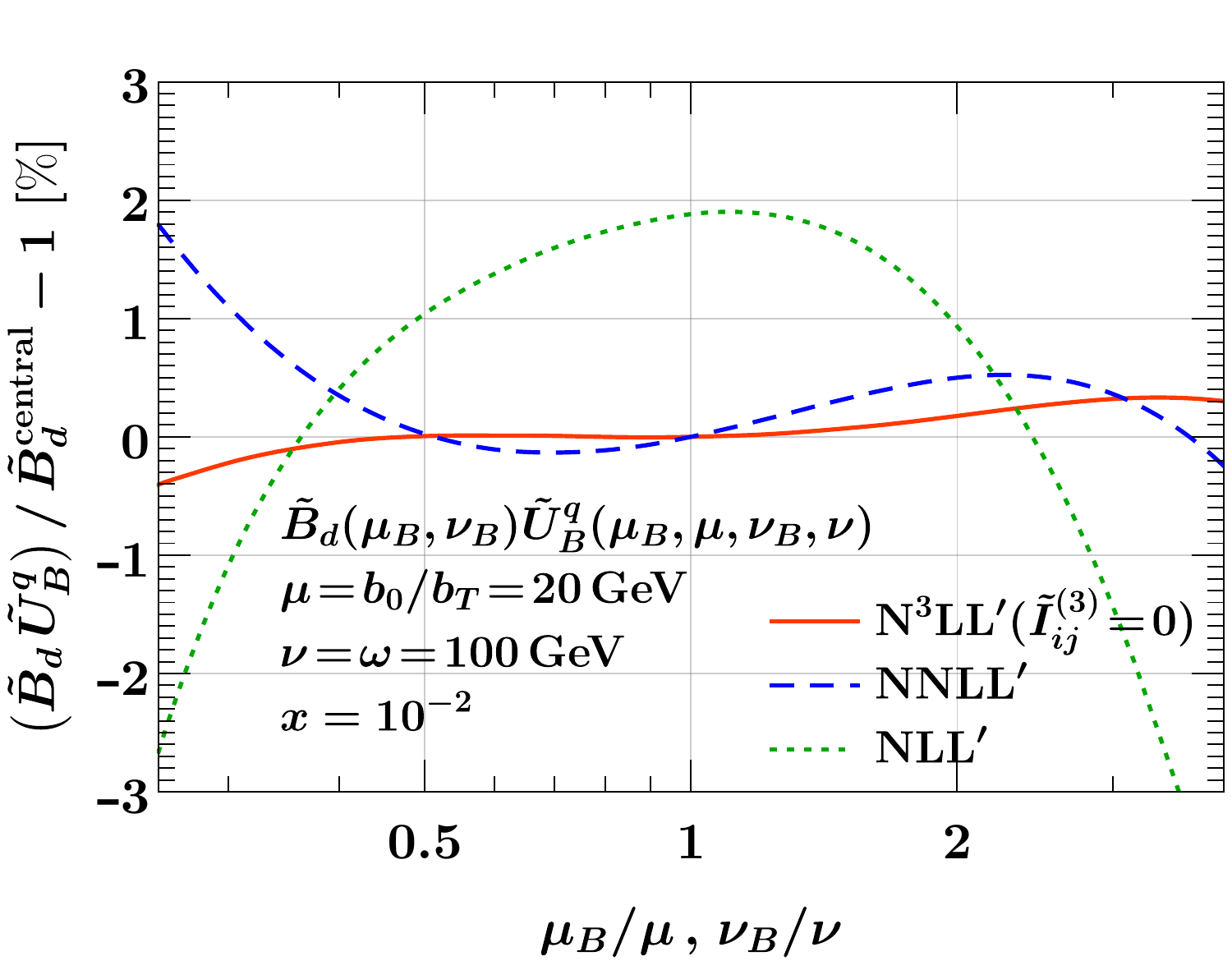}%
\hfill%
\includegraphics[width=\WidthTwoSubfigs]{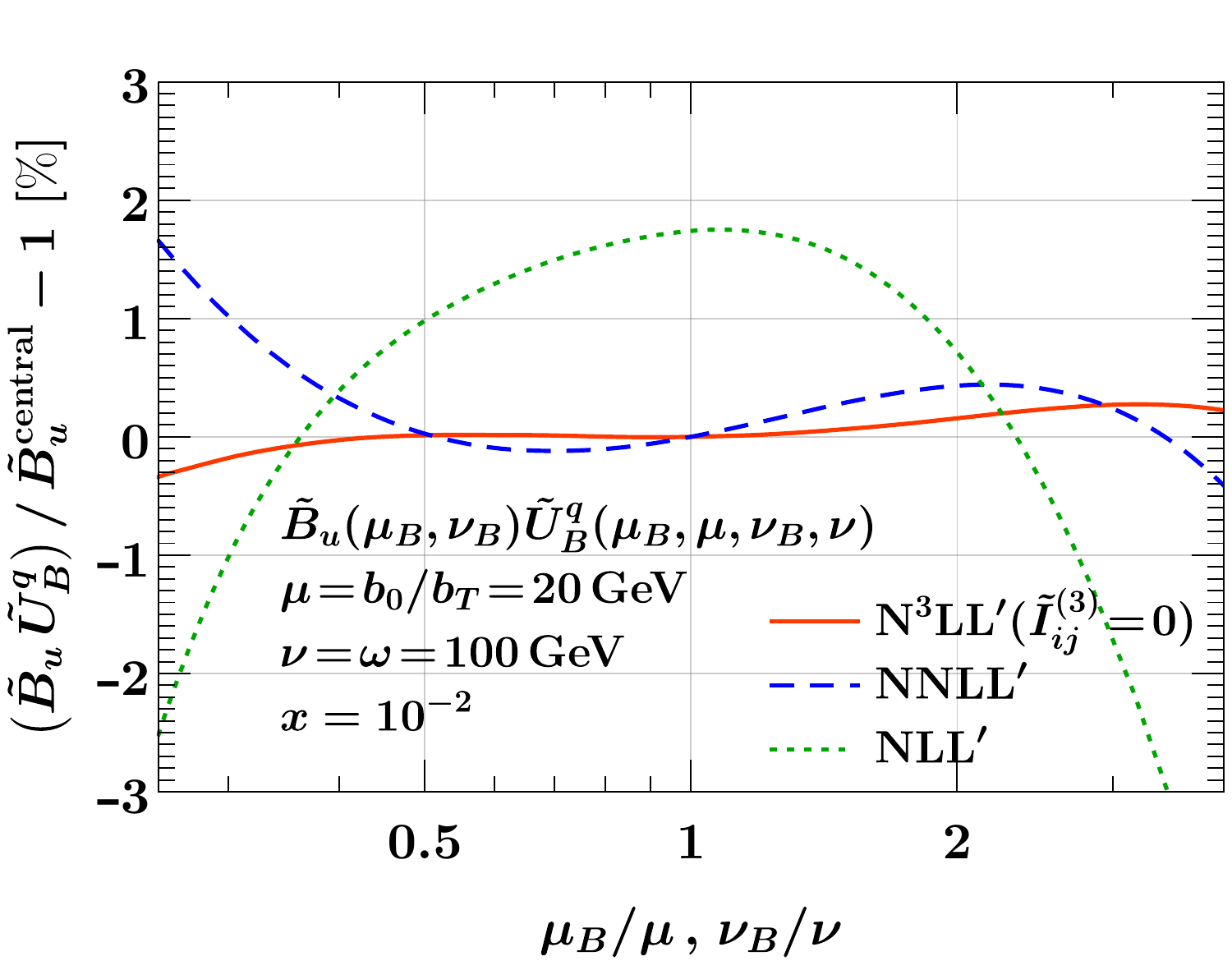}%
\\
\includegraphics[width=\WidthTwoSubfigs]{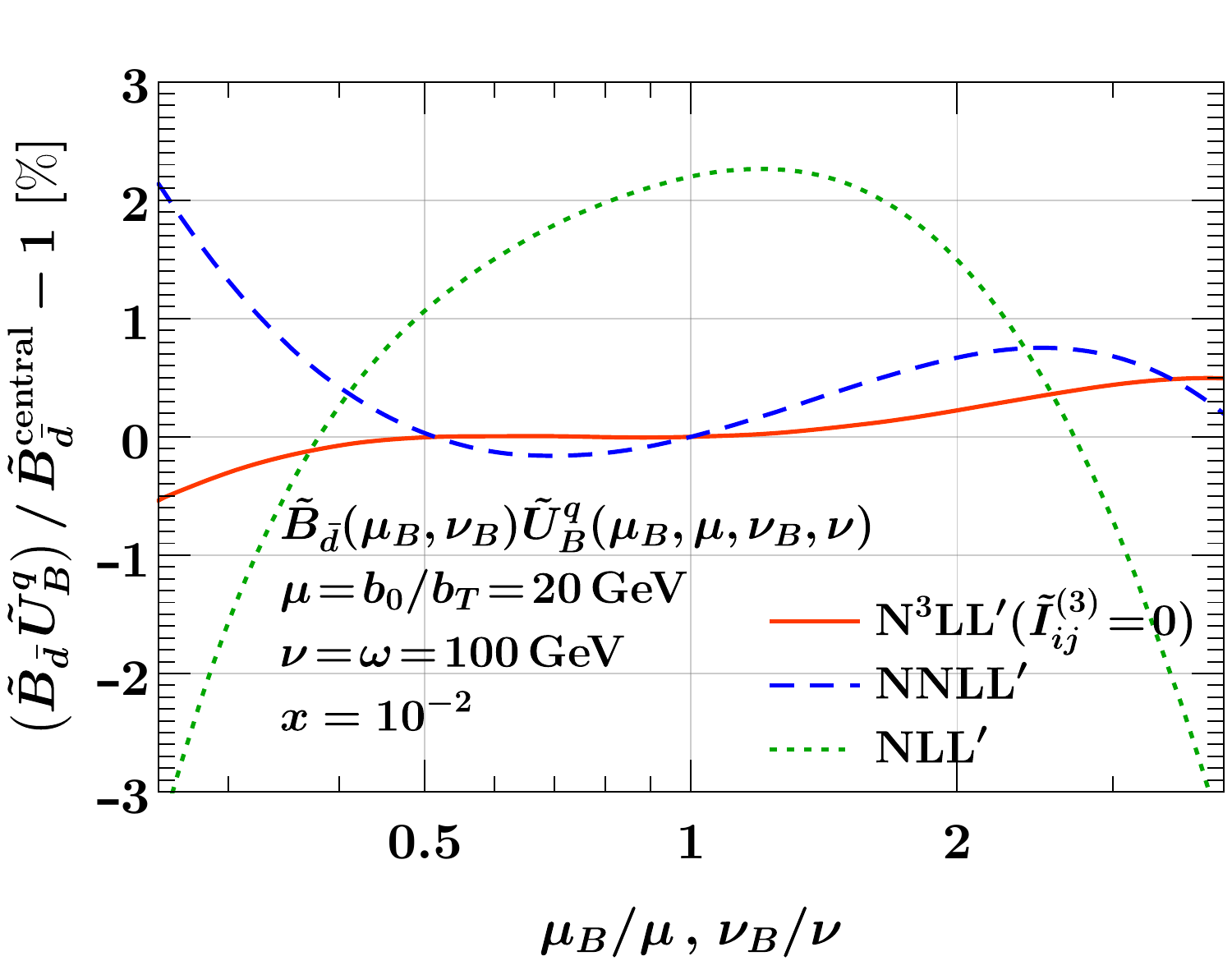}%
\hfill%
\includegraphics[width=\WidthTwoSubfigs]{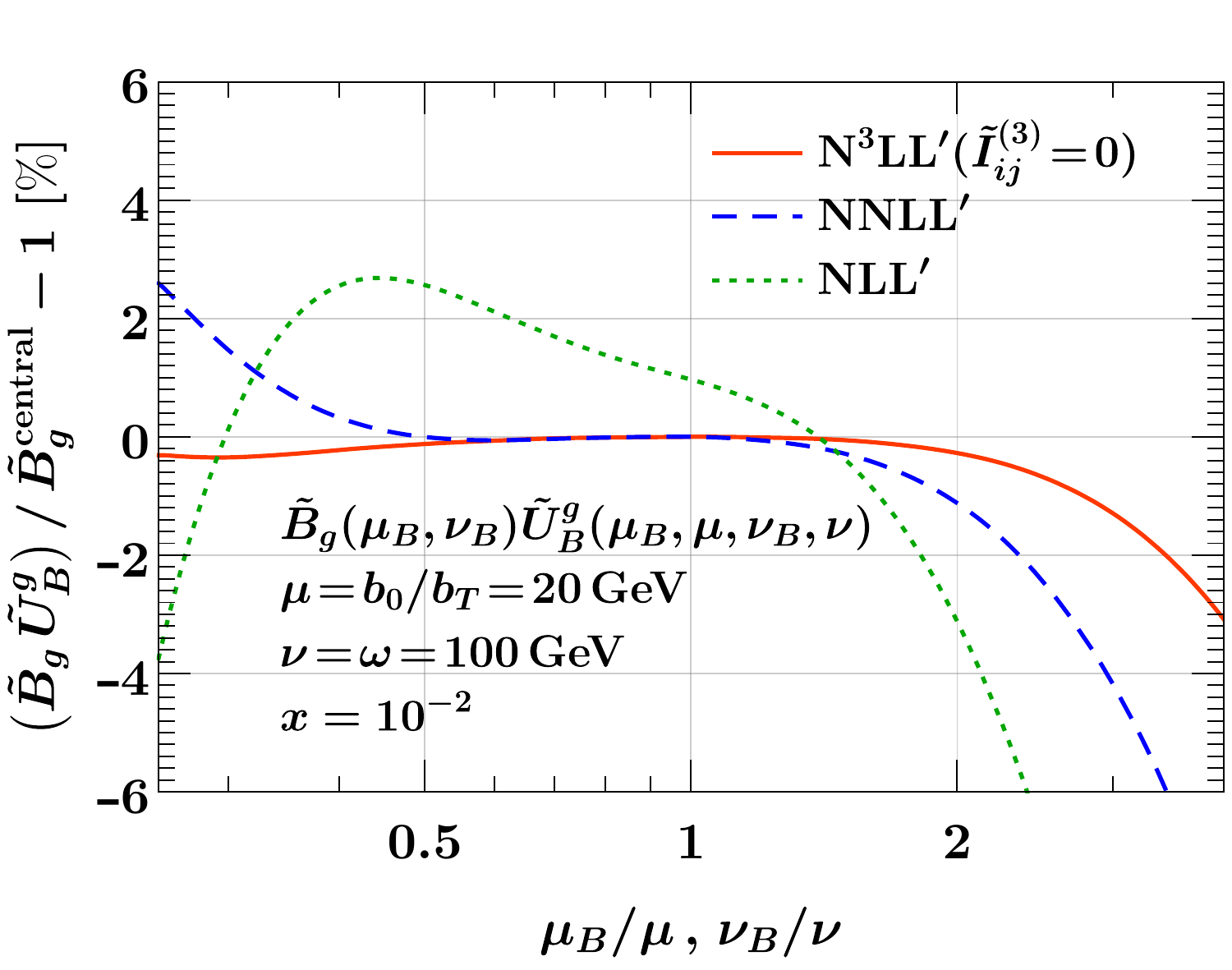}%
\caption{Residual scale dependence of the resummed $q_T$ beam function in
Fourier-space for $i = d$ (top left), $u$ (top right), $\bar{d}$ (bottom left), and $g$ (bottom right).
Shown are the relative deviations from the NNLL$'$ result $\tB_i^\central$ at the central scales $(\mu_B, \nu_B) = (\mu, \nu) = (b_0/b_T, \omega)$.}
\label{fig:scale_dependence_qT_beam}
\end{figure*}

In \fig{scale_dependence_qT_beam}, we show the residual dependence on $(\mu_B, \nu_B)$
at NLL$'$ (dotted green), NNLL$'$ (dashed blue), and N$^3$LL$'$ with the unknown
$\cI^\three_{ij}(z) = 0$ (solid orange) as the relative difference to the central
NNLL$'$ result at $(\mu_B, \nu_B) = (\mu,\nu) = (b_0/b_T, \omega)$
for $b_0/b_T = 20 \GeV$ and $\omega = 100 \GeV$.
As for $\Tau_0$, we use four-loop running of $\as$ and
\texttt{MMHT2014nnlo68cl}~\cite{Harland-Lang:2014zoa} NNLO PDFs throughout.
In all cases, the scale dependence is substantially reduced at each order.
We again anticipate that this qualitative behavior continues to hold
when the full result for $\tI^\three_{ij}(z)$ is included.

\subsection{Beam function coefficients in the eikonal limit}
\label{sec:qT_beam_finite}

We now proceed to extract the three-loop beam function coefficients in the $z\to 1$
limit from consistency relations with known soft matrix elements.
For the $q_T$ beam function, these consistency relations
arise from factorization theorems for the triple-differential
cross section $\df \sigma_{pp\to L} / \df Q^2 \df Y \df q_T$ that enable the joint $q_T$
and soft threshold resummation~\cite{Li:1998is, Laenen:2000ij, Kulesza:2002rh, Kulesza:2003wn}.
In terms of the momentum fractions $x_{a,b}$ defined in \eq{Born},
the soft threshold limit is equivalent to taking both $x_a \to 1$ and $x_b \to 1$.
As $x_{a,b} \to 1$, initial state radiation is constrained to have energy $\lesssim \lambda_- \lambda_+ Q$,
where
\begin{equation}
\lambda_-^2 \sim 1-x_a
\qquad \text{and} \qquad
\lambda_+^2 \sim 1-x_b
\end{equation}
are power-counting parameters that encode the distance from the kinematic endpoint.

The all-order factorization relevant for different hierarchies in $q_T/Q$ and
the threshold constraint $\lambda_- \lambda_+$ was derived in \refscite{Li:2016axz, Lustermans:2016nvk}.
Some consequences of the resulting consistency relations have already been explored
in \refscite{Li:2016axz, Lustermans:2016nvk}. In fact, the exponential regulator is \emph{defined} by its action on the refactorized pieces in these consistency relations.
In the following, we briefly review the relevant factorization theorems
and derive the all-order structure that arises for the $q_T$ beam function in the eikonal limit.

\paragraph{$q_T/Q \ll \lambda_- \lambda_+ \sim 1$}

In this regime, initial-state radiation is not yet subject to the threshold constraint,
and the standard $q_T$ factorization theorem \eq{qt_factorization} holds.
It receives power corrections $\ord{q_T^2/Q^2}$, but captures the exact dependence on $x_{a,b}$ via the beam functions.

\paragraph{$q_T/Q \ll \lambda_- \lambda_+ \ll 1$}

For this hierarchy, the factorization takes a form similar to \eq{qt_factorization},
but real collinear radiation into the final state is constrained in energy
by $1-x_{a,b} \ll 1$.
The leftover radiation in this limit is described by intermediate collinear-soft
modes~\cite{Bauer:2011uc, Procura:2014cba} in terms of $n_{a,b}$-collinear-soft
functions $\tilde{\cS}_i(k, b_T, \mu, \nu)$. They are matrix elements of collinear-soft
Wilson lines and depend on the small additional momentum $k = k^\mp$ available from either one of the (threshold) PDFs and on the color charge of the colliding partons.
The factorization theorem in this regime reads~\cite{Li:2016axz, Lustermans:2016nvk}
\begin{align} \label{eq:factorization_qT_threshold_regime_2}
\frac{\df \tilde{\sigma}(\vec{b}_T)}{\df Q^2\, \df Y}
&= \sum_{a,b} H_{ab}(Q^2, \mu)
\int\! \df k^-\,
\tilde{\cS}_i(k^-, b_T, \mu, \nu)\,
f^\thr_a \Bigl[x_a \Bigl(1 + \frac{k^-}{\omega_a} \Bigr), \mu \Bigr]
\nn \\ &\quad \times
\int\! \df k^+ \,
\tilde{\cS}_i(k^+, b_T, \mu, \nu) \,
f^\thr_b \Bigl[ x_b \Bigl(1 + \frac{k^+}{\omega_b} \Bigr), \mu \Bigr] \,
\tilde{S}_i(b_T, \mu, \nu)
\nn \\ &\quad
\times \Bigl[ 1 + \ORd{\frac{1}{b_T^2 \lambda_-^2 \lambda_+^2 Q^2}, \lambda_-^2, \lambda_+^2} \Bigr]
\,.\end{align}
Collinear-soft emissions do not contribute angular momentum,
so the polarization indices for gluon-induced processes become trivial in this limit
and we suppress them in the following.

\paragraph{$q_T/Q \sim \lambda_- \lambda_+ \ll 1$}

In this regime, the threshold constraint dominates and all radiation is forced to be soft.
The recoil against soft radiation with transverse momentum $\vec{k}_T = - \vec{q}_T$
is encoded in the fully-differential threshold soft function $S^\thr_i(k^-, k^+, \vec{k}_T)$.
In terms of its Fourier transform with respect to $\vec{k}_T$, $\tilde{S}^\thr_i(k^-, k^+, b_T)$,
the factorization theorem reads
\begin{align} \label{eq:factorization_qT_threshold_regime_3}
\frac{\df \tilde{\sigma}(\vec{b}_T)}{\df Q^2\, \df Y}
&= \sum_{a,b} H_{ab}(Q^2, \mu)
\int\! \df k^- \df k^+
f^\thr_a \Bigl[x_a \Bigl(1 + \frac{k^-}{\omega_a} \Bigr), \mu \Bigr]
f^\thr_b \Bigl[ x_b \Bigl(1 + \frac{k^+}{\omega_b} \Bigr), \mu \Bigr]
\nn \\ &\quad \times
\tilde{S}^\thr_i(k^-, k^+, b_T, \mu)\,
\bigl[ 1 + \ord{\lambda_-^2, \lambda_+^2} \bigr]
\,.\end{align}
Notably, the fully-differential threshold soft function is free of rapidity divergences
because they are regulated by the threshold constraint.
(This is the starting point of the exponential regularization procedure.)
The fully-differential soft function was calculated to $\ord{\as^2}$ in \refcite{Li:2011zp},
albeit in a different context,
and to $\ord{\as^3}$ in \refcite{Li:2016ctv}.
By construction, it satisfies
\begin{equation}
\int \! \df^2 \vec{k}_T \, S^\thr_i(k^-, k^+, \vec{k}_T, \mu)
= \tilde{S}^\thr_i(k^-, k^+, b_T = 0, \mu) = S^\thr_i(k^-, k^+, \mu)
\,,\end{equation}
where $S^\thr_i(k^-, k^+, \mu)$ is the double-differential threshold soft function
appearing in \eq{factorization_soft_threshold_qm_qp}.

\paragraph{Consistency relations}

Consistency between \eqs{qt_factorization}{factorization_qT_threshold_regime_2} implies
that the $x \to 1$ limit of the $q_T$ beam function is captured by the collinear-soft function~\cite{Li:2016axz, Lustermans:2016nvk},
\begin{equation} \label{eq:qT_beam_eikonal_limit_hadronic}
\tilde B_i(x, \vec{b}_T, \mu, \nu)
= \int \! \df k \, \tilde{\cS}_i(k, b_T, \mu, \nu) \, f_i^\thr\Bigl[x\Bigl(1 + \frac{k}{\omega}\Bigr), \mu\Bigr] \,
\bigl[ 1 + \ord{1-x} \bigr]
\,.\end{equation}
This is the analog of \eq{TauN_beam_eikonal_limit_hadronic} for $q_T$,
but this time relates the eikonal limit of the beam function to an exclusive collinear-soft matrix element
instead of the inclusive threshold soft function.
At the partonic level, \eq{qT_beam_eikonal_limit_hadronic} implies~\cite{Li:2016axz, Lustermans:2016nvk}
\begin{equation} \label{eq:qT_beam_eikonal_limit_partonic}
\tilde{\cI}_{ij}(z, b_T, \mu, \nu/\w)
= \delta_{ij} \, \omega \, \tilde{\cS}_{i}\bigl[\omega(1-z), b_T, \mu, \nu\bigr] \,
\bigl[ 1 + \ord{1-z} \bigr]
\,.\end{equation}
Note that \eq{qT_beam_eikonal_limit_partonic} is true for any rapidity regulator
as long as the same regulator is used on both sides.
The consistency between \eqs{factorization_qT_threshold_regime_2}{factorization_qT_threshold_regime_3} implies~\cite{Li:2016axz, Lustermans:2016nvk}
\begin{equation} \label{eq:refactorization_soft}
\tilde{S}_i^\thr(k^-, k^+, b_T, \mu)
= \tilde{\cS}_i(k^-, b_T, \mu, \nu) \, \tilde{\cS}_i(k^+, b_T, \mu, \nu) \,
\tilde{S}_i(b_T, \mu, \nu) \,
\Bigl[ 1 + \ORd{\frac{1}{b_T^2 k^- k^+}} \Bigr]
\,,\end{equation}
which again holds for any choice of rapidity regulator.
In particular, the left-hand side has no rapidity divergences,
so the dependence on the rapidity regulator cancels between the terms on the right-hand side.
Together, \eqs{qT_beam_eikonal_limit_hadronic}{refactorization_soft} uniquely determine
the eikonal limit of the beam function in any given rapidity regulator scheme
in terms of the fully-differential soft function (which is independent of the scheme)
and the $q_T$ soft function $\tilde{S}_i(b_T, \mu, \nu)$ (which determines the scheme).
Furthermore, the scheme ambiguity amounts to moving terms from the soft function boundary
coefficients into the coefficient of $\delta(1-z)$ in the beam function coefficients.
Since $\delta(1-z)$ is a leading-power contribution as $z \to 1$,
it follows that up to lower-order cross terms, all scheme-dependent terms in the beam function
are contained in the leading eikonal terms predicted by \eq{qT_beam_eikonal_limit_partonic}.

\paragraph{Extraction of the finite terms}

For the exponential regulator, the relation between the fully-differential
and standard TMD soft function is particularly simple,
leading to an all-order result for the collinear-soft function
in terms of the rapidity anomalous dimension, see \app{csoft_function}.
Inserting this result into \eq{qT_beam_eikonal_limit_partonic},
we find for the eikonal limit of the $b_T$-space beam function matching coefficient $\tilde{\cI}_{ij}$
in the exponential regulator scheme,
\begin{align} \label{eq:qT_beam_eikonal_finite_terms_all_order}
\tilde{\cI}_{ij}(z, b_T, \mu, \nu/\w)
= \delta_{ij} \, \frac{\w}{\nu} \cV_{\tilde{\gamma}_\nu^i(b_T, \mu)/2}\Bigl[\frac{\w}{\nu}(1-z)\Bigr] \,
\bigl[ 1 + \ord{1-z} \bigr]
\,,\end{align}
where the plus distribution $\cV_a(x)$ is defined in \eq{cVa_definition}.
The simplicity of this result is a direct consequence of the specific rapidity regulator,
i.e., one may equally well have imposed this form of the eikonal limit as the renormalization condition.
Nonetheless, when combined with the soft function to a given order,
the scheme dependence cancels and leaves behind a unique set of terms
that capture the threshold limit of the singular cross section in \eq{qt_factorization}.
We note that a close relation between the rapidity anomalous dimension
and the eikonal limit of the beam function is a scheme-independent feature~\cite{Lustermans:2016nvk},
and was also conjectured for the $\delta$-regulator in \refcite{Echevarria:2016scs}.

It is straightforward to expand \eq{qT_beam_eikonal_finite_terms_all_order} in $\as$
to obtain the finite terms in the matching coefficient at any given fixed order
using \eqs{gammaNu_FO}{cVa_series_expansion}.
Up to two loops we have
\begin{align} \label{eq:qT_beam_eikonal_finite_terms_two_loops}
\tilde{I}_{ij}^\one(z)
&= \Ordsq{(1-z)^0}
\,, \nn \\
\tilde{I}_{ij}^\two(z)
&= \delta_{ij} \, \frac{\tilde{\gamma}_{\nu\,1}^i}{2} \, \cL_0(1-z) + \Ordsq{(1-z)^0}
\,,\end{align}
in agreement with the full two-loop result~\cite{Luebbert:2016itl},
and where we have used that $\tilde{\gamma}_{\nu\,0}^i = 0$.
Including terms up to six loops for illustration, we find
\begin{align} \label{eq:csoft_prediction_six_loops}
\tilde{I}_{ij}^{(3)}(z)
&= \delta_{ij} \, \frac{\tilde{\gamma}_{\nu\,2}^i}{2} \, \cL_0(1-z) + \Ordsq{(1-z)^0}
\,, \nn \\
\tilde{I}_{ij}^{(4)}(z)
&= \delta_{ij} \, \frac{\tilde{\gamma}_{\nu\,3}^i}{2} \, \cL_0(1-z)
+ \frac{(\tilde{\gamma}_{\nu\,1}^i)^2}{4} \Bigl[ \cL_1(1-z) - \frac{\zeta_2}{2} \delta(1-z) \Bigr]
+ \Ordsq{(1-z)^0}
\,, \nn \\
\tilde{I}_{ij}^{(5)}(z)
&= \delta_{ij} \, \frac{\tilde{\gamma}_{\nu\,4}^i}{2} \, \cL_0(1-z)
+ \frac{\tilde{\gamma}_{\nu\,1}^i\tilde{\gamma}_{\nu\,2}^i}{2} \Bigl[ \cL_1(1-z) - \frac{\zeta_2}{2} \delta(1-z) \Bigr]
+ \Ordsq{(1-z)^0}
\,, \nn \\
\tilde{I}_{ij}^{(6)}(z)
&= \delta_{ij} \, \frac{\tilde{\gamma}_{\nu\,5}^i}{2} \, \cL_0(1-z)
+ \frac{(\tilde{\gamma}_{\nu\,2}^i)^2 + 2\tilde{\gamma}_{\nu\,1}^i\tilde{\gamma}_{\nu\,3}^i}{4}
  \Bigl[ \cL_1(1-z) - \frac{\zeta_2}{2} \delta(1-z) \Bigr]
\nn \\ &\quad
+ \frac{(\tilde{\gamma}_{\nu\,1}^i)^3}{8} \Bigl[ \frac{\cL_2(1-z)}{2} - \frac{\zeta_2}{2}\cL_0(1-z) + \frac{\zeta_3}{3}\delta(1-z) \Bigr]
+ \Ordsq{(1-z)^0}
\,.\end{align}
We again stress that these expressions are a direct consequence of the renormalization condition in the exponential regulator scheme and must be combined with the soft function in the same scheme
to obtain a scheme-independent result.
It is interesting to note that starting at four loops,
\eq{qT_beam_eikonal_finite_terms_all_order} does in fact predict a term proportional to $\delta(1-z)$ in the beam function matching coefficient
due to the inverse Fourier transform to $k^{\pm}$ back from the conjugate $b^\pm$ space,
where the regularization procedure is applied.

\subsection{Estimating beam function coefficients beyond the eikonal limit}
\label{sec:qT_beam_beyond_eikonal}

As in \sec{tau0_beam_beyond_eikonal}, we can use the eikonal limit of the beam function
coefficients to study to what extent it can be used to approximate the full result
and/or estimate the uncertainty due to the missing terms beyond the eikonal limit.

\begin{figure*}
\centering
\includegraphics[width=\WidthTwoSubfigs]{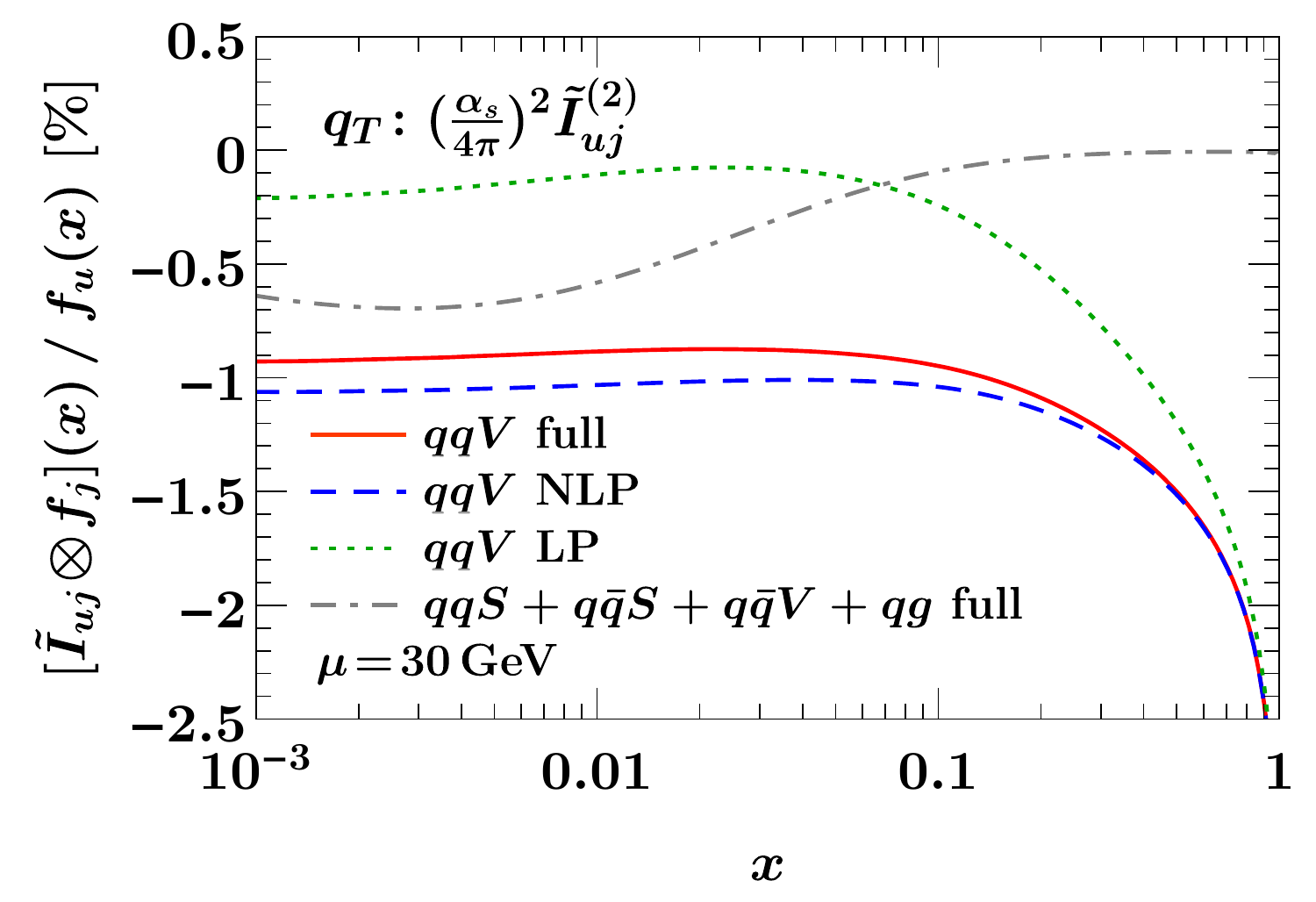}%
\hfill%
\includegraphics[width=\WidthTwoSubfigs]{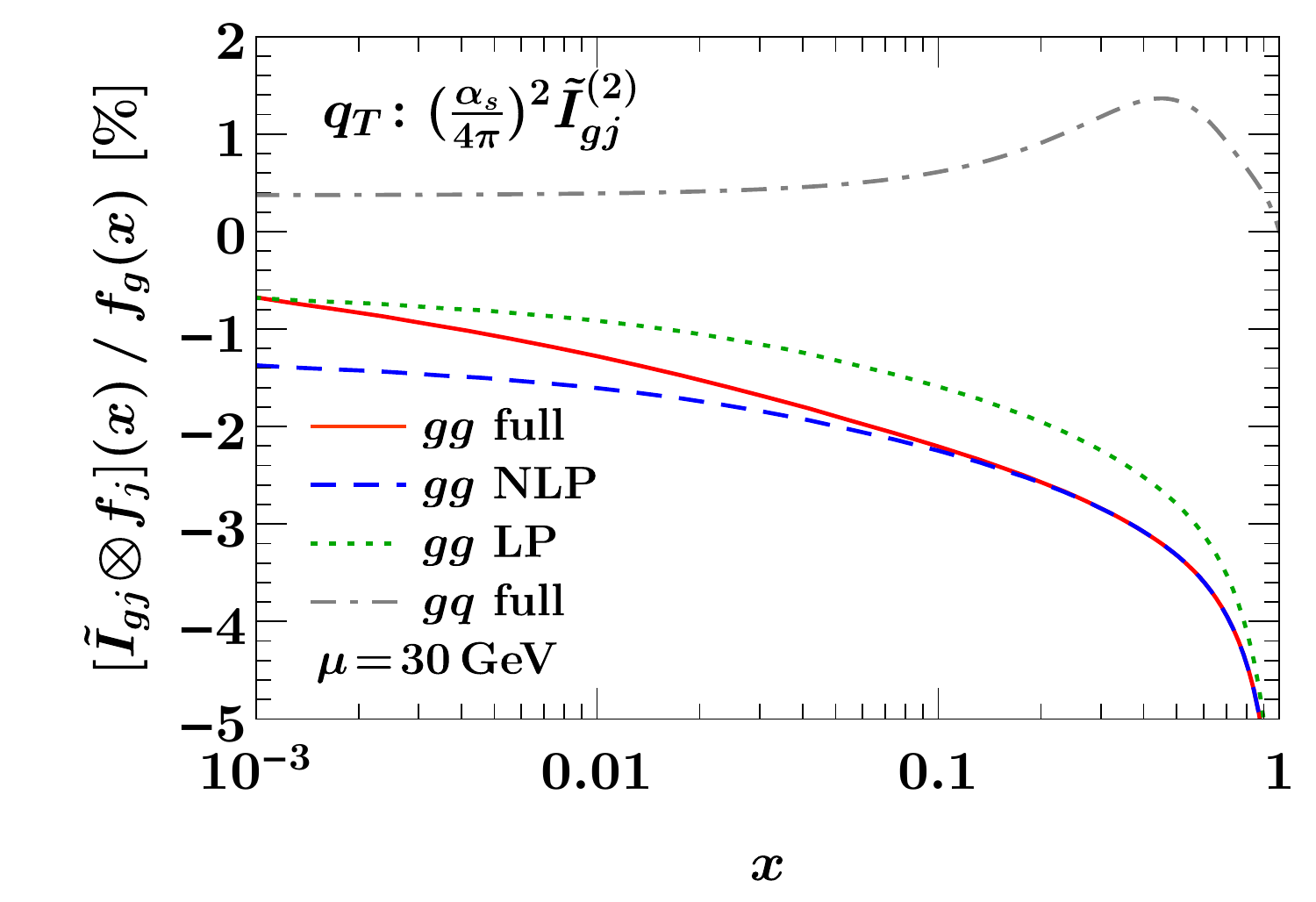}%
\caption{Comparison of the full beam function coefficients to their leading eikonal (LP)
and next-to-eikonal (NLP) expansion at NNLO. The $u$-quark channel is shown on the left
and the gluon channel on the right. In both cases we also show the sum of all
nondiagonal partonic channels for comparison.}
\label{fig:qTKernelpowerexp}
\end{figure*}

In \fig{qTKernelpowerexp}, we compare the full $q_T$ beam function coefficient (solid)
to its eikonal (LP dotted green) and next-to-eikonal (NLP dashed blue) expansions at NNLO
for the $u$-quark and gluon channels. Since the NLO coefficients are not singular,
we do not show the corresponding NLO results.
We always show the convolution $(I_{ij}\otimes f_j)(x) / f_i(x)$ with the appropriate
PDF $f_j$ and normalize to the PDF $f_i(x)$, corresponding to the LO result,
where $i=u$ for the $u$-quark case and $i=g$ for the gluon case.
With this normalization,
the shape gives an indication of the rapidity dependence of the beam function coefficient
relative to the LO rapidity dependence induced by the shape of the PDFs.
We also include the appropriate powers of $\alpha_s/(4\pi)$ at each order, so the overall
normalization shows the percent impact relative to the LO result.
For definiteness, the renormalization scale entering the PDFs is chosen as $\mu = 30~\GeV$.

In both flavor-diagonal contributions, denoted as $qqV$ and $gg$, the eikonal limit
correctly reproduces the divergent behavior as $x\to1$, but is off away from very large $x$.
Including the next-to-eikonal terms yields a sizable shift from the eikonal limit,
and provides a very good approximation in the shown $x$ region.
In the gluon channel, one can see a rise of the full kernel towards small $x$,
arising from an overall $1/z$ divergence in the coefficient $I_{gg}^{(2)}(z)$,
which is not captured by the expansion around $z\to1$. If desired, one could also
include the leading $z\to0$ behavior of the coefficients, which for simplicity
is not done here. For illustration, we also show the total
contribution from all other corresponding nondiagonal channels (gray dot-dashed).
In both cases, they are numerically subdominant to the flavor-diagonal channel
and also much flatter in $x$, since they only start at NLP.

Similar to the $\Tau_0$ coefficients in \sec{tau0_beam_beyond_eikonal},
we now wish to make an ansatz for the unknown three-loop NLP terms
to get an estimate of their size. A peculiar feature of the $q_T$ coefficients
is that up to three loops, its eikonal limit contains no logarithmic distributions
$\cL_n(1-z)$ with $n>0$, but only $\cL_0(1-z)$. In contrast, the NLP NNLO coefficient
does contain a double logarithm $\ln^2(1-z)$. Based on this observation, we make the
following ansatz for the N$^{n}$LO beam coefficient,
\begin{align} \label{eq:qT_ansatz}
\tilde I_{ij,\mathrm{approx}}^{(n)}(z) = \tilde I_{ij}^{(n)\,\mathrm{LP}}(z)
 &+ \Bigl[ X_1 \Gamma_0^i \ln^2(1-z) + X_2 \gamma_X^i \ln(1-z) \Bigr] \tilde I_{ij,\mathrm{reg}}^{(n-1)}(z)
 \nn\\&
 - X_3 (1-z) \tilde I_{ij}^{(n)\,\mathrm{LP}}(z)
\,.\end{align}
Here, $\tilde I_{ij,\mathrm{reg}}^{(n)}$ refers to the full regular (non-eikonal) piece
of the beam coefficient at $\cO(\as^n)$. At NLO, there is no NLP term, so at this order
we simply define the regular piece to be the appropriate color factor. More explicitly, we use
\begin{align}
 \tilde I_{ij,\mathrm{reg}}^{(1)}(z) &= -\delta_{ij} C_i
\,, \qquad
 \tilde I_{ij,\mathrm{reg}}^{(2)}(z)
 = \tilde I_{ij}^{(2)}(z) - \delta_{ij} \frac{\tilde\gamma_{\nu\, 1}^i}{2} \cL_0(1-z)
\,.\end{align}
The ansatz in \eq{qT_ansatz} dresses the lower-order regular kernel with two additional
logarithms $\ln(1-z)$. The coefficients of these logarithms are chosen such that at the
central choices $X_1 = X_2 = 1$, they reproduce the known double and single logarithms at NNLO.
The effective noncusp anomalous dimension $\gamma_X^i$ needed to achieve this is given by
\begin{align}
\gamma_X^g = 3 C_A - \beta_0
\,,\qquad
\gamma_X^q = 10 (C_F - C_A)
\,.\end{align}
The size of these additional logarithms can be probed by varying the coefficients $X_{1,2}$
by $\pm1$ around the central choice.
Furthermore, we add the eikonal limit $\tilde I_{ij}^{(n)\,\mathrm{LP}}$
suppressed by one power of $(1-z)$ to estimate the pure NLP constant.
Its coefficient $X_3$ is varied by $\pm1$ around the central choice $X_3 = 0$.

Since the $X_i$ probe independent structures, we can consider them as
uncorrelated. Hence, we add the impacts $\Delta_i$ on the final result of their
variation in quadrature
\begin{align}
\Delta = \Delta_{1} \oplus \Delta_{2} \oplus \Delta_{3} = \sqrt{\Delta_{1}^2 + \Delta_{2}^2 + \Delta_{3}^2}
\,.\end{align}

\begin{figure*}
\centering
\includegraphics[width=\WidthTwoSubfigs]{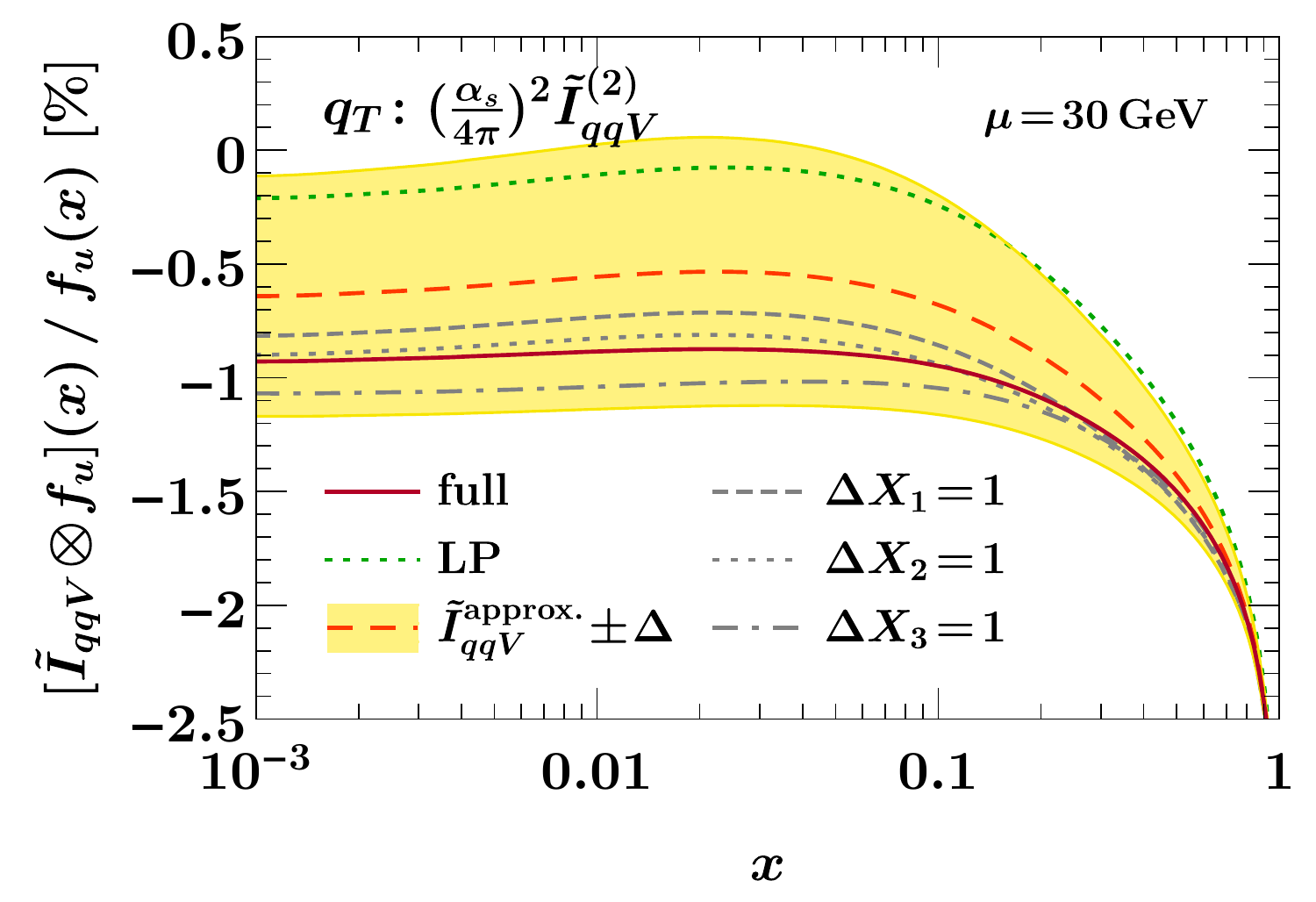}%
\hfill%
\includegraphics[width=\WidthTwoSubfigs]{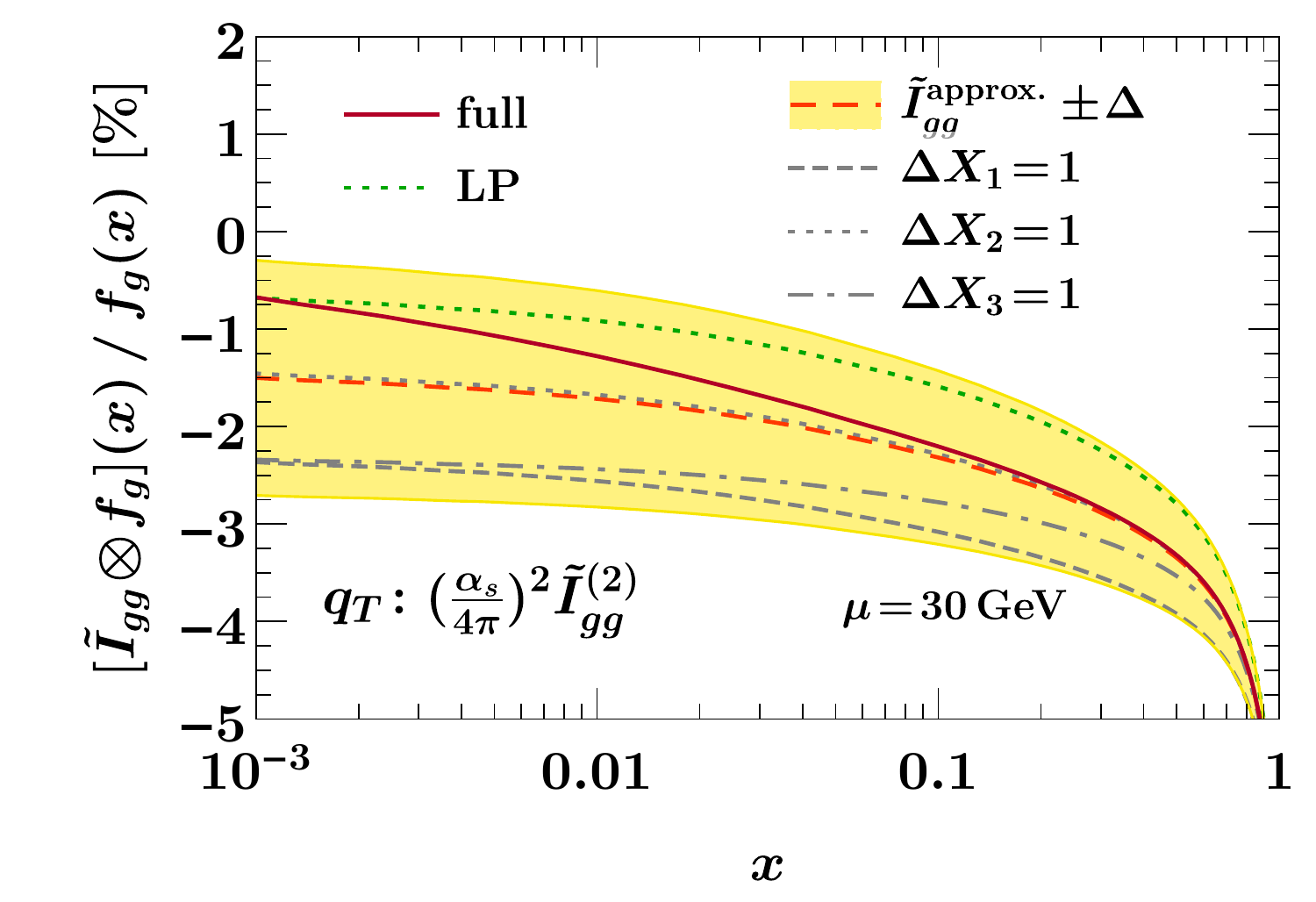}%
\\
\includegraphics[width=\WidthTwoSubfigs]{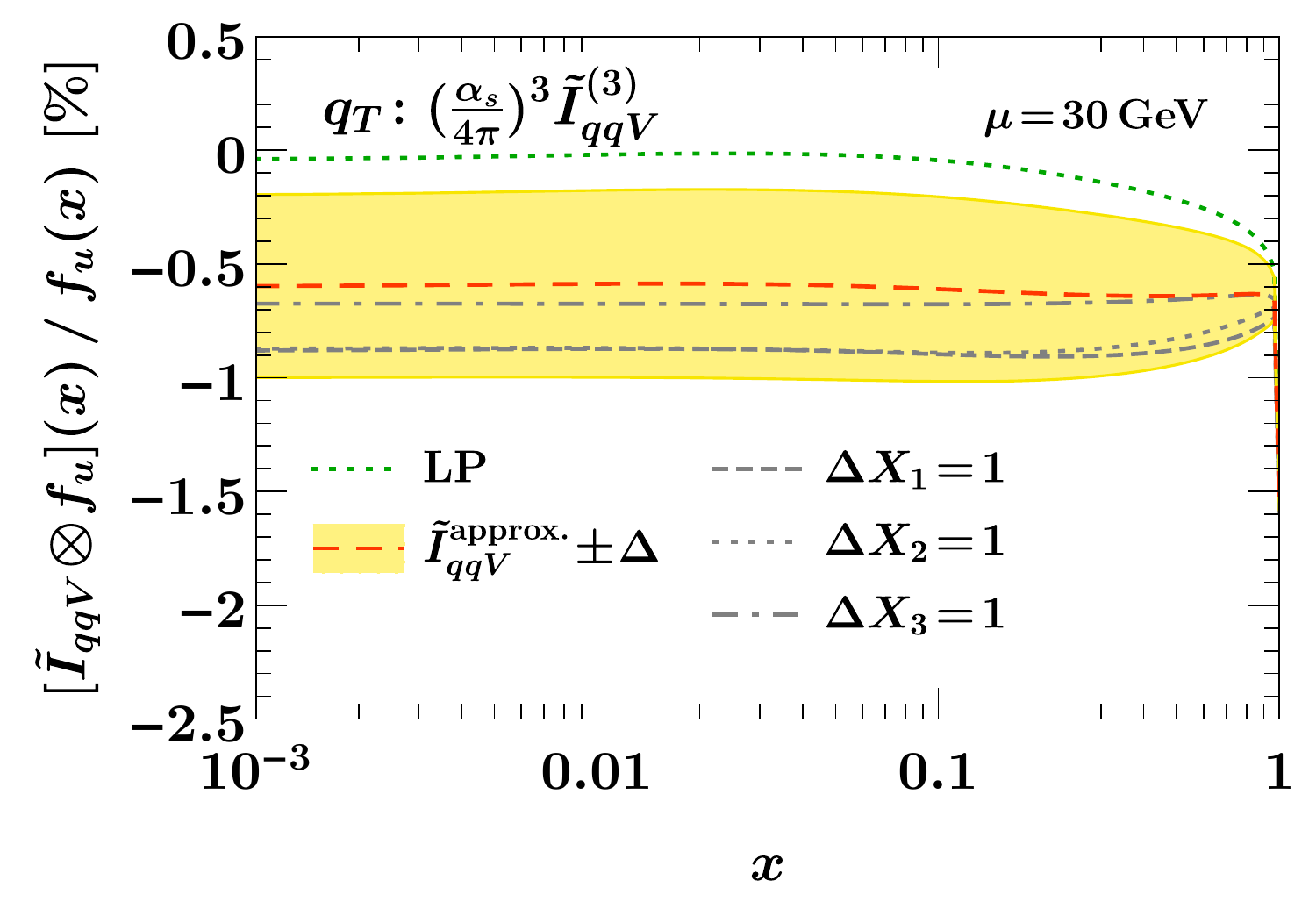}%
\hfill%
\includegraphics[width=\WidthTwoSubfigs]{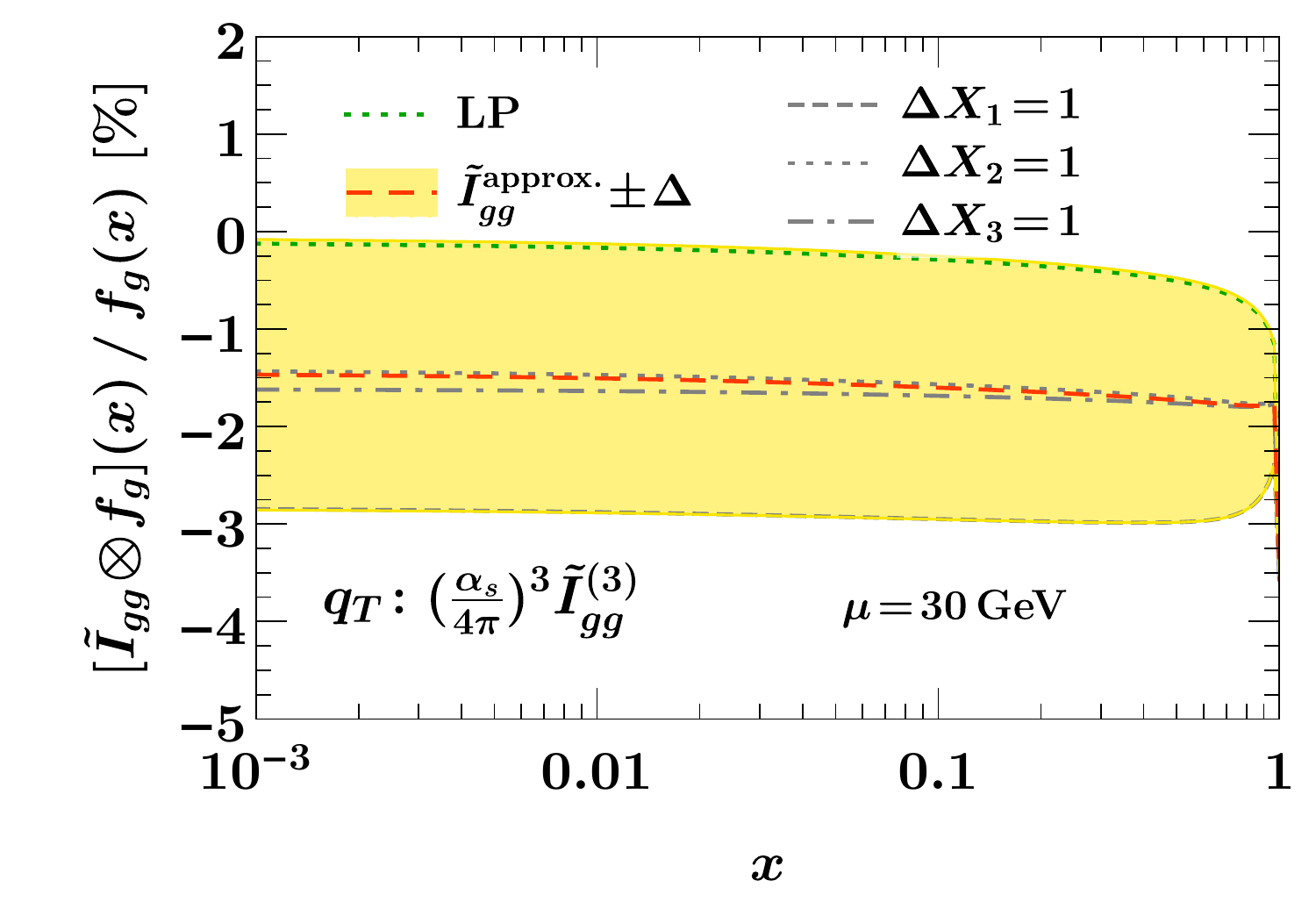}%
\caption{Approximate ansatzes for the NNLO (top) and N$^3$LO (bottom) kernels, in the $u$-quark (left) and gluon (right) channels.}
\label{fig:qTKernelApprox}
\end{figure*}

In \fig{qTKernelApprox}, we show the approximate kernel at NNLO (top)
and N$^3$LO (bottom) for the $u$-quark (left) and gluon (right) channels.
The dashed orange line shows the central result from our ansatz and the yellow band
its estimated uncertainty. The gray lines show the impact of the individual
variations of the $X_i$ as indicated.
In the top panel (NNLO), we also show the known
full two-loop results (red dashed). It shows that the ansatz including uncertainties
approximates the true result relatively well, even for the gluon case in the shown $x$ region.
In particular, the rather large shift from LP to the approximate NLP result is needed
to correctly capture the full result within uncertainties.

At N$^3$LO, we see again that the approximate result gives rise to a sizable shift
from the pure eikonal limit, which by itself is a very small correction.
This large shift arises on the one hand because the LP limit only contains $\cL_0(1-z)$
with a rather small coefficient $\gamma_{\nu\,2}^i$, while the NLP now contains up to
$\ln^4(1-z)$. The fact that the uncertainty bands are of similar
size at NNLO and N$^3$LO reflects their numerical importance
and that relatively little is known about the NLP structure,
which also motivates an exact calculation of the three-loop coefficients.

Finally, we briefly comment on the treatment of the unknown three-loop
beam function coefficients in \refcite{Cieri:2018oms}, where the $q_T$
subtraction was first applied at N$^3$LO for Higgs production.
There, the employed approximation was $\tilde I_{gg}^{(3)}(z) = \tilde C_{N3} \, \delta(1-z)$,
with $\tilde C_{N3}$ fixed such that the inclusive cross section is correctly reproduced.
This effectively absorbs the averaged effect of the actual $z$ dependence into
an effective $\delta(1-z)$ coefficient. From our results we know the exact
$\delta(1-z)$ coefficient, and so our approximate results give an independent estimate of the
actual rapidity dependence and total size of these unknown terms.

\FloatBarrier

\section{\texorpdfstring{N$^3$LO}{N3LO} subtractions}
\label{sec:subtractions}

The factorization in \eq{fact_generic} fully describes the limit $\tau\to0$ and
thus captures the singular structure of QCD in this limit. Hence, it can be used
to construct a subtraction method for fixed-order calculations. In principle,
this works for any resolution variable $\tau$ and any process for which a
corresponding factorization is known~\cite{Catani:2007vq, Zhu:2012ts,
Gao:2012ja, Catani:2014qha, Gao:2014nva, Gao:2014eea, Gangal:2014qda,
Boughezal:2015dva, Gaunt:2015pea}. The subtractions can be formulated
differential in $\tau$ or as a global $\tau$ slicing, which we briefly review in
the following. For a more extensive discussion we refer to
\refcite{Gaunt:2015pea}.

Our starting point is to write the inclusive cross section as the integral over the
differential cross section in $\tau$,
\begin{equation}
\sigma(X)
= \int\!\df\tau\, \frac{\df\sigma(X)}{\df\tau}
\,, \qquad
\sigma(X, \tau_\cut)
= \int^{\tau_\cut} \!\df\tau\, \frac{\df\sigma(X)}{\df\tau}
\,,\end{equation}
where the second relation defines the cumulant in $\tau_\cut$.
Here, $X$ denotes any measurements performed,
which can include $Q$ and $Y$ of the color singlet $L$ but also additional
measurements or cuts on its constituents. For $\tau\to 0$, the cross sections
scales like $\sim 1/\tau$, so performing the $\tau$ integral
requires knowing the full analytic distributional structure involving $\delta(\tau)$
and $\cL_n(\tau)$, which encodes the cancellation of real and virtual IR divergences.
To separate out the singular structure in $\tau$, we introduce a subtraction term,
\begin{align} \label{eq:tau_subtraction}
\sigma(X)
&= \sigma^\sub(X, \tau_\off)
+ \int\!\df\tau\, \biggl[ \frac{\df\sigma(X)}{\df\tau}
- \frac{\df\sigma^\sub(X)}{\df\tau} \theta(\tau < \tau_\off) \biggr]
\,,\end{align}
where $\df\sigma^\sub(X)/\df\tau$ captures all singularities for $\tau\to 0$,
\begin{equation} \label{eq:sub_sing}
\frac{\df\sigma(X)}{\df\tau}
= \frac{\df\sigma^\sing(X)}{\df\tau}\,[1 + \ord{\tau}]
\,,\qquad
\frac{\df\sigma^\sub(X)}{\df\tau}
= \frac{\df\sigma^\sing(X)}{\df\tau}\,[1 + \ord{\tau}]
\,,\end{equation}
and $\sigma^\sub(X, \tau_\off)$ is the integrated subtraction term,
\begin{equation}
\sigma^\sub(X, \tau_\off)
= \int\!\df\tau\, \frac{\df\sigma^\sub(X)}{\df\tau} \theta(\tau < \tau_\off)
\,.\end{equation}
By construction, the integrand in square brackets in \eq{tau_subtraction} contains at
most integrable singularities for $\tau\to 0$ and so the integral can be performed numerically. Hence, the
full cross section $\df\sigma(X)/\df\tau$ is only ever evaluated at finite $\tau > 0$,
which means it can be obtained from a calculation of the corresponding $ab\to L+1$-parton
process at one lower order. In practice, one always has a small IR cutoff $\delta$
on the $\tau$ integral,
\begin{equation} \label{eq:tau_subtraction_delta}
\sigma(X)
= \sigma^\sub(X, \tau_\off)
+ \int_{\delta} \!\df\tau\, \biggl[ \frac{\df\sigma(X)}{\df\tau}
- \frac{\df\sigma^\sub(X)}{\df\tau} \theta(\tau < \tau_\off) \biggr]
+ \Delta\sigma(X, \delta)
\,,\end{equation}
where the last term contains the integral over $\tau\leq\delta$,
\begin{equation}
\Delta\sigma(X, \delta) = \sigma(X, \delta) - \sigma^\sub(X, \delta)
\sim \ord{\delta}
\,.\end{equation}
which is neglected for $\delta\to 0$.

The above is a \emph{differential} $\tau$-subtraction scheme, where the
parameter $\tau_\off\sim 1$ determines the range over which the subtraction
acts. The key advantage of formulating the subtractions in terms of a physical
resolution variable $\tau$, is that the subtraction terms are given by the
singular limit of a physical cross section. Hence, they are precisely given by the
factorization formula for $\tau\to 0$, which is also the basis for the
resummation in $\tau$. In fact, this form of the subtraction is routinely used
when the resummed and fixed-order results are combined via an additive matching.
In this case, $\tau_\off$ corresponds to the point where the $\tau$ resummation
is turned off, and the term in square brackets in
\eq{tau_subtraction} is the nonsingular cross section that is added to the pure
resummed result.
Differential $\Tau_0$ subtractions are used in this way in the \geneva\ Monte Carlo
to combine the fully-differential NNLO calculation together with the NNLL$'$ $\Tau_0$
resummation with a parton shower~\cite{Alioli:2012fc, Alioli:2013hqa, Alioli:2015toa}.
The differential subtractions at N$^3$LO are a key ingredient for using
this method to combine N$^3$LO calculations with parton showers.

In contrast to a fully local subtraction scheme, all singularities are projected
onto the resolution variable $\tau$, so the subtractions are local in $\tau$
but nonlocal in the additional radiation phase space that is integrated over.
As discussed in \refcite{Gaunt:2015pea}, the subtractions can be made
more local by considering a factorization theorem that is differential in
more variables. For example, the combined $q_T$ and $\Tau_0$
resummation~\cite{Procura:2014cba, Lustermans:2019plv} offers the possibility
to construct double-differential $q_T-\Tau_0$ subtractions.

The key point of the differential subtraction is that $\delta$ can in principle
be made arbitrarily small, because the integrand of the $\tau$ integral is
nonsingular, which also means that the numerically expensive small $\tau$ region
does not need to be sampled with weight $1/\tau$. On the other hand, by letting
$\delta = \tau_\cut$ be a small but finite cutoff and setting $\tau_\off =
\tau_\cut$, \eq{tau_subtraction_delta} turns into a global $\tau$ subtraction or
\emph{slicing},
\begin{equation} \label{eq:tau_slicing}
\sigma(X)
= \sigma^\sub(X, \tau_\cut)
+ \int_{\tau_\cut} \!\df\tau\, \frac{\df\sigma(X)}{\df\tau}
+ \Delta\sigma(X, \tau_\cut)
\,.\end{equation}
The practical advantage of the slicing method is that it allows one to readily
turn an existing $L+1$-jet N$^{n-1}$LO calculation into a N$^n$LO calculation for $L$,
and so most implementations use this approach~\cite{Grazzini:2008tf, Catani:2009sm, Grazzini:2017mhc, Boughezal:2015dva, Campbell:2016jau, Boughezal:2016wmq, Heinrich:2017bvg, Cieri:2018oms,
Campbell:2019gmd}.
The main disadvantage is that the cancellation of the divergences now only happens
after the integration over $\tau$.
This makes the $L+1$-jet calculation very demanding, both in terms of
computational expense and numerical stability, because the $1/\tau$-divergent
integral of $\df\sigma(X)/\df\tau$ must be computed with sufficient accuracy
down to sufficiently small $\tau_\cut$, which in practice limits how small
one can take $\tau_\cut$.
Since the integral is divergent, one cannot let $\tau_\cut\to 0$
even in principle, so one always has a leftover systematic uncertainty from the
neglected power corrections $\Delta\sigma(X, \tau_\cut)$.

The numerical efficiency of the subtractions can be improved by including the
power corrections in the subtractions for both $\Tau_0$~\cite{Moult:2016fqy, Moult:2017jsg, Ebert:2018lzn,
Boughezal:2016zws, Boughezal:2018mvf, Boughezal:2019ggi} and $q_T$~\cite{Ebert:2018gsn, Cieri:2019tfv}.
The size of the missing power corrections also strongly depends on the precise definition of $\Tau_0$~\cite{Moult:2016fqy, Moult:2017jsg, Ebert:2018lzn}.
The hadronic definition in \eq{Tau0_2} exhibits power corrections that grow like $e^{\abs{Y}}$
at large $Y$, which is not the case for the leptonic definition.
The power corrections also depend on the Born measurement $X$.
In particular, additional selection or isolation cuts on the color-singlet
constituents typically enhance the power corrections from $\ord{\tau}$ to
$\ord{\sqrt{\tau}}$~\cite{Ebert:2019zkb}.

\subsection{Subtraction terms}

The singular terms needed for the subtractions only depend on the Born phase
space, so we can write them as
\begin{equation}
\frac{\df\sigma^\sing(X)}{\df\tau}
= \int\!\df \Phi_0\, \frac{\df\sigma^\sing(\Phi_0)}{\df\tau}\, X(\Phi_0)
\,,\end{equation}
where $\Phi_0 \equiv \Phi_0(\kappa_a, \kappa_b, \w_a, \w_b)$ denotes the full Born phase space,
including the parton labels $\kappa_{a,b}$, the total color-singlet momentum $q^\mu$ parametrized in terms of
$\w_{a,b}$ as in \eqs{Born_process}{Born} as well as the internal
phase space of $L$. The $X(\Phi_0)$ denotes the
measurement function that implements the measurement $X$ on a Born configuration.

The singular terms are defined such that their $\tau$ dependence is minimal and
given by
\begin{align} \label{eq:diff_subtractions}
\frac{\df\sigma^\sing(\Phi_0)}{\df\tau}
&= \cC_{-1}(\Phi_0)\,\delta(\tau) + \sum_{n\geq 0} \cC_n(\Phi_0)\, \cL_n(\tau)
\nn \\
&= \sum_{m \geq 0} \biggl[\cC_{-1}^{(m)}(\Phi_0)\,\delta(\tau) + \sum_{n = 0}^{2m-1} \cC_n^{(m)}(\Phi_0) \cL_n(\tau) \biggr] \Bigl(\frac{\alpha_s}{4\pi} \Bigr)^m
\,.\end{align}
Their integral over $\tau \leq \tau_\cut$ immediately follows as
\begin{align} \label{eq:integrated_subtractions}
\sigma^\sing(\Phi_0, \tau_\cut)
&= \cC_{-1}(\Phi_0) + \sum_{n\geq 0} \cC_n(\Phi_0)\, \frac{\ln^n \tau_\cut}{n+1}
\nn \\
&= \sum_{m \geq 0} \biggl[\cC_{-1}^{(m)}(\Phi_0)
 + \sum_{n = 0}^{2m-1} \cC_n^{(m)}(\Phi_0) \frac{\ln^n \tau_\cut}{n+1} \biggr] \Bigl(\frac{\alpha_s}{4\pi} \Bigr)^m
\,.\end{align}
The differential subtractions
are given by using \eq{diff_subtractions} for $\tau > 0$, which amounts to dropping
the $\cC_{-1}(\Phi_0)\delta(\tau)$ term and using $\cL_n(\tau>0) = \ln^{n-1}(\tau)/\tau$.
The integrated subtractions are directly given by \eq{integrated_subtractions}.

The precise definition of the $\cC_n(\Phi_0)$ coefficients depends on the
normalization of the dimensionless variable $\tau$ or equivalently on the boundary
condition of the $\cL_n(\tau)$.
Rescaling $\tau \to \lambda\tau$ moves contributions from $\cC_n(\Phi_0)$ to
$\cC_{m<n}(\Phi_0)$. This freedom was used in \refcite{Gaunt:2015pea} to
absorb all terms with $n \geq 0$ in \eq{integrated_subtractions} into a $\cC_{-1}(\Phi_0, \Tau_\off)$
by taking $\tau \equiv \Tau_0/\Tau_\off$.
Here, we prefer to keep the cutoff dependence explicit as in \eq{integrated_subtractions}
and take
\begin{align}
\tau &\equiv \frac{\Tau_0}{Q}
\qquad\text{(for $\Tau_0$)}
\,,\qquad
\tau \equiv \frac{q_T^2}{Q^2}
\qquad\text{(for $q_T$)}
\,.\end{align}

The $m$-loop subtraction coefficients $\cC_n^{(m)}(\Phi_0)$ directly follow from expanding
\eq{Tau0_fact} for $\Tau_0$ or \eq{qt_factorization} for $q_T$ to $m$th order in $\as$.
For the three-loop coefficients this yields
\begin{align} \label{eq:subtraction_coeffs}
\cC_{-1}^\three(\Phi_0)
&= H^\three(\Phi_0)\, f_a(x_a)\, f_b(x_b)
+ \sum_{m = 1}^3 H^{(3-m)}(\Phi_0)\, \bigl[B_a(x_a) B_b(x_b) S\bigr]_{-1}^{(m)}
\,, \nn \\
\cC_{n\geq 0}^\three(\Phi_0)
&= \sum_{k = 1}^3 H^{(3-k)}(\Phi_0)\, \bigl[B_a(x_a) B_b(x_b) S\bigr]_n^{(k)}
\,,\end{align}
where for simplicity we have suppressed the dependence on $\mu$ and the distinction
of the $\Tau_0$ vs. $q_T$ beam and soft functions.
The virtual three-loop corrections to the Born process
are contained in $H^\three(\Phi_0)$, which only enters in $\cC_{-1}^\three$.
The $m$-loop soft/collinear contribution $[BBS]_n^{(m)}$ follows from inserting
the fixed-order expansions of the respective beam and soft function, reexpanding their
product to $m$th order and picking out the coefficients of $\delta(\tau)$ and
$\cL_n(\tau)$.
The three-loop boundary coefficients of the beam and soft functions only enter
in $\cC_{-1}^\three$ and thus are needed for the integrated subtraction terms but
not the differential ones. Note also that most of the process and $\Phi_0$
dependence resides in the hard coefficients, while the soft/collinear contributions
only depend on $x_{a,b}$ and the parton types,
\begin{equation}
\bigl[B_a(x_a) B_b(x_b) S\bigr]_n^{(m)}
= \int\!\frac{\df z_a}{z_a} \frac{\df z_b}{z_b}\,
\sum_{i,j} \bigl[\cI_{ai}(z_a) \cI_{bi}(z_b) S\bigr]_n^{(m)}
f_i\Bigl(\frac{x_a}{z_a}\Bigr) f_j\Bigl(\frac{x_b}{z_b}\Bigr)
\,.\end{equation}
The results for the subtraction coefficients $\cC_n(\Phi_0)$
in \eq{subtraction_coeffs} up to three loops for both $\Tau_0$ and $q_T$ have been
implemented in the C++ library \scetlib~\cite{scetlib} and will be made publicly available.

Note that evaluating \eq{subtraction_coeffs} for $\Tau_0$ requires rescaling and convolving the
plus distributions in the beam and soft functions, as discussed in
\refcite{Gaunt:2015pea}. For $q_T$, expanding the $\bt$-space result
$\df\tilde\sigma^\sing(\bt)$ yields
powers of the $\bt$-space logarithm $L_b^n$ up to $n \leq 6$. Their
Fourier transform, given in \tab{LnbtopT} in \app{plus_distr_qt}, yields simple $\delta(\qt)$ and
$\cL_n(\qt, \mu)$, which are easily rescaled to $\delta(\tau)$ and $\cL_n(\tau)$.

Note also that the original $q_T$ subtraction method in \refcite{Catani:2007vq} was based on the
$q_T$ resummation framework of \refcite{Bozzi:2005wk}, where the canonical $\bt$-space
logarithms are replaced by
\begin{equation}
L_b
\to \tilde L_b \equiv \ln\Bigl(\frac{b_T^2 \mu^2}{b_0^2} + 1 \Bigr)
\,.\end{equation}
This form is also used e.g.\ in \refscite{Grazzini:2017mhc, Cieri:2018oms}.
While using $\tilde L_b$ has certain advantages in the context of $q_T$ resummation,
it is unnecessary for the purpose of $q_T$ subtractions, since $L_b$ and $\tilde L_b$
yield the same singular terms and only differ by power corrections.
A drawback of using $\tilde L_b$ here is that the
Fourier transform of $\tilde L_b^n$ yields complicated expressions in $q_T$ space,
see appendix B in \refcite{Bozzi:2005wk}, whose cumulants are not known analytically
and must be performed numerically.

\section{Conclusions}
\label{sec:conclusions}

We have studied the three-loop structure of beam and soft functions
for both $0$-jettiness $\Tau_0$ and transverse momentum $q_T$.
These functions are defined as collinear proton matrix elements and soft vacuum
matrix element, measuring the small light-cone momentum (for $\Tau_0$) or
total transverse momentum (for $q_T$) of all soft and collinear emissions,
and thus are universal objects probing the infrared structure of QCD.

The all-order structure of the beam and soft functions is governed by renormalization
group equations, which we have employed to derive their full three-loop structure.
For the currently unknown scale-independent boundary coefficients
$I^\three_{ij}(z)$ of the N$^3$LO beam functions, we employ consistency between different
factorization limits to derive their leading eikonal limit $I^\three_{ij}(z\to1)$,
i.e.\ the full singular limit of the beam functions as $z\to1$,
and estimate the size of the unknown terms beyond the eikonal limit.
All results of this paper will be made available in the C++ library \scetlib~\cite{scetlib}.

Our results provide important ingredients required for the resummation of $\Tau_0$ and $q_T$
at N$^3$LL$'$ and N$^4$LL order. In particular, they are important for extending
the $q_T$ and $\Tau_0$ subtraction methods to N$^3$LO, for which we provide the
complete set of differential subtraction terms at three loops, which are
for example necessary for extending the matching of
fixed-order calculations to parton showers to N$^3$LO$+$PS.
The integrated subtraction terms are not yet fully known at three loops,
but the obtained eikonal limit allows us to provide a first approximation for a full
three-loop subtraction, and will be a useful cross check once the full
$q_T$ and $\Tau_0$ beam functions become available.

\paragraph{Note added:} As discussed in the introduction,
since this paper first appeared, the full three-loop integrated
subtraction terms have become available. Specifically,
results for the three-loop $\Tau_0$ quark beam function
in the generalized large-$N_c$ approximation
have appeared in \refcite{Behring:2019quf} and the complete result has
been calculated in \refcite{Ebert:2020unb}.
The three-loop beam functions for $q_T$
have been calculated in \refscite{Luo:2019szz, Ebert:2020yqt}.
In all cases, the full calculations have confirmed our predictions of the
eikonal terms at three loops.

\acknowledgments
We thank Goutam Das for fruitful discussions,
V.~Ravindran for providing us with the results of \refcite{Ravindran:2006bu, Ahmed:2014uya},
and Hua Xing Zhu for providing us with a numerical version of the results in \refcite{Li:2016ctv}.
We also thank Marius Wiesemann and Leandro Cieri for clarifying discussions
regarding \refscite{Grazzini:2017mhc, Cieri:2018oms}.
J.M.\ and F.T.\ thank the MIT Center for Theoretical Physics for hospitality
and M.E.\ thanks the DESY theory group for hospitality.
This work was supported in part by the Office of Nuclear Physics of the U.S.\
Department of Energy under Contract No.\ DE-SC0011090,
the Alexander von Humboldt Foundation through a Feodor Lynen Research Fellowship,
the Deutsche Forschungsgemeinschaft (DFG) under Germany's Excellence
Strategy -- EXC 2121 ``Quantum Universe'' -- 390833306,
and the PIER Hamburg Seed Project PHM-2019-01.
M.E.\ and F.T.\ also thank the Mainz Institute of
Theoretical Physics of the DFG Cluster of Excellence PRISMA$^+$ (Project ID 39083149)
for hospitality while portions of this work were performed.

\appendix

\section{Plus distributions and Fourier transforms}
\label{app:plusDist}

Here, we summarize the definitions and relations for plus distributions.

\subsection{One-dimensional plus distributions}
\label{app:plus_distr_TauN}

Following \refcite{Ligeti:2008ac}, we denote plus distributions as
\begin{align} \label{eq:plusdef_standard}
\cL_{n}(x)
&= \biggl[ \frac{\theta(x) \ln^n x}{x}\biggr]_+
= \lim_{\epsilon \to 0} \frac{\df}{\df x} \Bigl[ \theta(x-\epsilon) \frac{\ln^{n+1} x}{n+1} \Bigr]
\,, \nn \\
\cL^{a}(x) &= \biggl[ \frac{\theta(x)}{x^{1-a}} \biggr]_+
= \lim_{\epsilon \to 0} \frac{\df}{\df x} \Bigl[ \theta(x-\epsilon) \frac{x^a - 1}{a} \Bigr]
\,,\end{align}
such that
\begin{equation}
\cL_n(x > 0) = \frac{\ln^n x}{x}
\,, \quad
\cL^a(x > 0) = \frac{1}{x^{1-a}}
\,, \qquad
\int_0^1 \! \df x \, \cL_n(x) = \int_0^1 \! \df x \, \cL^a(x) = 0
\,.\end{equation}
For distributions with dimensionful arguments we define
\begin{alignat}{3} \label{eq:cL_n_k}
\cL_{n}(k,\mu) &= \frac{1}{\mu} \cL_n\Bigl(\frac{k}{\mu}\Bigr)
\,, \qquad
\cL_{n}(t, \mu^2) = \frac{1}{\mu^2} \cL_n\Bigl(\frac{t}{\mu^2}\Bigr)
\,.\end{alignat}

Using $\cL^a(x)$ we further define the distribution
\begin{align} \label{eq:cVa_definition}
\cV_a(x)
= \frac{e^{-\gamma_E a}}{\Gamma(1+a)} \bigl[ a \cL^a(x) + \delta(x) \bigr]
\,, \qquad
\cV_a(k, \mu)
= \frac{1}{\mu} \cV_a \Bigl( \frac{k}{\mu} \Bigr)
\,,\end{align}
which satisfies the group property
\begin{equation} \label{eq:cVa_group_property}
(\cV_a \cV_b)(k, \mu) =
\int \! \df k'\,
\mathcal{V}_a(k - k',\mu) \, \mathcal{V}_b(k',\mu) = \mathcal{V}_{a+b}(k,\mu)
\,, \qquad
\mathcal{V}_0(k, \mu) = \delta(k)
\,.\end{equation}
The $\mu$ dependence of $\cV_a(k, \mu)$ is given by
\begin{equation} \label{eq:cVa_shift_boundary_condition}
\cV_a(k, \mu) = \Bigl( \frac{\mu'}{\mu} \Bigr)^a \, \cV_a(k, \mu')
\,, \qquad
\mu \frac{\df}{\df \mu} \cV_a(k, \mu) = -a \cV_a(k, \mu)
\,.\end{equation}
Expanding $\cV_a(k, \mu)$ in powers of $a$ we find
\begin{align} \label{eq:cVa_series_expansion}
\cV_a(k, \mu)
&= \delta(k) + a\,\cL_0(k, \mu) + \frac{a^2}{2!}\bigl[2\cL_1(k, \mu) - \zeta_2\delta(k) \bigr]
\nn \\ & \quad
+ \frac{a^3}{3!} \bigl[3\cL_2(k, \mu) - 3\zeta_2\cL_0(k, \mu) + 2\zeta_3 \delta(k) \bigr]
+ \ord{a^4}
\,.\end{align}
The Fourier transformation of $\cV_a(k, \mu)$ is given by
\begin{align} \label{eq:cVa_fourier_transform}
\int \! \df k \, e^{-\img k y} \, \cV_a(k, \mu) = e^{-a L_y}
\,, \quad
\int \! \frac{\df y}{2\pi}\, e^{\img k y} \, e^{-a L_y} = \cV_a(k, \mu)
\,, \quad
L_y = \ln(\img y \mu e^{\gamma_E})
\,.\end{align}

\subsection{Two-dimensional plus distributions for \texorpdfstring{$\qt$}{qT}}
\label{app:plus_distr_qt}

\begin{table}
\centering
\begin{tabular}{c|l}
\hline\hline
$L_b^n$ & $\text{FT}^{-1}[L_b^n]$ \\ \hline
$1$
& $\delta^{(2)}(\qt)$
\\ \hline
$L_b$
& $-\cL_{0}(\qt, \mu)$
 \\ \hline
$L_b^2$
& $+2 \cL_{1}(\qt, \mu)$
 \\ \hline
$L_b^3$
& $-3 \cL_{2}(\qt, \mu) - 4 \zeta_3 \delta^{(2)}(\qt)$
 \\ \hline
$L_b^4$
& $+4 \cL_{3}(\qt, \mu) + 16 \zeta_3 \cL_{0}(\qt, \mu)$
 \\ \hline
$L_b^5$
& $ -5 \cL_{4}(\qt, \mu) - 80 \zeta_3 \cL_{1}(\qt, \mu) - 48 \zeta_5 \delta^{(2)}(\qt)$
 \\ \hline
$L_b^6$
& $+6 \cL_{5}(\qt, \mu) + 240 \zeta_3 \cL_{2}(\qt, \mu) + 288 \zeta_5 \cL_{0}(\qt, \mu) + 160 \zeta_3^2 \delta^{(2)}(\qt)$
 \\ \hline\hline
\end{tabular}
\caption{Fourier transform of $L_b^n = \ln^n(b_T^2 \mu^2 / b_0^2)$ to $\qt$ space
for $n \le 6$, as given by \eq{FT_2d}.}
\label{tab:LnbtopT}
\end{table}

Following \refcite{Ebert:2016gcn}, we define two-dimensional plus distributions
in $\qt$ as
\begin{equation} \label{eq:Ln_qt}
\cL_n(\qt, \mu)
= \frac{1}{\pi \mu^2} \cL_n\biggl(\frac{q_T^2}{\mu^2}\biggr)
\,,\end{equation}
where $\cL_n(x)$ is defined as above in \eq{plusdef_standard}, such that
\begin{equation}
\int_{|\qt| \le \mu} \df^2\qt \, \cL_n(\qt,\mu)
 = \pi \int_0^{\mu^2} \df q_T^2 \,  \frac{1}{\pi \mu^2} \cL_n\biggl(\frac{q_T^2}{\mu^2}\biggr)
 = 0
\,.\end{equation}
The cumulant for a generic cut $\abs{\qt} \leq q_T^\cut$ follows to be
\begin{equation} \label{eq:Ln_cumulant}
 \int_{|\qt| \le q_T^\cut} \df^2\qt \, \cL_n(\qt,\mu)
 = \frac{\theta(q_T^\cut)}{n+1} \ln^{n+1} \frac{(q_T^\cut)^2}{\mu^2}
\,.\end{equation}
The Fourier transformation of $\cL_n(\qt, \mu)$ and its inverse are~\cite{Ebert:2016gcn}
\begin{align} \label{eq:FT_2d}
\int\!\df^2\qt\, e^{-i \qt \cdot \bt} \cL_n(\qt,\mu)
&= \frac{1}{n+1} \sum_{k=0}^{n+1} (-1)^k \binom{n+1}{k} R_2^{(n+1-k)} L_b^k
\,, \\ \nn
\int\! \frac{\df^2\bt}{(2\pi)^2}\, e^{i \qt \cdot \bt} L_b^n
&= \sum_{k=0}^{n-1} (-1)^{k+1} n \binom{n-1}{k} R_2^{(n-k-1)} \cL_k(\qt,\mu) +  R_2^{(n)} \delta^{(2)}(\qt)
\,,\end{align}
where $L_b$ is the usual logarithm in Fourier space
\begin{equation}
L_b = \ln(b_T^2 \mu^2 / b_0^2)
\,,\qquad b_0 = 2 e^{-\gamma_E}
\,,\end{equation}
and the coefficients $R_2^{(n)}$ in \eq{FT_2d} are given by
\begin{equation} \label{eq:dR2}
R_2^{(n)}
= \frac{\df^n}{\df a^n}\, e^{2\gamma_E a}\, \frac{\Gamma(1+a)}{\Gamma(1-a)}\, \bigg\rvert_{a=0}
\,.\end{equation}
Up to N$^3$LO, we require the Fourier transforms of $L_b^n$
with $n \le 6$, which are summarized in \tab{LnbtopT}.

\section{Threshold soft function}
\label{app:threshold_soft}

Here we discuss the double-differential threshold soft function $S_i^\thr(k^-, k^+, \mu)$,
which appears in the soft threshold factorization for the inclusive cross section in \eq{factorization_soft_threshold_qm_qp} and determines the eikonal limit of the
$\Tau_0$ beam function in \eq{TauN_beam_eikonal_limit_result}.
We give its complete N$^3$LO result in \app{threshold_soft_three_loops}
in terms of a convenient plus distribution basis defined in \app{threshold_soft_basis}.
In \app{threshold_soft_method}, we discuss how the three-loop coefficients
are extracted from the known three-loop results for the closely-related inclusive
threshold soft function.

\subsection{Plus distribution basis}
\label{app:threshold_soft_basis}

A key property of the threshold soft function is that is invariant under the
simultaneous rescaling $k^- \mapsto k^- e^{+y}$ and $k^+ \mapsto k^+ e^{-y}$,
see \eq{threshold_soft_rescaling}. To make this property manifest, we define a
basis of plus distributions in $k^\pm$ that individually satisfy this property,
\begin{align}
\frac{\theta(k^-)\theta(k^+)}{\mu^2} \Bigl(\frac{k^- k^+}{\mu^2}\Bigr)^{-1+a}
&= \biggl[\frac{\delta(k^-)}{a} +\sum_{n=0}^{\infty} \frac{a^n}{n!} \cL_n(k^-,\mu)\biggr] \biggl[\frac{\delta(k^+)}{a} +\sum_{m=0}^{\infty} \frac{a^m}{m!} \cL_m(k^+,\mu)\biggr]
\nn \\
&\equiv \frac{\delta(k^-, k^+)}{a^2} + \sum_{n=0}^{\infty} \frac{a^{n-1}}{n!}\cL_{n}(k^-,k^+,\mu)
\,.\end{align}
Note that the leading $\delta(k^-, k^+)$ term multiplies a double pole in $a$.
The second line implicitly defines the $\cL_n(k^-, k^+, \mu)$ by the expansion
of the first line in powers of $a$.
They are by construction invariant under rescaling, because the left-hand side is.
Explicitly, they are given by
\begin{align}
\delta(k^-, k^+) &= \delta(k^-) \, \delta(k^+)
\,, \nn \\
\cL_{n}(k^-, k^+, \mu) &= \delta(k^-)\, \cL_{n}(k^+,\mu) + \cL_{n}(k^-,\mu)\, \delta(k^+)
\nn \\ & \quad
+ n \sum_{m = 0}^{n-1} \binom{n-1}{m} \cL_{m}(k^-, \mu) \, \cL_{n-1-m}(k^+, \mu)
\,.\end{align}

\subsection{Three-loop result}
\label{app:threshold_soft_three_loops}

The threshold soft function satisfies the all-order RGE
\begin{align} \label{eq:threshold_soft_rge}
\mu \frac{\df}{\df \mu} S^\thr_i(k^-, k^+, \mu)
&= \int \! \df \ell^- \df \ell^+ \, \gamma_\thr^i(k^- - \ell^-, k^+ - \ell^+, \mu) \, S^\thr_i(\ell^-, \ell^+, \mu)
\,, \nn \\
\gamma_\thr^i(k^-, k^+, \mu)
&= -2 \GammaC^i[\as(\mu)] \, \cL_0(k^-, k^+, \mu) + \gamma_\thr^i[\as(\mu)] \, \delta(k^-, k^+)
\,.\end{align}
Expanding the threshold soft function in $\alpha_s$ as
\begin{equation}
S^\thr_i(k^-, k^+, \mu)
= \sum_{n=0}^\infty S^{\thr(n)}_i(k^-, k^+, \mu)\, \Bigl[\frac{\as(\mu)}{4\pi}\Bigr]^n
\,,\end{equation}
and suppressing all arguments for brevity, $S_{i}^{\thr(n)} \equiv S_{i}^{\thr(n)}(k^-, k^+, \mu)$,
$\cL_n \equiv \cL_n (k^-, k^+, \mu)$, $\delta \equiv \delta(k^-, k^+)$,
the three-loop solution of \eq{threshold_soft_rge} takes the form
\begin{align} \label{eq:threshold_soft_n3lo}
S_i^{\thr\zero}
&= \delta
\,, \nn \\
S_i^{\thr\one}
&= \cL_1 \, \Gamma_0^i
- \cL_0 \, \frac{\gamma_{\thr\,0}^{i}}{2}
+ \delta \, s^{\thr\one}_i
\,, \nn \\
S_i^{\thr\two}
&= \cL_3 \, \frac{(\Gamma_0^i)^2}{2}
 - \cL_2 \, \frac{\Gamma_0^i}{2} \Bigl(\beta_0 + \frac{3}{2}\gamma_{\thr\,0}^{i} \Bigr)
\nn \\ &\quad
+ \cL_1 \Bigl[
   - 2\zeta_2 (\Gamma_0^i)^2
   + \Bigl(\beta_0 + \frac{\gamma_{\thr\,0}^{i}}{2} \Bigr) \frac{\gamma_{\thr\,0}^{i}}{2}
   + \Gamma_1^i
   + \Gamma_0^i \, s^{\thr\one}_i
\Bigr]
\nn \\ &\quad
+ \cL_0 \Bigl[
     \Gamma_0^i\bigl(2\zeta_3 \Gamma_0^i + \zeta_2 \gamma_{\thr\,0}^{i} \bigr)
   - \frac{\gamma_{\thr\,1}^{i}}{2}
   - \Bigl(\beta_0 + \frac{\gamma_{\thr\,0}^{i}}{2}\Bigr) s^{\thr\one}_i
\Bigr]
+ \delta \, s^{\thr\two}_i
\,, \nn \\
S_i^{\thr\three}
&= \cL_5\, \frac{(\Gamma^i_0)^3}{8}
 - \cL_4\,
   \frac{5}{8}(\Gamma^i_0)^2 \Bigl(\frac{2}{3}\beta_0 + \frac{\gamma_{\thr\,0}^i}{2} \Bigr)
\nn \\ & \quad
+ \cL_3\, \Gamma^i_0 \Bigl[
   - 2\zeta_2 (\Gamma^i_0)^2
   + \frac{\beta_0^2}{3}
   + \Bigl(\frac{5}{3}\beta_0 + \frac{\gamma_{\thr\,0}^i}{2}\Bigr) \frac{\gamma_{\thr\,0}^i}{2}
   + \Gamma_1^i
   + \frac{\Gamma_0^i}{2} s_i^{\thr\one}
\Bigr]
\nn \\ & \quad
+ \cL_2\,
\biggl\{
   (\Gamma^i_0)^2 \Bigl[
      5\zeta_3 \Gamma^i_0
      + 3\zeta_2 (\beta_0 + \gamma^i_{\thr\,0})
   \Bigr]
   - \Bigl(\beta_0 + \frac{3}{4}\gamma^i_{\thr\,0}\Bigr)\Bigl(\beta_0\frac{\gamma^i_{\thr\,0}}{2} + \Gamma^i_1\Bigr)
\nn \\ & \qquad \qquad
   - \frac{(\gamma^i_{\thr\,0})^3}{16}
   - \frac{\Gamma^i_0}{2} \Bigl[
      \beta_1 + \frac{3}{2}\gamma^i_{\thr\, 1}
       + \Bigl(4\beta_0 + \frac{3}{2}\gamma^i_{\thr\,0} \Bigr) s_i^{\thr\one}
   \Bigr]
\biggr\}
\nn \\ & \quad
+ \cL_1\,
\biggl\{
   (\Gamma^i_0)^2 \bigl[4\zeta_4\Gamma^i_0 - \zeta_3(6\beta_0 + 4\gamma^i_{\thr\,0}) \bigr]
   - \zeta_2 \Gamma^i_0 \bigl[
      (3\beta_0 + \gamma^i_{\thr\,0})\gamma^i_{\thr\,0}
      + 4\Gamma^i_1
   \bigr]
\nn \\ & \qquad \qquad
   + \beta_0 \gamma^i_{\thr\,1}
   + \frac{\gamma^i_{\thr\,0}}{2} (\beta_1 + \gamma^i_{\thr\,1})
   + \Gamma^i_2
\nn \\ & \qquad \qquad
   + \Bigl[
      -2\zeta_2(\Gamma^i_0)^2
      + 2\beta_0^2
      + \Bigl(3\beta_0 + \frac{\gamma^i_{\thr\,0}}{2}\Bigr) \frac{\gamma^i_{\thr\,0}}{2} + \Gamma^i_1
   \Bigr] s_i^{\thr\one}
   + \Gamma^i_0\, s_i^{\thr\two}
\biggr\}
\nn \\ & \quad
+ \cL_0\,
\biggl\{
   (\Gamma^i_0)^2 \bigl[
      - \Gamma^i_0 (8 \zeta_2 \zeta_3 - 6 \zeta_5)
      + 2\zeta_4 (\beta_0 - \gamma^i_{\thr\,0})
   \bigr]
   + \zeta_3\Gamma^i_0 \Bigl[
      \Bigl(\beta_0 + \frac{\gamma^i_{\thr\,0}}{2} \Bigr) \gamma^i_{\thr\,0}
\nn \\ & \qquad \qquad
      + 4\Gamma^i_1
   \Bigr]
   + \zeta_2 \bigl(\gamma^i_{\thr\,0} \Gamma^i_1 + \Gamma^i_0 \gamma^i_{\thr\,1} \bigr)
   - \frac{\gamma^i_{\thr\,2}}{2}
   + \Bigl[
      (\Gamma^i_0)^2 2\zeta_3
      + \Gamma^i_0\zeta_2 (2\beta_0 + \gamma^i_{\thr\,0} )
\nn \\ & \qquad \qquad
      - \Bigl(\beta_1 + \frac{\gamma^i_{\thr\,1}}{2}\Bigr)
   \Bigr] s_i^{\thr\one}
   - \Bigl(2\beta_0 + \frac{\gamma^i_{\thr\,0}}{2}\Bigr) s_i^{\thr\two}
\biggr\}
+ \delta \, s^{\thr\three}_i
\,.\end{align}

Consistency of the factorization theorems in \eq{Tau0_fact},
\eqref{eq:factorization_generalized_threshold_qm_qp}, and~\eqref{eq:factorization_soft_threshold_qm_qp}
implies
\begin{equation} \label{eq:consistency_rules_supreme}
2\gamma_B^i(\as) + \gamma_S^i(\as) = 2\gamma_f^i(\as) + \gamma_\thr^i(\as) = \gamma_f^i(\as) + \gamma_B^i(\as)
\,,\end{equation}
because the hard function is the same in all cases.
Here, $\gamma_f^i(\as)$ is the coefficient of $\delta(1-z)$ in the PDF anomalous dimension \eq{DGLAP}.
Solving \eq{consistency_rules_supreme} for $\gamma_\thr^i(\as)$, we find
\begin{equation} \label{eq:threshold_soft_noncusp}
\gamma_\thr^i(\as) = - \gamma_S^i(\as)
\,, \qquad
\gamma_{\thr\,n}^i = - \gamma_{S\,n}^i
\,,\end{equation}
where the soft anomalous dimension coefficients $\gamma_{S\,n}^i$
are given in \eq{gammaS_n}.

The boundary coefficients $s_i^{\thr(n)}$, which are defined as the coefficients of
$\delta(k^-,k^+)$ in \eq{threshold_soft_n3lo},
are given by~\cite{Anastasiou:2014vaa, Li:2014afw}%
\footnote{%
We note that the coefficient of $C_i C_A$ in the two-loop finite term disagrees
with the $\vec{b}_T \to 0$ limit of the fully-differential soft function as
reported in terms of $k^\pm$ and $\vec{b}_T$ in \refcite{Li:2011zp}. This color
structure only enters at two loops and thus is unaffected by non-Abelian
exponentiation. We were unable to resolve this difference, but tend to attribute
it to a typographical error in \refcite{Li:2011zp} because \refscite{Li:2016axz,
Li:2016ctv} agreed with the pure position-space result of \refcite{Li:2011zp} in
terms of $b^\pm$ and $\vec{b}_T$.}
\begin{align} \label{eq:threshold_soft_finite_terms}
s^{\thr\one}_i &= -C_i\, 2\zeta_2
\,, \nn \\
s^{\thr\two}_i
&= C_i \biggl[
   C_i\, 21\zeta_4 + C_A \Bigl(\frac{208}{27} - 4 \zeta_2 - 10 \zeta_4 \Bigl) + \beta_0 \Bigl(\frac{164}{27} - 5 \zeta_2 - \frac{10 \zeta_3}{3}\Bigr)
\biggr]
\,, \nn \\
s^{\thr\three}_i
&= C_i \biggl[
   C_i^2 \Bigl(\frac{640}{3}\zeta_3^2 - \frac{499}{6}\zeta_6 \Bigr)
   + C_i C_A \Bigl(
      -\frac{416}{27}\zeta_2 - \frac{512}{9}\zeta_3 + \frac{188}{3}\zeta_4
      + 224 \zeta_3^2 - 77 \zeta_6
   \Bigr)
\nn \\ & \quad\hspace{4ex}
   + C_i \beta_0 \Bigl(
      -\frac{328}{27}\zeta_2 - \frac{448}{9}\zeta_3 + \frac{235}{3}\zeta_4
      + \frac{308}{3} \zeta_2 \zeta_3 - 64 \zeta_5
   \Bigr)
\nn \\ & \quad\hspace{4ex}
   + C_A^2 \Bigl(
      \frac{115895}{324} - \frac{45239}{486}\zeta_2 - \frac{23396}{81}\zeta_3
      - \frac{334}{3}\zeta_4  + 240 \zeta_2 \zeta_3 - 224 \zeta_5
      + \frac{1072}{9}\zeta_3^2 + \frac{4348}{27}\zeta_6
   \Bigr)
\nn \\ & \quad\hspace{4ex}
   + C_A \beta_0 \Bigl(
      - \frac{363851}{2916} + \frac{1043}{486}\zeta_2 - \frac{140}{81}\zeta_3
      + \frac{230}{9}\zeta_4 - \frac{164}{3}\zeta_2 \zeta_3 + \frac{632}{9}\zeta_5
   \Bigr)
\nn \\ & \quad\hspace{4ex}
   + \beta_0^2 \Bigl(
      - \frac{64}{729} - \frac{34}{3} \zeta_2 - \frac{20}{27}\zeta_3
      + \frac{31}{3}\zeta_4
   \Bigr)
\nn \\ & \quad\hspace{4ex}
   + \beta_1 \Bigl(
      \frac{42727}{972} - \frac{275}{18}\zeta_2 - \frac{1636}{81}\zeta_3
      - \frac{76}{9} \zeta_4 + \frac{40}{3} \zeta_2 \zeta_3 - \frac{112}{9}\zeta_5
   \Bigr)
\biggr]
\,.\end{align}
We have also checked that inserting the above coefficients into \eq{threshold_soft_n3lo}
and expanding against the Drell-Yan hard function,
we reproduce the three-loop soft-virtual partonic cross section in
\refscite{Ravindran:2006bu, Ahmed:2014uya}
in terms of $1 - z_a = k^-/(Qe^{+Y})$ and $1 - z_b = k^+/(Qe^{-Y})$.

\subsection{Extraction method}
\label{app:threshold_soft_method}

The double-differential threshold soft function
depends on the total lightcone momentum components $k^\pm$ of the soft hadronic final state.
Equivalently, its Fourier transform
\begin{align} \label{eq:threshold_soft_fourier_transform}
\hat{S}_i^\thr(b^+, b^-, \mu)
= \int \! \df k^- \df k^+ \, e^{+\img (k^-b^+ + k^+b^-)/2} \, S_i^\thr(k^-, k^+, \mu)
\,,\end{align}
depends on the time-like separation $(b^-n^\mu + b^+\bar{n}^\mu)/2$
between the Wilson lines in the soft matrix element.

Importantly, $\hat{S}_i^\thr(b^+, b^-, \mu)$ only depends on the product $b^+ b^-$
by the rescaling relation \eq{threshold_soft_rescaling},
and thus only depends on $b^+ b^- \mu^2$ by dimensional analysis.
On the other hand, the dependence on $\mu$ is fully predicted by the RGE \eq{threshold_soft_rge},
which in position space reads
\begin{align} \label{eq:threshold_soft_rge_conjugate_space}
\mu \frac{\df}{\df \mu} \hat{S}_i^\thr(b^+, b^-, \mu)
= \Bigl\{ 2\GammaC^i[\as(\mu)] \, L_\thr(b^+, b^-, \mu) + \gamma^i_\thr[\as(\mu)] \Bigr\}\,
\hat{S}_i^\thr(b^+, b^-, \mu)
\,.\end{align}
This implies that at any given order in perturbation theory,
$\hat{S}_i^\thr(b^+, b^-, \mu)$ is a polynomial in
\begin{equation}
L_\thr(b^+, b^-, \mu) \equiv \ln \Bigl( -\frac{b^+ b^- \mu^2 e^{2\gamma_E}}{4} - \img 0 \Bigr)
\,.\end{equation}
The relevant Fourier transforms between $L_\thr^n$ and $\cL_n(k^-, k^+, \mu)$
follow from the one-dimensional Fourier transforms in appendix~B of \refcite{Ebert:2016gcn},
accounting for the relative factors of $-1/2$ in the Fourier exponent in \eq{threshold_soft_fourier_transform}.

A factorization analogous to \eq{factorization_soft_threshold_qm_qp}
holds for the inclusive cross section $\df \sigma / \df Q^2$,
where the corresponding inclusive threshold soft function $S_i^\thr(k^0, \mu)$
only depends on the total energy $k^0$ of soft radiation.
In particular, $S_i^\thr(k^0, \mu)$ is the process-independent soft contribution
to the inclusive partonic cross section $\sigma_{ab}(z)$ in the soft-virtual limit $z \to 1$,
where $1-z = 2k^0/Q$.
In position space, the inclusive threshold soft function $\hat{S}_i^\thr(b^0, \mu)$ is defined
in terms of Wilson lines separated by $b^0 (n^\mu + \bar{n}^\mu)/2$, i.e., strictly along the time axis.
This is a special case of \eq{threshold_soft_fourier_transform},
so the two position-space threshold soft functions are simply related by
\begin{equation} \label{eq:threshold_soft_relate_incl}
\hat{S}_i^\thr(b^0, b^0, \mu) = \hat{S}_i^\thr(b^0, \mu)
\,.\end{equation}
This is of course equivalent to integrating over the longitudinal momentum $k^3$ of soft radiation.
We stress that \eq{threshold_soft_relate_incl} cannot be used to approximate
$\hat{S}_i^\thr(b^+, b^-, \mu)$ by taking $b^+ = b^-$ in general.
This is because in \eq{factorization_soft_threshold_qm_qp} the $k^+$ and $k^-$ dependences
are separately convolved with the PDFs and thus the rescaling property
\eq{threshold_soft_rescaling} is lost at the level of the cross section.
See also Appendix~D of \refcite{Lustermans:2019cau} for further discussion
of this point.

Inserting \eq{threshold_soft_relate_incl} into \eq{threshold_soft_rge_conjugate_space}
implies that both threshold soft functions have the same noncusp anomalous dimension
given by \eq{threshold_soft_noncusp}.
Moreover, the position-space boundary coefficients of the double-differential soft function
at $L_\thr = 0$, i.e., at $\mu = \mu_\ast \equiv +\img 2 e^{-\gamma_E}/b^0$,
are equal to the inclusive ones at the same scale.
Hence, the double-differential threshold
soft function can be constructed from the inclusive one.

The inclusive threshold soft function was calculated to three loops in
\refscite{Anastasiou:2014vaa, Li:2014afw}.
Here we use the results of \refcite{Li:2014afw}, where the three-loop soft function
for $i = g$ is reported in exponentiated form,
\begin{align} \label{eq:threshold_soft_relate_incl_finite_terms}
\hat{S}_i^\thr( b^0, \mu_\ast )
= \exp\biggl\{ \frac{C_i}{C_A} \Bigl[
    \frac{\as(\mu_\ast)}{4\pi}   c_{1\,\text{\refcite{Li:2014afw}}}^H
   \!+ \frac{\as^2(\mu_\ast)}{(4\pi)^2} \Delta c_{2\,\text{\refcite{Li:2014afw}}}^H
   \!+ \frac{\as^3(\mu_\ast)}{(4\pi)^3} \Delta c_{3\,\text{\refcite{Li:2014afw}}}^H
   \Bigr]
   \!+\! \ord{\as^4}
\biggr\}
.\end{align}
We have also exploited Casimir scaling to three loops to restore the dependence on $C_i$.
Comparing \eq{threshold_soft_relate_incl_finite_terms}
to the position-space solution of \eq{threshold_soft_rge_conjugate_space} at $L_\thr = 0$,
we obtain \eq{threshold_soft_finite_terms} for the momentum-space boundary coefficients
after performing the inverse Fourier transform.

\section{Collinear-soft function for the exponential regulator}
\label{app:csoft_function}

In this appendix we derive the all-order expression for the collinear-soft function
using the exponential regulator, which leads to
\eq{qT_beam_eikonal_finite_terms_all_order} in the main text.

We start by defining the complete Fourier transform of the fully-differential
threshold soft function
\begin{equation}
\hat{S}^\thr_i(b^+, b^-, b_T, \mu)
= \int\!\df^4 k\, e^{+\img b\cdot k}\, S^\thr_i(k^-, k^+, k_T, \mu)
\,,\end{equation}
where $b^\mu = (b^+, b^-, \vec{b}_T)$ is the four-vector Fourier conjugate of
$k^\mu = (k^+, k^-, \vec{k}_T)$ with $b \cdot k = b^+ k^-/2 + b^- k^+/2 - \vec{b}_T \cdot \vec{k}_T$.
Correspondingly, we define the Fourier transform of $\tilde{\cS}_i(k^\pm, b_T, \mu, \nu)$
with respect to its lightcone momentum argument $k^\pm$ as
\begin{align}
\hat{\cS}_i(b^+, b_T, \mu, \nu)
= \int \! \df k^- \, e^{+\img k^- b^+/2} \, \tilde{\cS}_i(k^-, b_T, \mu, \nu)
\,,\end{align}
and analogously for $b^- \leftrightarrow b^+$ and $k^+ \leftrightarrow k^-$.
Fully in position space, the consistency relation \eq{refactorization_soft} reads
\begin{equation} \label{eq:refactorization_soft_position_space}
\hat{S}_i^\thr(b^+, b^-, b_T, \mu)
= \hat{\cS}_i(b^+, b_T, \mu, \nu) \, \hat{\cS}_i(b^-, b_T, \mu, \nu) \, \tilde{S}_i(b_T, \mu, \nu) \,
\Bigl[ 1 + \ORd{\frac{b^+ b^-}{b_T^2}} \Bigr]
\,.\end{equation}

In the exponential regulator scheme,
the regulated $q_T$ soft function is \emph{defined} as~\cite{Li:2016axz, Li:2016ctv}%
\footnote{Comparing eq.~(2) in \refcite{Li:2016ctv} to eq.~(33) in \refcite{Li:2016axz}
suggests that the latter has a spurious factor of $2$ in the denominator, noting that their $\tau = 1/\nu$.}
\begin{equation} \label{eq:exponential_regularization}
\tilde{S}_i(b_T, \mu, \nu')
= \lim_{\nu' \to \infty} \hat{S}^\thr_i\Bigl( \frac{\img b_0}{\nu'}, \frac{\img b_0}{\nu'}, b_T, \mu \Bigr)
\,,\end{equation}
where we use $\nu'$ to distinguish it from the scale at which we later wish to evaluate the collinear-soft function.
The prescription for taking the limit is to keep all nonvanishing terms.
In particular, a logarithmic dependence of the right-hand side on $\nu'$ is to be kept.
Inserting \eq{refactorization_soft_position_space}, we have
\begin{align}
\tilde{S}_i(b_T, \mu, \nu')
&= \lim_{\nu' \to \infty} \Bigl[ \hat{\cS}_i\Bigl(\frac{\img b_0}{\nu'}, b_T, \mu, \nu\Bigr) \,
\hat{\cS}_i\Bigl( \frac{\img b_0}{\nu'}, b_T, \mu, \nu\Bigr) \,
\tilde{S}_i(b_T, \mu, \nu)
+ \ORd{\frac{1}{\nu'^2 b_T^2}} \Bigr]
\nn \\
&= \tilde{S}_i(b_T, \mu, \nu) \lim_{\nu' \to \infty} \Bigl[ \hat{\cS}_i\Bigl(\frac{\img b_0}{\nu'}, b_T, \mu, \nu\Bigr) \, \hat{\cS}_i\Bigl( \frac{\img b_0}{\nu'}, b_T, \mu, \nu\Bigr) \Bigr]
\nn \\
&= \tilde{S}_i(b_T, \mu, \nu) \, \hat{\cS}_i\Bigl(\frac{\img b_0}{\nu'}, b_T, \mu, \nu\Bigr) \,
\hat{\cS}_i\Bigl( \frac{\img b_0}{\nu'}, b_T, \mu, \nu\Bigr)
\,.\end{align}
In the second line we moved the $q_T$ soft function out of the limit,
since it does not depend on $\nu'$, and dropped the power corrections.
On the third line we used that all dependence of the $\hat{\cS}_i$ on $\nu'$ is logarithmic,
so the limit is trivial.
Because the exponential regulator is symmetric under an interchange of collinear-soft directions, we find
\begin{align}
\hat{\cS}_i^2\Bigl(\frac{\img b_0}{\nu'}, b_T, \mu, \nu\Bigr)
= \frac{\tilde{S}_i(b_T, \mu, \nu')}{\tilde{S}_i(b_T, \mu, \nu)}
= \exp \Bigl[ \tilde{\gamma}_\nu^i(b_T, \mu) \ln \frac{\nu'}{\nu} \Bigr]
\,,\end{align}
where the second equality follows from solving the rapidity RGE of the soft function between $\nu$ and $\nu'$ at fixed $\mu$.
Assuming we are dealing with the $n_a$-collinear-soft function that depends on $b^+$,
we can analytically continue back to $\nu' = \img b_0/b^+ = 2 \img /(b^+ e^{\gamma_E})$, leaving
\begin{align}
\hat{\cS}_i(b^+, b_T, \mu, \nu)
= \exp\Bigl[ -\frac{1}{2} \tilde{\gamma}_\nu^i(b_T, \mu) \ln (-\img b^+ \nu e^{\gamma_E}/2) \Bigr]
\,.\end{align}
Evaluating the inverse Fourier transform using \eq{cVa_fourier_transform},
we find the following all-order relation for the momentum-space $n_a$-collinear-soft function
in the exponential regulator scheme,
\begin{equation}
\tilde{\cS}_i(k^-, b_T, \mu, \nu) = \cV_{\tilde{\gamma}_\nu^i(b_T, \mu)/2}(k^-, \nu)
\,,\end{equation}
and identically for the $n_b$-collinear one as a function of $k^+$.
In other words, the collinear-soft function in the exponential regulator scheme
is simply given by the rapidity RG evolution between its canonical rapidity scale $\nu_\cS \sim k^-$ and $\nu$,
with trivial boundary condition at $\nu_\cS$.
Inserting this result into \eq{qT_beam_eikonal_limit_partonic} leads to \eq{qT_beam_eikonal_finite_terms_all_order} in the main text.

\section{Perturbative ingredients}
\label{app:pertubativeingredients}

\subsection{Anomalous dimensions}
\label{app:anom_dims}

We expand the QCD $\beta$ function as
\begin{equation}
\mu \frac{\df \as(\mu)}{\df \mu}
= \beta[\as(\mu)]
\,, \qquad
\beta(\as) = -2\as \sum_{n = 0}^\infty \beta_n \Bigl( \frac{\as}{4\pi} \Bigr)^{n+1}
\,.\end{equation}
The coefficients up to three loops in the \MSbar~scheme are~\cite{Tarasov:1980au, Larin:1993tp}
\begin{align} \label{eq:beta_n}
\beta_0
&= \frac{11}{3}\,C_A - \frac{4}{3}\,T_F\,n_f
\,, \\
\beta_1
&= \frac{34}{3}\,C_A^2 - 2T_F\,n_f \Bigl(\frac{10}{3}\, C_A + 2 C_F\Bigr)
\,, \nn \\
\beta_2
&= \frac{2857}{54}\,C_A^3
 + 2T_F\,n_f \Bigl(- \frac{1415}{54}\,C_A^2 - \frac{205}{18}\,C_F C_A + C_F^2  \Bigr)
 + 4T_F^2\,n_f^2 \Bigl(\frac{79}{54}\, C_A  + \frac{11}{9}\, C_F \Bigr)
\nn \,.\end{align}
The cusp anomalous dimension and all noncusp anomalous dimensions are expanded as
\begin{equation}
\GammaC^i(\as) = \sum_{n = 0}^\infty \Gamma_n^i \Bigl( \frac{\as}{4\pi} \Bigr)^{n+1}
\,,\qquad
\gamma(\as) = \sum_{n = 0}^\infty \gamma_n \Bigl( \frac{\as}{4\pi} \Bigr)^{n+1}
\,.\end{equation}
The coefficients of the $\overline{\mathrm{MS}}$ cusp anomalous dimension to three loops are~\cite{Korchemsky:1987wg, Moch:2004pa, Vogt:2004mw}
\begin{align} \label{eq:Gamma_n}
\Gamma_0^i &= 4 C_i
\,,\nn\\
\Gamma_1^i
&= 4 C_i \biggl[
   C_A \Bigl(\frac{67}{9} - 2\zeta_2 \Bigr)
   - \frac{20}{9}\, T_F\, n_f
\biggr]
\,,\nn\\
\Gamma_2^i
&= 4 C_i \biggl\{
   C_A^2 \Bigl(\frac{245}{6} - \frac{268}{9}\zeta_2 + \frac{22}{3}\zeta_3 + 22\zeta_4 \Bigr)
\nn \\ & \qquad\quad
   + 2T_F\,n_f \biggl[
      C_A \Bigl(-\frac{209}{27} + \frac{40}{9}\zeta_2 - \frac{28}{3}\zeta_3 \Bigr)
      + C_F \Bigl(-\frac{55}{6} + 8\zeta_3 \Bigr)
   \biggr]
   - \frac{16}{27}\, T_F^2\, n_f^2
\biggr\}
\,,\end{align}
where $C_i=C_F$ for $i=q$ and $C_i=C_A$ for $i=g$.

\subsection{Ingredients for \texorpdfstring{$\Tau_0$}{Tau0}}
\label{app:ingredients_Tau0}

The quark beam function noncusp anomalous dimension coefficients to three loops are~\cite{Stewart:2010qs}
\begin{align} \label{eq:gammaBq_n}
\gamma_{B\,0}^q &= 6 C_F
\,, \nn \\
\gamma_{B\,1}^q
&= 2C_F \biggl[
   C_A \Bigl(\frac{73}{9} - 40 \zeta_3\Bigr)
   + C_F \Bigl(\frac{3}{2} - 12\zeta_2 + 24 \zeta_3 \Bigr)
   + \beta_0 \Bigl(\frac{121}{18} + 2\zeta_2 \Bigr)
\biggr]
\,, \nn \\
\gamma_{B\,2}^q
&= 2 C_F \biggl[
   C_A^2 \Bigl(
      \frac{52019}{162} - \frac{1682}{27}\zeta_2 - \frac{2056}{9}\zeta_3
      - \frac{820}{3}\zeta_4 + \frac{176}{3}\zeta_2\zeta_3 + 232 \zeta_5
   \Bigr)
\nn\\ & \quad\hspace{6ex}
   + C_A C_F \Bigl(
      \frac{151}{4} - \frac{410}{3}\zeta_2 + \frac{844}{3}\zeta_3
      - \frac{494}{3}\zeta_4 + 16\zeta_2\zeta_3 + 120 \zeta_5
   \Bigr)
\nn\\ & \quad\hspace{6ex}
   + C_F^2 \Bigl(
      \frac{29}{2} + 18\zeta_2 + 68 \zeta_3 + 144\zeta_4
      - 32\zeta_2\zeta_3 - 240 \zeta_5
   \Bigr)
\nn\\ & \quad\hspace{6ex}
   + C_A\beta_0 \Bigl(
      -\frac{7739}{54} + \frac{650}{27}\zeta_2
      - \frac{1276}{9}\zeta_3 + \frac{617}{3}\zeta_4
   \Bigr)
\nn\\ & \quad\hspace{6ex}
   + \beta_0^2 \Bigl(
      -\frac{3457}{324} + \frac{10}{3}\zeta_2 + \frac{16}{3}\zeta_3
   \Bigr)
   + \beta_1 \Bigl(
      \frac{1166}{27} - \frac{16}{3}\zeta_2 + \frac{52}{9}\zeta_3
       - \frac{82}{3}\zeta_4
   \Bigr)
\biggr]
\,.\end{align}
They have been confirmed recently by an explicit three-loop calculation of the jet
function~\cite{Bruser:2018rad}, see also \refcite{Banerjee:2018ozf}.

The gluon beam function noncusp anomalous dimension coefficients to three loops are~\cite{Berger:2010xi}
\begin{align} \label{eq:gammaBg_n}
\gamma_{B\,0}^g &= 2 \beta_0
\,, \nn \\
\gamma_{B\,1}^g
&= 2C_A \biggl[
   C_A \Bigl(\frac{91}{9} - 16\zeta_3 \Bigr)
   + \beta_0 \Bigl(\frac{47}{9} - 2\zeta_2 \Bigr)
\biggr]
+ 2\beta_1
\,, \nn \\
\gamma_{B\,2}^g
&= 2C_A \biggl[
   C_A^2 \Bigl(
      \frac{49373}{162} - \frac{944}{27}\zeta_2 - \frac{2260}{9}\zeta_3
      - 144\zeta_4 + \frac{128}{3}\zeta_2\zeta_3 + 112\zeta_5
   \Bigr)
\nn\\ & \quad\hspace{6ex}
   + C_A\, \beta_0 \Bigl(
      - \frac{6173}{54} - \frac{376}{27}\zeta_2 + \frac{140}{9}\zeta_3 + 117\zeta_4
   \Bigr)
   + \beta_0^2 \Bigl(
      -\frac{493}{81} - \frac{10}{3}\zeta_2 + \frac{28}{3}\zeta_3
   \Bigr)
\nn\\ & \quad\hspace{6ex}
   + \beta_1 \Bigl(
      \frac{1765}{54} - 2\zeta_2 - \frac{152}{9}\zeta_3 - 8\zeta_4
   \Bigr)
\biggr]
+ 2 \beta_2
\,. \end{align}

The soft noncusp anomalous dimension coefficients to three loops
follow from consistency by $\gamma_S^i(\as) = - 2\gamma_B^i(\as) -4\gamma_C^i(\as)$,
where the $\gamma_C^i(\as)$ are taken from \refcite{Ebert:2017uel}. They are the
hard noncusp anomalous dimensions and are known up to three loops from the quark
and gluon form factors~\cite{Kramer:1986sg, Matsuura:1987wt, Matsuura:1988sm, Harlander:2000mg, Gehrmann:2005pd, Moch:2005id, Moch:2005tm}. We obtain,
\begin{align} \label{eq:gammaS_n}
\gamma_{S\,0}^i &= 0
\,, \nn \\
\gamma_{S\,1}^i
&= 2C_i \biggl[
   C_A \Bigl(-\frac{64}{9} + 28\zeta_3 \Bigr)
   + \beta_0 \Bigl(-\frac{56}{9} + 2\zeta_2 \Bigr)
\biggr]
\,, \nn \\
\gamma_{S\,2}^i
&= 2C_i \biggl[
   C_A^2 \Bigl(
      -\frac{37871}{162} + \frac{620}{27}\zeta_2 + \frac{2548}{9}\zeta_3
       + 144\zeta_4 - \frac{176}{3}\zeta_2\zeta_3 - 192\zeta_5
   \Bigr)
\nn\\ & \quad\hspace{5ex}
   + C_A\, \beta_0 \Bigl(
      \frac{4697}{54} + \frac{484}{27}\zeta_2 + \frac{220}{9}\zeta_3 - 112\zeta_4
   \Bigr)
   + \beta_0^2 \Bigl(
      \frac{520}{81} + \frac{10}{3}\zeta_2 - \frac{28}{3}\zeta_3
   \Bigr)
\nn\\ & \quad\hspace{5ex}
   + \beta_1 \Bigl(
     - \frac{1711}{54} + 2\zeta_2 + \frac{152}{9}\zeta_3 + 8\zeta_4
   \Bigr)
\biggr]
\,. \end{align}
Finally, the soft function coefficients to two loops are~\cite{Stewart:2009yx, Kelley:2011ng, Monni:2011gb, Hornig:2011iu, Kang:2015moa}
\begin{align} \label{eq:Tau0_soft_n}
s_i^\zero &= 1
\,, \nn \\
s_i^\one &= C_i\, 2\zeta_2
\,, \nn \\
s_i^\two
&= C_i
\biggl[
   - C_i\, 27 \zeta_4
   + C_A \Bigl(-\frac{640}{27} + 8\zeta_2 + 44 \zeta_4 \Bigr)
   + \beta_0 \Bigl(-\frac{20}{27} - \frac{37}{3}\zeta_2 + \frac{58}{3} \zeta_3 \Bigr)
\biggr]
\,.\end{align}
The three-loop coefficient is still unknown.

\subsection{Ingredients for \texorpdfstring{$q_T$}{qT}}
\label{app:ingredients_qt}

In the exponential regulator, the noncusp anomalous dimension $\tilde\gamma_S^i$
of the $q_T$ soft function is equal to that of the threshold soft function $\gamma_\thr^i$,
which in turn is the negative of the $\Tau_0$ soft anomalous dimension $\gamma_S^i$.
As a result, we have
\begin{alignat}{9} \label{eq:gammaS_qt_n}
\tilde \gamma_S^i(\as) &= \gamma_{\rm thr}^i(\as) = -\gamma_S^i(\as)
\,, \qquad &
\tilde\gamma_{S\,n}^i &= - \gamma_{S\,n}^i
\,, \nn \\
\tilde\gamma_B^i(\as) &= \gamma_B^i(\as) + \gamma_S^i(\as)
\,, \qquad &
\tilde\gamma_{B\,n}^i &= \gamma_{B\,n}^i + \gamma_{S\,n}^i
\,.\end{alignat}
The result for $\tilde\gamma^i_B$ follows from RG consistency and the fact that the hard anomalous
dimension is the same for $q_T$ and $\Tau_0$.
The $\gamma_{S\,n}^i$ and $\gamma_{B\, n}^i$ coefficients are given in
\eqss{gammaBq_n}{gammaBg_n}{gammaS_n} above.

The rapidity anomalous dimensions coefficients, which enter the fixed-order
expansion of $\tilde\gamma_\nu^i$ in \eq{gammaNu_FO}, are known up to three
loops~\cite{Luebbert:2016itl, Li:2016ctv, Vladimirov:2016dll}. They are given by
\begin{align} \label{eq:gammaNuConstants}
\tilde\gamma_{\nu\,0}^i
&= 0
\,,\nn\\
\tilde\gamma_{\nu\,1}^i
&= 2 C_i \biggl[ C_A \Bigl( -\frac{64}{9} + 28\zeta_3 \Bigr) - \frac{56}{9} \beta_0 \biggr]
\,,\nn\\
\tilde\gamma_{\nu\,2}^i
&= 2 C_i \biggl[
   C_A^2 \Bigl( -\frac{37871}{162} + \frac{620}{27} \zeta_2 + \frac{2548}{9} \zeta_3 + 144 \zeta_4- \frac{176}{3} \zeta_2 \zeta_3 - 192 \zeta_5 \Bigr)
\nn \\ & \qquad\quad
   + C_A \beta_0 \Bigl( \frac{3865}{54} + \frac{412}{27} \zeta_2 + \frac{220}{9} \zeta_3 - 50 \zeta_4 \Bigr)
\nn \\ & \qquad\quad
   + \beta_0^2 \Bigl( - \frac{464}{81} - 8 \zeta_3 \Bigr)
   + \beta_1 \Bigl( -\frac{1711}{54}+ \frac{152}{9} \zeta_3  + 8 \zeta_4 \Bigr)
\biggr]
\,.\end{align}
The soft function coefficients are also known up to three loops~\cite{Luebbert:2016itl, Li:2016ctv},
and are given by
\begin{align}
\tilde s_i^\zero &= 1
\,,\nn\\
\tilde s_i^\one &= -C_i\, 2\zeta_2
\,,\nn\\
\tilde s_i^\two
&= C_i \biggl[C_i\, 5 \zeta_4 +
   C_A \Bigl( \frac{208}{27} - 4 \zeta_2 + 10 \zeta_4 \Bigr)
   + \beta_0 \Bigl( \frac{164}{27} - 5 \zeta_2 - \frac{14}{3} \zeta_3 \Bigr)
\biggr]
\,,\nn\\
\tilde s_i^\three
&= C_i \biggl[
   - C_i^2\, \frac{35}{6} \zeta_6
   + C_i C_A \Bigl( - \frac{416}{27}\zeta_2 + 20 \zeta_4  - 35 \zeta_6 \Bigr)
   + C_i \beta_0 \Bigl(-\frac{328}{27}\zeta_2 + 25 \zeta_4 + \frac{28}{3} \zeta_2 \zeta_3 \Bigr)
\nn \\ & \quad\hspace{4ex}
   + C_A^2 \Bigl(
      \frac{115895}{324} - \frac{51071}{486}\zeta_2
      - \frac{23396}{81}\zeta_3 - 58 \zeta_4
      + 240 \zeta_2 \zeta_3 - 224 \zeta_5
      + \frac{928}{9}\zeta_3^2 - \frac{3086}{27}\zeta_6
   \Bigr)
\nn \\ & \quad\hspace{4ex}
   + C_A \beta_0 \Bigl(
      -\frac{363851}{2916} + \frac{2987}{486}\zeta_2 - \frac{428}{81}\zeta_3
      + \frac{830}{9}\zeta_4 - \frac{220}{3} \zeta_2 \zeta_3 + \frac{1388}{9}\zeta_5
   \Bigr)
\nn \\ & \quad\hspace{4ex}
   + \beta_0^2 \Bigl(
      - \frac{64}{729} - \frac{34}{3}\zeta_2
      - \frac{140}{27} \zeta_3 - \frac{11}{3}\zeta_4
   \Bigr)
\nn \\ & \quad\hspace{4ex}
   + \beta_1 \Bigl(
      \frac{42727}{972} - \frac{275}{18}\zeta_2 - \frac{1744}{81} \zeta_3
      - \frac{76}{9}\zeta_4 + \frac{40}{3} \zeta_2 \zeta_3 - \frac{112}{9}\zeta_5
   \Bigr)
\biggr]
\,.\end{align}

\subsection{Mellin kernels and splitting functions}
\label{app:DGLAP}

We decompose the flavor dependence of a generic Mellin-convolution kernel $K_{ij}(z)$ as
\begin{alignat}{5} \label{eq:flavor_decomposition}
K_{q_i q_j}(z) = K_{\bq_i \bq_j}(z) &= \delta_{ij} K_{qqV}(z) + K_{qqS}(z) + K_{qq\Delta S}(z)
\,,\nn\\
K_{q_i \bq_j}(z) = K_{\bq_i q_j}(z) &= \delta_{ij} K_{q\bq V}(z) + K_{qqS}(z) - K_{qq\Delta S}(z)
\,,\nn\\
K_{q_i g}(z) = K_{\bq_i g}(z) &= K_{q g}(z)
\,, \nn \\[0.4em]
K_{gg}(z) &= K_{gg}(z)
\,,\nn\\
K_{g q_i}(z) = K_{g \bq_i}(z) &= K_{g q}(z)
\,.\end{alignat}
This decomposition is sufficient and unique to all orders
by the flavor and charge symmetries of QCD.
The $K_{qqV}$ and $K_{gg}$ contributions are already present at tree level,
the $K_{qg}$ and $K_{gq}$ channels start at one loop,
the $K_{qqS}$ and $K_{q\bq V}$ channels open up at two loops,
and the $K_{qq\Delta S}$ channel only receives contributions from topologies
at three loops and beyond. This decomposition also makes it straightforward to
evaluate and iterate sums over intermediate partons.
For example, for the convolution of two generic kernels $(K K')_{ij}(z)$, we have
\begin{align} \label{eq:flavor_sums}
(K K')_{gg}(z) &= (K_{gg} K'_{gg})(z)  + 2n_f (K_{gq} K'_{qg})(z)
\,, \nn \\
(K K')_{qg}(z) &= \bigl[ \bigl( K_{qqV} + K_{q\bq V} + 2n_f K_{qqS} \bigr) K'_{qg} \bigr](z)
   + (K_{qg} \, K'_{gg})(z)
\,, \nn \\
(K K')_{gq}(z) &= \bigl[ K_{gq} \bigl( K'_{qqV} + K'_{q\bq V} + 2n_f K'_{qqS} \bigr) \bigr](z)
   + (K_{gg} \, K'_{gq})(z)
\,, \nn \\
(K K')_{qqV}(z) &= (K_{qqV} K'_{qqV})(z) + (K_{q\bq V} K'_{q\bq V})(z)
\,, \nn \\
(K K')_{q\bq V}(z) &= (K_{qqV} K'_{q\bq V})(z) + (K_{q\bq V} K'_{qqV})(z)
\,, \nn \\
(K K')_{qqS}(z) &= \bigl[ K_{qqS} \bigl( K'_{qqV} + K'_{q\bq V}\bigr) \bigr](z)
   + \bigl[ \bigl( K_{qqV} + K_{q\bq V}\bigr) K'_{qqS} \bigr](z)
   \nn\\&\quad
   + 2 n_f \bigl(K_{qqS} \, K'_{qqS}\bigr)(z) + \bigl(K_{qg} \, K'_{gq}\bigr)(z)
\,, \nn \\
(K K')_{qq\Delta S}(z) &=
   \bigl[ K_{qq\Delta S} (K'_{qqV} - K'_{q\bq V})\bigr](z)
   +\bigl[ (K_{qqV} - K_{q\bq V}) K'_{qq\Delta S} \bigr](z)
   \nn\\&\quad
   + 2 n_f \bigl( K_{qq\Delta S} K'_{qq\Delta S}\bigr)(z)
\,,\end{align}
where $n_f$ is the number of active flavors, and the outer brackets on the right-hand side
indicate Mellin convolutions \emph{without} flavor sums.

The DGLAP splitting functions are defined as the anomalous dimension of the PDFs,
\begin{equation} \label{eq:DGLAP}
\mu \frac{\df}{\df\mu} f_i(x,\mu) = 2 \sum_j \int\! \frac{\df z}{z}\, P_{ij}(z,\mu)\, f_j\Bigl(\frac{x}{z}, \mu\Bigr)
\,.\end{equation}
We perturbatively expand them in powers of $\as/4\pi$, see \eq{gamma_expansion},
and decompose their flavor dependence as in \eq{flavor_decomposition}.
The DGLAP kernels have been calculated at three loops in \refscite{Moch:2004pa,Vogt:2004mw}.
Denoting the results of \refscite{Moch:2004pa, Vogt:2004mw} by a calligraphic $\cP$
to distinguish them from our $P_{ij}^{(n)}$, we can relate the two notations by
\begin{alignat}{3}
P_{qqV}^{(n)}(z) &= \frac{1}{2} \Bigl[ \cP_{ns}^{(n)+}(z) + \cP_{ns}^{(n)-}(z) \Bigr]
\,, \qquad &
P_{gg}^{(n)}(z) &= \cP_{gg}^{(n)}(z)
\,,\nn\\
P_{q\bq V}^{(n)}(z) &= \frac{1}{2} \Bigl[ \cP_{ns}^{(n)+}(z) - \cP_{ns}^{(n)-}(z) \Bigr]
\,, \qquad &
P_{g q}^{(n)}(z) &= \cP_{gq}^{(n)}(z)
\,,\nn\\
P_{qqS}^{(n)}(z) &= \frac{1}{2 n_f} \cP_{ps}^{(n)}(z)
\,, \qquad &
P_{q g}^{(n)}(z) &= \frac{1}{2 n_f} \cP_{qg}^{(n)}(z)
\,,\nn\\
P_{qq \Delta S}^{(n)}(z) &= \frac{1}{2 n_f} \cP_{ns}^{(n)s}(z)
\,.\end{alignat}

\addcontentsline{toc}{section}{References}
\bibliographystyle{jhep}
\bibliography{refs}

\end{document}